\documentclass[preprint,3p,12pt]{elsarticle}
\usepackage{mathrsfs}
\usepackage{amsmath}
\usepackage{stmaryrd}
\usepackage{bbding}
\usepackage{dcolumn}
\usepackage{graphicx}
\usepackage{amsfonts}
\usepackage{amssymb}
\usepackage{psfrag}
\usepackage{wrapfig}
\usepackage{subfigure}
\usepackage{makeidx}
\usepackage{bm}
\usepackage{epsf}
\usepackage{epsfig}
\usepackage{setspace}
\usepackage{graphicx}
\usepackage{epstopdf}
\usepackage{psfrag}
\usepackage{subfigure}
\usepackage{color}

\begin{document}
\title{An acoustic and shock wave capturing compact high-order gas-kinetic scheme with spectral-like resolution}

\author[HKUST1]{Fengxiang Zhao}
\ead{fzhaoac@connect.ust.hk}

\author[HKUST2]{Xing Ji}
\ead{xjiad@connect.ust.hk}

\author[HKUST1]{Wei Shyy}
\ead{weishyy@ust.hk}

\author[HKUST1,HKUST2,HKUST3]{Kun Xu\corref{cor}}
\ead{makxu@ust.hk}

\address[HKUST1]{Department of Mechanical and Aerospace Engineering, Hong Kong University of Science and Technology, Clear Water Bay, Kowloon, HongKong}
\address[HKUST2]{Department of Mathematics, Hong Kong University of Science and Technology, Clear Water Bay, Kowloon, HongKong}
\address[HKUST3]{Shenzhen Research Institute, Hong Kong University of Science and Technology, Shenzhen, China}
\cortext[cor]{Corresponding author}

\begin{abstract}
In this paper, a compact high-order gas-kinetic scheme (GKS) with spectral resolution will be presented and used in the simulation of acoustic and shock waves. For accurate simulation, the numerical scheme is required to have excellent dissipation-dispersion preserving property, while the wave modes, propagation characteristics, and wave speed of the numerical solution should be kept as close as possible to the exact solution of governing equations. For compressible flow simulation with shocks, the numerical scheme has to be equipped with proper numerical dissipation to make a crispy transition in the shock layer. Based on the high-order gas evolution model, the GKS provides a time accurate solution at a cell interface, from which both time accurate flux function and the time evolving flow variables can be obtained. The GKS updates explicitly both cell averaged conservative flow variables and the cell-averaged gradients by applying Gauss-theorem along the boundary of the control volume.  Based on the cell-averaged flow variables and cell-averaged gradients, a reconstruction with compact stencil  can be obtained. With the same stencil of a second-order scheme, a reconstruction up to 8th-order spacial accuracy  can be constructed, which include the nonlinear reconstruction for initial non-equilibrium state and linear reconstruction for the equilibrium one. The GKS unifies the nonlinear and linear reconstruction through a time evolution process at a cell interface from the non-equilibrium state to an equilibrium one. In the smooth acoustic wave region, the linear reconstruction will contribute mainly to the evolution. In the non-equilibrium shock region, the nonlinear reconstruction will play a dominant role. In the region between these two limits, the contribution from nonlinear and linear reconstructions depends on the weighting functions of $\exp(-\Delta t /\tau)$ and $(1- \exp(-\Delta t /\tau))$, where $\Delta t $ is the time step and $\tau$ is the particle collision time, which is enhanced in the shock region. As a result, both shock and acoustic wave can be captured accurately in GKS. The scheme is especially suitable for the simulation of shock and vortex interaction. In addition, the compact GKS uses multi-stage multi-derivative (MSMD) temporal discretization to improve time accuracy with less middle stages. The two stage fourth order method is implemented. The compact GKS has 8th-order spatial accuracy and 4th-order temporal accuracy. The compact GKS can use a large time step with CFL number $CFL \geq 0.3$ in acoustic simulation.
A series of numerical tests with acoustic and shock-vortex interaction are presented to demonstrate the validity of current scheme for
capturing both small amplitude sound wave and shock discontinuity. The same compact GKS provides the state-of-art numerical solutions in all test cases. 
\end{abstract}


\maketitle

\section{Introduction}

High-order and high-resolution schemes are needed in many applications related to the flows with small-scale structure and complex interaction,
such as turbulence flow, shock and boundary layer interactions, and shock and vortex interactions.
In the past decades, great effort has been paid on the development of high-order schemes for compressible Euler and Navier-Stokes equations
 with the co-existing smooth and discontinuous solutions \cite{harten1,harten2,shu1988efficient,liu,jiang}.
The reconstruction schemes of essentially non-oscillatory (ENO) \cite{harten2,shu1988efficient} and weighted essentially non-oscillatory (WENO) \cite{liu,jiang} have received the most attention.
ENO and WENO schemes can effectively distinguish the smooth and discontinuous shock in the reconstruction and maintain a uniformly high-order accuracy without obvious spurious oscillations in the unresolved region.
The core of WENO scheme is to design smooth indicators and to  obtain adaptive nonlinear convex combination of lower order polynomials.
WENO schemes can achieve very high-order accuracy in the smooth region and maintain essentially non-oscillatory property around discontinuities.
In order to improve the accuracy and reduce the numerical dissipation of WENO schemes, the modified schemes, such as WENO-M and WENO-Z schemes \cite{WENO-M,WENO-Z}, have been developed. The hybrid schemes of combing high-order linear schemes and nonlinear WENO have been proposed as well \cite{hill,taylor,ren,zhang}. Most effort of these works is about the selection of optimal stencils and the design of weighting functions.

High-resolution schemes have also been proposed for simulating smooth flow and wave propagation \cite{lele,mahesh}.
Due to high order of accuracy and small numerical domain of dependence, the compact schemes have better resolution for capturing solution at high wavenumber. Other advantages of compact scheme include their simple implementation and easy numerical boundary treatment on unstructured meshes \cite{WangQ}.
Compact schemes have obvious advantages for smooth flow and acoustics problem.
In the compact finite volume schemes based on Riemann solvers and compact finite difference schemes, the unknown spatial derivatives are used in spatial dicretization on compact stencils. The resolution of compact scheme is improved compared with non-compact scheme based on the same stencil. However, the resolution becomes poor when wavenumber closes to $\pi$. Another shortage is that the unknown derivatives are obtained by solving a linear system connecting the whole computational domain, which many not be efficiently solved for unsteady problems.
And the linear system is not theoretically valid for the solutions in the vicinity of discontinuities.
In order to allow compact schemes to compute discontinuous shocks, limiters and nonlinear reconstruction are introduced \cite{WangQ,WCNS}.
Other compact schemes include discontinuous Galerkin (DG) methods \cite{reed,cockburn1,cockburn2}, flux reconstruction (FR) scheme \cite{FR}, and so on. For the DG scheme, high-order polynomials determined through additional degrees of freedom (DOF) in each cell are evolved based on their
distinguishable governing equations from the weak formulations. The DOF can be the derivatives of physical variables, such as the direct
evolution equations for the gradient of flow variables in the HWENO scheme \cite{qiu1,qiu2}. FR scheme can achieve high-order spatial discretization through inner solution points on each cell and flux correction procedure. Both schemes introduce the same number of additional degree of freedom on each cell. And the limiting procedure or trouble cell detection are needed in order to capture discontinuous solutions.

High-order and high-resolution schemes have been successfully applied for compressible flow simulations. Some schemes have been used to study  acoustics problems, such as the WENO schems and compact finite difference methods for sound wave generation, shock-vortex interactions, and acoustic and jet flow propagation \cite{zhang2013vor,shock-vor,zhang2005shockvor,bai, zhang2015jet}.
But, problems are remained by applying current high-order and high-resolution schemes to acoustics computation.
The difficulties of computational acoustics include the disparity of acoustic wave and mean flow, large spectral bandwidth, long propagation distance, and distinct and well separated length scales of flows \cite{tam1995AIAA}.
Although WENO schemes can successfully solve the flows with discontinuous shock, its numerical dissipation is too large for acoustics problem.
Thus, a large number of mesh points is needed for acoustic computation \cite{zhang2013vor}.
The previous compact schemes have limited CFL number ($CFL\leqslant 0.1$) in order to maintain high-order temporal accuracy \cite{bai,tam1993,DGacoustic2019}. For those compact schemes associated with coupled linear systems, the efficiencies can be
severely deteriorated on unstructured mesh for unsteady flow simulations.
Compact finite volume schemes based on Riemann solvers and compact finite difference schemes with dispersion relation optimization have good resolution for a large range of wavenumber, while the resolution becomes poor when wavenumber approaches to $\pi$.
Additionally, nonlinear limiters are needed in the above compact schemes in case of strong shock waves.
It seems hard to find out an optimal limiting strategy to make high-order and high-resolution schemes work effectively
in both smooth and discontinuous regions.
Boundary condition implementation is another difficulty in computational acoustics.
With limited computational domain, numerical treatment of acoustics wave crossing through the boundaries of computational domain has to be
properly designed.
The radiation and outflow boundary conditions are proposed to make the acoustics and flow disturbances leave the domain with minimal reflections \cite{tam1993}. Besides, sponge zones can also be used in acoustics computations to absorb and minimize reflections from computational boundaries.

The compact high-order gas-kinetic scheme (GKS) has been developed on the same stencil of a second-order scheme \cite{CGKSAIA}. The compact GKS has been successfully applied to compressible flow simulations with high resolution, high order accuracy,
and excellent robustness for both smooth and discontinuous solutions under large CFL number ($CFL\geq0.5$) for inviscid flow.
In this paper, the compact high-order GKS will be further studied and used in the computation of acoustics waves.
In the compact GKS, due to the time-accurate evolution model at a cell interface both the cell averages and their cell-averaged gradients can be uniformly obtained at next time step through the time-dependent numerical fluxes and flow variables themselves from the same time-accurate gas distribution function at a cell bounadry.
The update of gradient in GKS is coming from the evolution solution at the cell interface, such as
$$({\bar W}_j^{n+1})_x = \frac{1}{\Delta x} \int_{x_{j-1/2}}^{x_{j+1/2}} \frac{\partial W^{n+1}}{\partial x} dx = \frac{1}{\Delta x} (W_{j+1/2}^{n+1} - W_{j-1/2}^{n+1}),$$
which is essentially different from DG method for the update of the similar degrees of freedom \cite{cockburn1,cockburn2}. In the current compact GKS, at the cell interface the strong or physical evolution solution is provided from the high-order gas evolution model in comparison with the
1st-order Riemann solver.
The cell averaged variables and their gradients are coming from the same cell interface evolution solution.
The cell averaged flow variables are obtained from the conservation laws through the cell interface fluxes, and
the cell averaged gradients are updated by the divergence theorem on the interface flow variables \cite{xu2}.
Therefore, besides the cell averaged flow variables, their slopes can be directly used in compact reconstruction, and a class of 6th- and 8th-order compact GKS have been developed with the use of standard WENO reconstruction without additional limiting process \cite{CGKSAIA}.
So far, the 8th-order compact GKS has the highest order of accuracy and the best resolution.
At the same time, the 8th-order GKS has similar robustness as the lower-order schemes for strong shock capturing and keep the high-order accuracy in the smooth region.

In addition to compact high-order reconstruction, the special properties of GKS also make the scheme suitable for the study of acoustics problems \cite{xu1,pan0,ji1}.
Firstly, the high-order time accurate numerical flux make the current compact GKS use a larger CFL number than that used by other compact high-order and high-resolution schemes based on Riemann solver in acoustic wave propagation, such as many test cases in the paper.
Secondly, the GKS uses an evolution process
from the initial non-equilibrium to the final equilibrium state in the determination of time-dependent interface flow variables.
In the compact reconstruction, the WENO-type nonlinear reconstruction is used for the initial non-equilibrium state and the linear reconstruction is
adopted in the determination of the equilibrium one. As a result, both nonlinear and linear reconstructions are unified in the GKS
evolution process. In the smooth flow region the linear reconstruction will contribute mostly in the determination of interface gas distribution function and the spectral-like resolution can be obtained. At the same time, in the discontinuous region, the nonlinear reconstruction
will persist in the determination of gas evolution and the non-equilibrium state will be kept to provide appropriate dissipation for the shock capturing.
The GKS will automatically identify the non-equilibrium and equilibrium regions and use the appropriate physics for the corresponding flow evolution. Even with an initial nonlinear reconstruction in the smooth flow region, the contribution from nonlinear reconstruction will decay exponentially as $\exp(-\Delta t/\tau)$ and the linear reconstruction will take over in the determination of the interface distribution function,
especially in the case with a large CFL number.
In the discontinuous region, the enhanced collision time has  $\tau \sim \Delta t$, the nonlinear reconstruction remains.
 As a result, both discontinuous shock and smooth aero-acoustic waves can be naturally captured by the compact GKS
 through its dynamic adaptation.
Thirdly, based on the time-dependent flux function and its time derivative, the multi-stage multi-derivative (MSMD) temporal discretization
 can be used \cite{hairer}. For example, the two-stage fourth-order (S2O4) temporal discretization has been well developed and validated
in the computation  \cite{christlieb,li,pan1,liAIA}. Since only one intermediate stage is needed in the S2O4 method instead of three intermediate stages in the conventional fourth-order Runge-Kutta method \cite{shu1988efficient}, the current compact GKS is efficient by reducing two
intermediate states and their corresponding reconstructions.
Fourthly, the interface evolution solution of the compact GKS provides both inviscid and viscous terms and the Navier-Stokes solutions
are directly obtained.
Fifthly, the GKS is a multidimensional scheme, both gradients in the normal and tangential directions of a cell interface will participate the
gas evolution at a cell interface. It paves the way for a smooth transition from a finite volume scheme to a Lax-Wendroff-type central difference scheme in the smooth flow region.
For the Riemann solver, the wave is always propagating in the normal direction of a cell interface.
 To have a multi-dimensional property for a numerical scheme is paramountly important for reducing mesh orientation effect on the physical solution and keeping the correct physical wave propagation direction.

This paper is organized as follows. The GKS and MSMD method will be introduced in Section 2.
Section 3 is about the review of compact high-order reconstructions.
In Section 4, the compact GKS will be tested in a wide range of acoustic problems from smooth sound wave propagation to shock-vortex interactions.
In order to validate the robustness of the compact GKS, the 8th-order compact GKS will be used in the simulation of the flow with shock interaction.
The last section is the conclusion.

\section{Gas-kinetic scheme and two-stage fourth-order time discretization}
The GKS mainly provides a time-accurate gas evolution model from an initial data with possible discotninuities \cite{xu2,xu1}.
The high-order GKS combines the high-order spatial reconstruction and various types of temporal discretization \cite{liQB2010high,pan0,ji1,ji2,pan4}. GKS has been used in compressible multi-component flow \cite{pan2}, compressible DNS at high Mach number \cite{cao2019}, and turbulence simulation \cite{cao2019implicit}.
A brief introduction of GKS and the special features will be presented in this section.

\subsection{Gas-kinetic scheme}
The gas-kinetic evolution model in GKS is based on the BGK equation \cite{BGK-1},
\begin{equation}\label{bgk}
f_t+\textbf{u}\cdot\nabla f=\frac{g-f}{\tau},
\end{equation}
where $f$ is the gas distribution function, $g$ is the corresponding equilibrium state that $f$ approaches, and $\tau$ is particle collision time.
The equilibrium state $g$ is a Maxwellian distribution,
\begin{equation*}
\begin{split}
g=\rho(\frac{\lambda}{\pi})^{\frac{K+2}{2}}e^{-\lambda((u-U)^2+(v-V)^2+\xi^2)},
\end{split}
\end{equation*}
where $\lambda =m/2kT $, and $m, k, T$ are the molecular mass, the Boltzmann constant, and temperature, respectively.
$K$ is the number of internal degrees of freedom, i.e. $K=(4-2\gamma)/(\gamma-1)$ for two-dimensional flow,
and $\gamma$ is the specific heat ratio. $\xi$ is the internal variable with $\xi^2=\xi^2_1+\xi^2_2+...+\xi^2_K$.
Due to the conservation of mass, momentum and energy during particle collisions, $f$ and $g$ satisfy the compatibility condition,
\begin{equation}\label{compatibility}
\int \frac{g-f}{\tau}\pmb{\psi} \mathrm{d}\Xi=0,
\end{equation}
at any point in space and time, where $\pmb{\psi}=(\psi_1,\psi_2,\psi_3,\psi_4)^T=(1,u,v,\displaystyle \frac{1}{2}(u^2+v^2+\xi^2))^T$, $\text{d}\Xi=\text{d}u\text{d}v\text{d}\xi_1...\text{d}\xi_{K}$.

The macroscopic mass $\rho$, momentum ($\rho U, \rho V$), and energy $\rho E$ can be evaluated from the gas distribution function,
\begin{equation}\label{g-to-convar}
{\textbf{W}} =
\left(
\begin{array}{c}
\rho\\
\rho U\\
\rho V\\
\rho E\\
\end{array}
\right)
=\int f \pmb{\psi} \mathrm{d}\Xi.
\end{equation}
The corresponding fluxes for mass, momentum, and energy in $i$-th direction is given by
\begin{equation}\label{g-to-flux}
{\textbf{F}_i} =\int u_i f \pmb{\psi} \mathrm{d}\Xi,
\end{equation}
with $u_1 = u$ and $u_2 = v$ in the 2D case.

Based on the BGK equation, the GKS provides a time accurate evolution solution $f$ at a cell interface \cite{xu1}.
On the mesh size scale, the conservations of mass, momentum and energy in a control volume become
\begin{equation}\label{semifvs}
\frac{\text{d}\textbf{W}_{ij}}{\text{d}t}=-\frac{1}{\Delta x}
(\textbf{F}_{i+1/2,j}(t)-\textbf{F}_{i-1/2,j}(t))-\frac{1}{\Delta y}
(\textbf{G}_{i,j+1/2}(t)-\textbf{G}_{i,j-1/2}(t)),
\end{equation}
where $\textbf{W}_{ij}$ is the cell averaged conservative variables, $\textbf{F}_{i\pm 1/2,j}(t)$ and $\textbf{G}_{i,j\pm 1/2}(t)$
are the time dependent fluxes at cell interfaces in $x$ and $y$ directions. $\textbf{F}_{i\pm 1/2,j}(t)$ and $\textbf{G}_{i,j\pm 1/2}(t)$ can be discretized at the numerical quadrature points along the cell interface.
Due to the connection among the flow variables $\textbf{W}$, the fluxes $\textbf{F}$ and $\textbf{G}$, and the distribution function $f$,
the central point of GKS is to construct a time-dependent gas distribution function $f$ at the cell interface.
The integral solution  of BGK equation is \cite{xu2},
\begin{equation}\label{integral1}
\begin{split}
f(x_{i+1/2},y_{j_\ell},t,u,v,\xi)=&\frac{1}{\tau}\int_0^t g(x',y',t',u,v,\xi)e^{-(t-t')/\tau}\mathrm{d}t' \\
&+e^{-t/\tau}f_0(-ut,-vt,u,v,\xi),
\end{split}
\end{equation}
where $(x_{i+1/2}, y_{j_\ell})=(0,0)$ is the numerical quadrature point at the cell interface for flux evaluation, and $x_{i+1/2}=x'+u(t-t')$ and $y_{j_\ell}=y'+v(t-t')$ are the particle trajectory.
Here $f_0$ is the initial state of gas distribution function $f$ at $t=0$.
The integral solution basically states a physical process from the particle free transport in $f_0$ in the kinetic scale to the hydrodynamic flow evolution in the integration of $g$.
The contributions from $f_0$ and $g$ in the determination of $f$ at the cell interface depend on the ratio of time step to the local particle collision time, i.e., $\exp(-t/\tau)$.
For the NS solution, the determination of $f_0$ depends only on the initial reconstructions of macroscopic flow variables, because
the gas distribution function for the NS solution can be evaluated from the Chapman-Enskog expansion.
For the current high-order GKS, the high-order nonlinear WENO reconstruction with compact stencils will be used in the determination of $f_0$,
and the compact linear reconstruction is adopted in the determination of $g$.
Therefore, the above integral solution not only incorporates a physical evolution process from initial discontinuous non-equilibrium state
to a continuous equilibrium one, but also unifies the nonlinear  and the linear reconstructions  in the evolution process.
This fact is crucially important for the current scheme to capture both nonlinear shock and linear acoustic wave accurately
in a single computation with a dynamic adaptation.

In order to obtain the solution $f$, both $f_0$ and $g$ in Eq.(\ref{integral1}) need to be modeled \cite{xu2,liQB2010high}.
For the current compact high-order GKS, the simplified third-order gas distribution function is used \cite{zhou},
\begin{align}\label{3rd-simplify-flux}
f(x_{i+1/2},y_{j_\ell},t,u,v,\xi)&=g_0+{\overline{A}} g_0t+\frac{1}{2}\overline{a}_{tt}g_0t^2\nonumber\\
&-\tau[(\overline{a}_1u+\overline{a}_2v+\overline{A})g_0+(\overline{a}_{xt}u+\overline{a}_{yt}v+\overline{a}_{tt})g_0t]\nonumber\\
&-e^{-t/\tau}g_0[1-(\overline{a}_{1}u+\overline{a}_{2}v)t]\nonumber\\
&+e^{-t/\tau}g_l[1-(a_{1l}u+a_{2l}v)t]H(u)\nonumber\\
&+e^{-t/\tau}g_r[1-(a_{1r}u+a_{2r}v)t](1-H(u)),
\end{align}
where the terms related to $g_0$ are from the integral of the equilibrium state and the terms related to $g_l$ and $g_r$ are from the initial term $f_0$ in the Eq.(\ref{integral1}). All the coefficients in Eq.(\ref{3rd-simplify-flux}) can be determined from the initially reconstructed  macroscopic flow variables.
Based on the above time accurate gas distribution function at a cell interface, the flow variables $\textbf{W}^{n+1}$ at next time step on the
cell interface can be obtained.
Since a second-order flux function is accurate enough for a 4th-order time accuracy through the two-stage fourth-order (S2O4) time discretization, the numerical flux at a cell interface can be evaluated from a second-order time-accurate distribution function which is obtained directly by removing the third-order terms in Eq.(\ref{3rd-simplify-flux}), such as the terms of $\overline{a}_{tt}, \overline{a}_{xt}$, and $\overline{a}_{yt}$.

\subsection{Two-stage fourth-order time discretization}
Two-stage fourth-order (S2O4) method \cite{li,pan1,seal} is adopted in the current scheme to achieve a fourth-order temporal accuracy.
Since the above time accurate gas distribution function at a cell interface has a complicated dependence on time in the non-smooth region,
the treatment proposed in \cite{li,pan1} is used to extract a linearly time-dependent flux function.
The same idea is used to obtain the time derivatives of $f$ at a cell interface \cite{ji2}.
Theoretically, even higher-order MSMD methods could be chosen to achieve higher-order temporal accuracy \cite{chan}, while the fourth-order in time seems to be an optimal choice for the high-order compact GKS.

For conservation laws, the semi-discrete finite volume scheme Eq.(\ref{semifvs}) is rewritten as
\begin{align*}
\frac{\text{d} \textbf{W}_{ij}}{\text{d} t}:=\mathcal{L}(\textbf{W}_{ij}),
\end{align*}
where $\mathcal{L}(\textbf{W}_{ij})$ is the numerical operator for spatial derivative of fluxes.
A fourth-order temporal accurate solution for $\textbf{W}(t)$ at $t=t_n
+\Delta t$ can be obtained by
\begin{equation}\label{step-hyper-1}
\begin{split}
\textbf{W}^*&=\textbf{W}^n+\frac{1}{2}\Delta t\mathcal
{L}(\textbf{W}^n)+\frac{1}{8}\Delta t^2\frac{\partial}{\partial
t}\mathcal{L}(\textbf{W}^n),\\
\textbf{W}^{n+1}&=\textbf{W}^n+\Delta t\mathcal
{L}(\textbf{W}^n)+\frac{1}{6}\Delta t^2\big(\frac{\partial}{\partial
	t}\mathcal{L}(\textbf{W}^n)+2\frac{\partial}{\partial
	t}\mathcal{L}(\textbf{W}^*)\big),
\end{split}
\end{equation}
where $\mathcal{L}$ and $\frac{\partial}{\partial t}\mathcal{L}$ are related to the fluxes and the time derivatives of
the fluxes which are both evaluated from the time-dependent gas distribution function $f(t)$.
The middle state $\textbf{W}^*$ is obtained at time $t^* = t^n + \Delta t /2$.

With the time accurate gas distribution function $f(t)$, along the same line of MSMD the gas distribution function $f$
at $t^{n+1}$ at a cell interface can be obtained,
\begin{align*}
f^* =& f^n + \frac{1}{2} \Delta t f_t^n + \frac{1}{8} (\Delta t)^2 f_{tt}^n, \\
f^{n+1}=& f^n +\Delta t f_t^n + \frac{1}{6}\Delta t^2( f_{tt}^n + 2 f_{tt}^* ),
\end{align*}
where $f^*$ is for the middle state at time $t^* = t^n + \Delta t /2$. The derivatives at $t^n$ and $t^*$ can be determined
from the time-dependent distribution function in Eq.(\ref{3rd-simplify-flux}).
Therefore, the evaluated $f^{n+1}$ in the above equation has a fourth order accuracy with local truncation error $O(\Delta t^5)$ \cite{ji2}.
Therefore, based on the $ f_{i+1/2}^{n+1} $ at a cell interface, the flow variables $\textbf{W}_{i+1/2}^{n+1}$ can be explicitly obtained,
\begin{align*}
\textbf{W}_{i+1/2}^{n+1} = \int \pmb{\psi} f_{i+1/2}^{n+1} \mathrm{d} \Xi.
\end{align*}
The time accurate solution $\textbf{W}^{n+1}_{i+1/2}$ depends on the high-order gas evolution model, which is critically important to develop
high-order compact scheme, because more information is provided locally to do a compact reconstruction.
In comparison with the schemes based on the Riemann solver, even though the interface values exist as well in the Riemann solution,
this solution has no time accuracy and cannot be used in the reconstruction in the next time step.

\section{Compact high-order GKS}
The compact high-order GKS, including 6th-order and 8th-order schemes, have been proposed in \cite{CGKSAIA}. In this section, a review and overview of the compact GKS will be presented. The compact GKS is constructed on the same stencil of a second-order scheme.
The high-order compact GKS seems have the same robustness as the second order scheme for capturing both smooth and discontinuous solutions with
 a CFL number on the order ($CFL \geq 0.5$) for the inviscid flow. The high order accuracy and high resolution of the compact GKS will be further validated
 in the current paper for the shock and acoustic wave calculations.

\subsection{Overview of compact GKS}
Compact schemes can achieve higher order accuracy and better resolution than that of non-compact schemes \cite{lele}.
 To get a high-order scheme with the stencil of a second-order scheme is pursued continuously in CFD community in the past decades.
In order to implement high-order reconstruction on a compact stencil explicitly, besides cell averages additional local flow variables need to be provided.
The compact high-order GKS is based on the time-accurate gas evolution model \cite{xu2}. Over the time interval $[t_n,t_{n+1}]$, the gas distribution function $f(x_{i+1/2},t)$ at the cell interface $x_{i+1/2}$ is known from the time-accurate gas evolution model, and the numerical flux $\textbf{F}_{i+1/2}(t)$ and flow variable $\textbf{W}_{i+1/2}(t)$ at the interface can be determined by
\begin{equation}\label{micro-to-macro}
\begin{split}
{{\textbf{F}_{i+1/2}(t)}} =\int uf(x_{i+1/2},t)\pmb{\psi} \mathrm{d} \Xi,\\
{{\textbf{W}_{i+1/2}(t)}} =\int f(x_{i+1/2},t)\pmb{\psi} \mathrm{d} \Xi.
\end{split}
\end{equation}
As a result, the cell averages can be updated through fluxes in the finite volume scheme.
The cell-averaged slopes $\textbf{W}^{'}_i$ can be updated as well from Gauss-theorem
\begin{align}\label{compact-deriv-1}
\textbf{W}^{'}_i=\frac{1}{\Delta x}(\textbf{W}_{i+1/2}(t)-\textbf{W}_{i-1/2}(t)),
\end{align}
where the cell averaged slope is defined on the cell as
\begin{align}\label{compact-deriv}
\textbf{W}^{'}_i\equiv \frac{1}{\Delta x}\int_{I_i}\frac{\partial{\textbf{W}}(t)}{\partial{x}} \mathrm{d} x.
\end{align}
In 2D case, the divergence theorem can be applied on a closed boundary of the control volume to obtain the averaged gradients of the
flow variables in the cell. With the updates of both cell averaged flow variables and their gradients,
the compact GKS can be expressed in an abstract form
\begin{align}\label{fv-abstract-compact}
\overline{E}_h(t)\cdot \{\textbf{W}_i,\textbf{W}^{'}_i\}\equiv A_h \cdot E(f(t)) \cdot R(\cdot;\textbf{W}_{i-k},\cdots,\textbf{W}_{i+k},\textbf{W}^{'}_{i-k},\cdots,\textbf{W}^{'}_{i+k}),
\end{align}
where $E(f(t))$ is the time-accurate gas evolution model, $A_h$ is a spatial cell averaging operator, and $\overline{E}_h(t)$ is the numerical operator corresponding to the compact GKS.
The compact GKS updates the cell averages and their slopes from the same gas evolution model.
Even with compact high-order reconstruction, the GKS has a numerical domain of dependence close to the physical one with a  reasonable
CFL number $0.5$, and uses one intermediate stage for the 4th-order of time accuracy.
The high order of accuracy and high resolution properties of the compact GKS will be presented in the next section.

\subsection{Compact linear reconstruction}
Compact reconstruction used in GKS is presented for one dimensional case.
The linear compact reconstruction is given first.
Instead of obtaining a smooth high-order polynomial on each cell, the reconstruction will be done for the
interface values which include point-wise values and their slopes at the interface.

The compact stencil for the reconstruction at $x_{i+1/2}$ is $S_{i+1/2}=\{I_{i-1},I_{i},I_{i+1},I_{i+2}\}$.
On each cell $I_k, k=i-1,\cdots,i+2$, the cell averages and their slopes of conservative or characteristic variables
are known and denoted as $Q_{i}$ and $Q^{'}_{i}$.
The linear compact 8th-order reconstruction can be determined uniquely \cite{CGKSAIA}, and the details are given as,
\begin{equation}\label{recons-8th-val}
\begin{split}
P^7(x_{i+1/2})=\frac{1}{420}(&25Q_{i-1}+185Q_{i}+185Q_{i+1}+25Q_{i+2}+ \\
  &6\Delta xQ^{'}_{i-1}+102\Delta xQ^{'}_{i}-102\Delta xQ^{'}_{i+1}-6\Delta xQ^{'}_{i+2}),\\
P^7_x(x_{i+1/2})   =\frac{1}{108\Delta x}(&-14Q_{i-1}-270Q_{i}+270Q_{i+1}+14Q_{i+2}- \\
                      &3\Delta xQ^{'}_{i-1}-99\Delta xQ^{'}_{i}-99\Delta xQ^{'}_{i+1}-3\Delta xQ^{'}_{i+2}),\\
P^7_{xx}(x_{i+1/2})=\frac{1}{4\Delta x^2}(&-4Q_{i-1}+4Q_{i}+4Q_{i+1}-4Q_{i+2}- \\
                      &\Delta xQ^{'}_{i-1}-9\Delta xQ^{'}_{i}+9\Delta xQ^{'}_{i+1}+\Delta xQ^{'}_{i+2}).
\end{split}
\end{equation}
Based on the compact stencil $S_{i+1/2}$, a series of compact reconstruction can be developed theoretically.
For example, the compact 6th-order reconstruction has also been constructed as well \cite{CGKSAIA}.
However, the compact 8th-order scheme has better resolution and accuracy, and keeps almost the same robustness as the
lower order schemes. In this paper, the 8th-order compact GKS will be reviewed and used in the shock and acoustic wave computations.

\begin{figure}[!htb]
\centering
\includegraphics[width=0.55\textwidth]{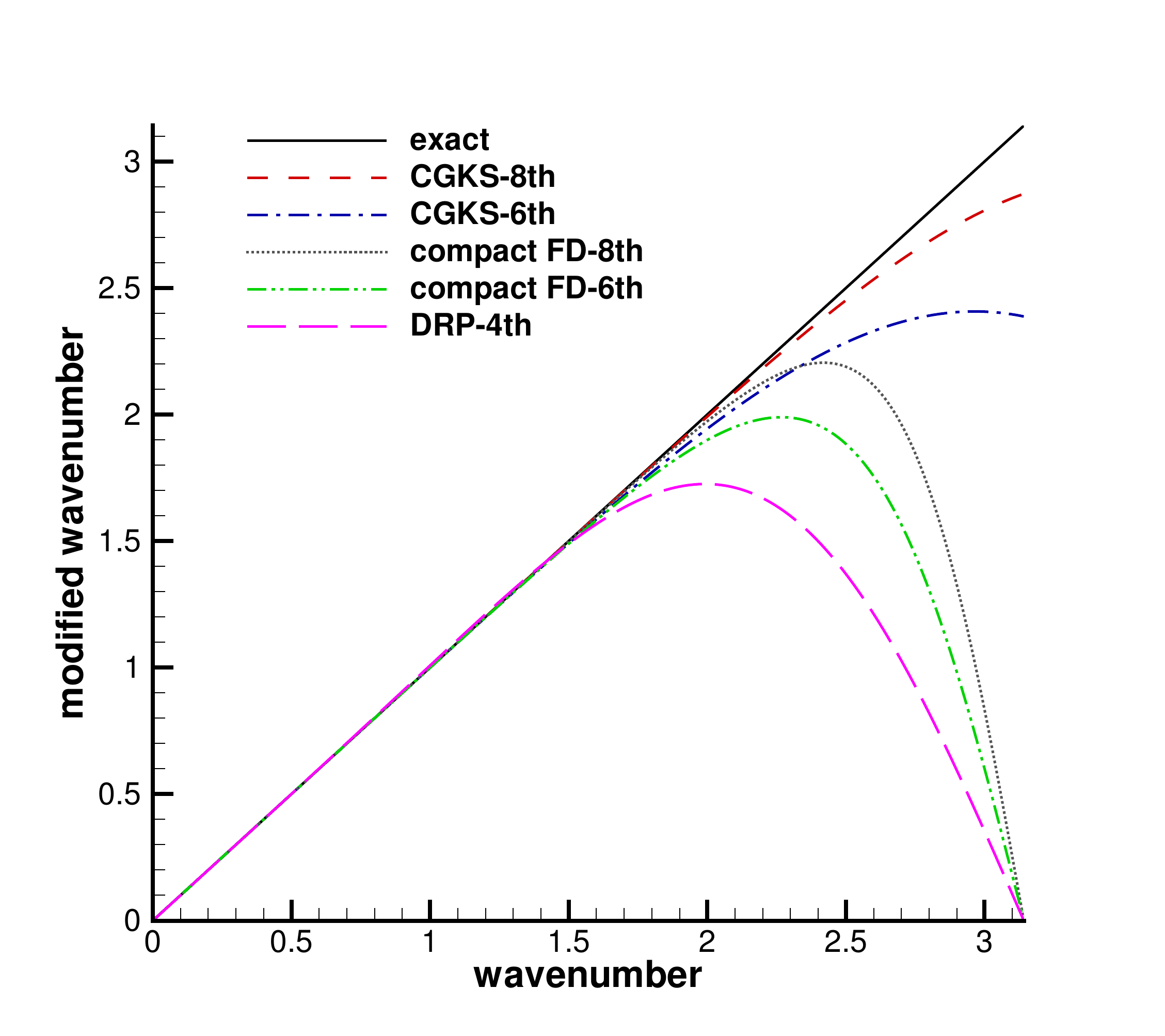}
\caption{\label{1d-resolution}  Plots of modified wavenumber and phase speed vs wavenumber for different schemes.
Compact FD-6th is the 6th-order tridiagonal scheme studied by Lele \cite{lele},
and the scheme has the best resolution in the series of 6th-order finite scheme;
compact FD-8th is the 8th-order pentadiagonal scheme \cite{lele}; DRP-4th scheme is proposed in \cite{tam1993}, and the dispersion property of the scheme is optimized.}
\end{figure}

The spatial resolution of the compact GKS is presented by Fourier analysis \cite{CGKSAIA}. The linear compact $8$th-order scheme
has a spectral-like resolution in terms of the definition of Lele \cite{lele}.
Fig.\ref{1d-resolution} shows the modified wavenumber for different schemes.
The compact $6$th-order GKS, compact finite difference schemes of Lele \cite{lele} and dispersion-relation-preserving (DRP) scheme \cite{tam1993} are also included.
Since the derivatives used in the compact finite difference schemes of Lele are coupled with point-wise values,
their resolution becomes poor at large wavenumber in comparison with the compact GKS, where the both cell-averaged values and their slopes are independently provided in the scheme.
Different from the previous dispersion analysis \cite{lele, DGacoustic2019, Numer-dispersion}, the high resolution in the current scheme is a  result from both spatial reconstruction and high-order evolution model, where both cell averages and slopes are independently given by the same
time evolving gas distribution function.
Although the current compact $8$th-order GKS has the same form of discretization in terms of spatial derivatives as that in the compact $8$th-order scheme of Lele \cite{lele}, the slopes in the scheme of Lele intrinsically depend on cell averaging values. The cell averages and their slopes are not fully independent in the traditional compact schemes. Therefore, the compact $8$th-order GKS has a better resolution
than that in other compact schemes.

\begin{figure}[!htb]
\centering
\includegraphics[width=0.55\textwidth]{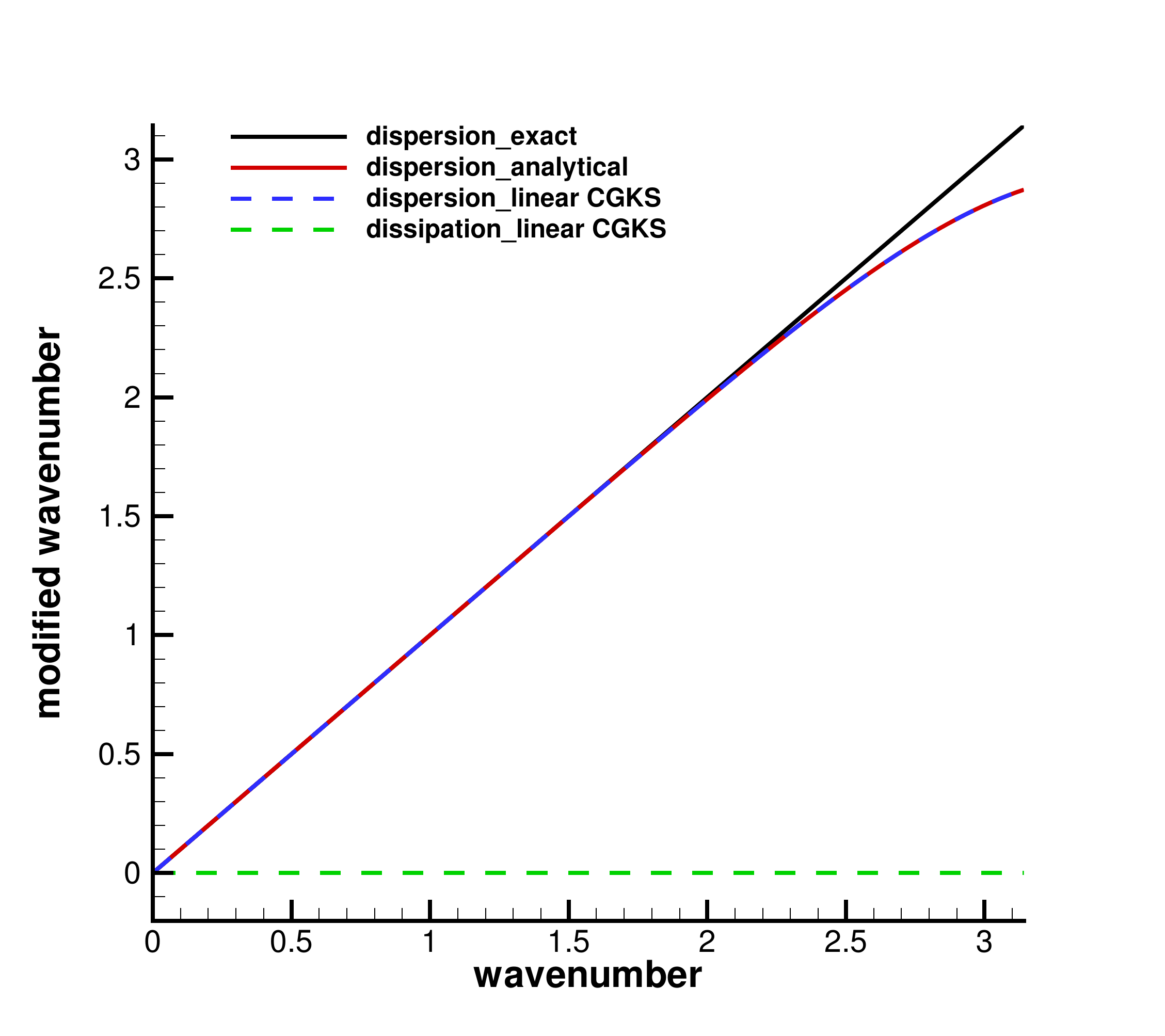}
\caption{\label{1d-resolution-numer} Numerical dispersion and dissipation properties of compact 8th-order GKS, which are consistent with the analytical ones. }
\end{figure}

To further understand the high resolution of the compact GKS and the essential differences from other compact schemes, such as DG \cite{reed,cockburn1,cockburn2} and FR \cite{FR} methods with additional internal DOF, the numerical dispersion property of the compact $8$th-order GKS with linear reconstruction is evaluated using the method in \cite{Numer-dispersion}. The dispersion and dissipation properties of the current scheme are shown in Fig.\ref{1d-resolution-numer}. For linear reconstruction, the numerical dispersion and dissipation properties are consistent with the analytical ones.
In GKS, the cell averages in Eq.(\ref{step-hyper-1}) and the cell-averaged derivatives in  Eq.(\ref{compact-deriv-1}) are  determined at $t^{n+1}$ simultaneously from the same gas evolution solution at the interface.
Intrinsically, a single high-order evolution model at a cell interface in GKS determines the spatial discretization and dispersion property.
However, for these compact schemes with inner DOF, there are additional evolution models independently for the update of these DOF.

The analytical and numerical dispersion properties of the compact $8$th-order GKS are consistent with the sampling theorem where more than two independent values are needed to sample a wave with a wavenumber of $\pi$. In the compact $8$th-order GKS, there are two values on each cell. Therefore, the wave with a wavenumber of $\pi$ is sampled by $4$ known values and the spatial discretization can have a good resolution.

\subsection{Compact nonlinear reconstruction}

In compressible gas dynamics, the acoustics, shock waves, and shear layers can co-exist.
In order to avoid numerical oscillation and spurious wave generation around discontinuity and maintain high-order accuracy,
the compact nonlinear WENO-type reconstruction is derived and it goes back to the linear one in the smooth region.
For the capturing of possible discontinuity at a cell interface, the point-wise values at the left and right sides of a cell interface
have to be evaluated separately.
For simplicity, the WENO reconstruction procedure is given in detail for the construction of the left side value at the cell interface $x=x_{i+1/2}$.
The procedure for the right side value can be obtained similarly according to the symmetric property.
The left side value by WENO reconstruction is given by the candidate polynomials as follows
\begin{align}\label{right-left}
Q_{i+1/2}=\sum_{k=0}^{l}\delta_{k}q_{k,i+1/2},
\end{align}
where $Q_{i+1/2}$ is the left point-wise value in the 8th-order WENO reconstruction,
$l$ is the number of candidate polynomials, $\delta_{k}$ is the WENO weight, and $q_{k,i+1/2}$
is the point-wise value of the candidate polynomial $q_{k}(x)$ at $x_{i+1/2}$.

For the nonlinear reconstruction the sub-stencils have to be defined first.
The determination of candidate polynomials is very important for the quality of the scheme.
When a discontinuity exists in the large stencil $S_{i+1/2}$, a smooth candidate polynomials will play an important role in reconstruction.
In such a case, the averaged slopes appearing in the sub-stencil should be kept away from the interface in case of possible discontinuities,
and the candidate polynomial becomes sufficiently smooth.
Based on the above consideration, the following sub-stencils are determined,
\begin{align*}
S_0&=\{Q_{i-1},Q_{i},Q^{'}_{i-1}\} \leftrightarrow q_{0}(x), \\
S_1&=\{Q_{i-1},Q_{i},Q_{i+1}\} \leftrightarrow q_{1}(x), \\
S_2&=\{Q_{i},Q_{i+1},Q_{i+2},Q^{'}_{i+2}\} \leftrightarrow q_{2}(x), \\
S_3&=\{Q_{i-1},Q_{i},Q_{i+1}\} \leftrightarrow q_{3}(x), \\
S_4&=\{Q_{i},Q_{i+1},Q_{i+2}\} \leftrightarrow q_{4}(x), \\
S_5&=\{Q_{i-1},Q_{i},Q_{i+1},Q^{'}_{i-1},Q^{'}_{i}\} \leftrightarrow q_{5}(x), \\
S_6&=\{Q_{i},Q_{i+1},Q_{i+2},Q^{'}_{i+1},Q^{'}_{i+2}\} \leftrightarrow q_{6}(x),
\end{align*}
where $q_{2}(x)$ is a cubic polynomial, $q_{5}(x)$ and $q_{6}(x)$ are fourth-order polynomials, and others are quadratic polynomials.
And $q_{k,i+1/2}$ can be uniquely determined as
{\setstretch{1.5}
\begin{align}\label{sub-stencil}
\begin{cases}
q_{0,i+1/2}&=\displaystyle\frac{1}{6}(-7Q_{i-1}+13Q_{i}-4hQ^{'}_{i-1}), \\
q_{1,i+1/2}&=\displaystyle\frac{1}{6}(-Q_{i-1}+5Q_{i}+2Q_{i+1}), \\
q_{2,i+1/2}&=\displaystyle\frac{1}{24}(5Q_{i}+32Q_{i+1}-13Q_{i+2}+6hQ^{'}_{i+2}), \\
q_{3,i+1/2}&=\displaystyle\frac{1}{6}(-Q_{i-1}+5Q_{i}+2Q_{i+1}),\\
q_{4,i+1/2}&=\displaystyle\frac{1}{6}(2Q_{i}+5Q_{i+1}-Q_{i+2}),\\
q_{5,i+1/2}&=\displaystyle\frac{1}{30}(10Q_{i-1}+19Q_{i}+Q_{i+1}+3hQ^{'}_{i-1}+21hQ^{'}_{i}),\\
q_{6,i+1/2}&=\displaystyle\frac{1}{30}(Q_{i}+19Q_{i+1}+10Q_{i+2}-21hQ^{'}_{i+1}-3hQ^{'}_{i+2}).
\end{cases}
\end{align}
}

In the smooth region, the convex combination with $\delta_{k}=d_k$ recovers the reconstruction in
Eq.\eqref{recons-8th-val}, which can be the condition to get the linear weights \cite{jiang}.
The linear weights $d_{k}$ of 8th-order reconstruction are
\begin{align*}
d^8_{0}&=\frac{3}{98},~d^8_{1}=\frac{5}{98},~d^8_{2}=\frac{4}{49},~d^8_{3}=\frac{7}{98}~d^8_{4}=\frac{7}{98},~d^8_{5}=\frac{17}{49},~d^8_{6}=\frac{17}{49}.
\end{align*}

The WENO-Z nonlinear weights are used in the current compact scheme and they are defined as \cite{WENO-Z}
\begin{equation}\label{WENO-Z-w}
\delta_{k}=\frac{\alpha_{k}}{\sum_{m=0}^{l}\alpha_{m}},
~~\alpha_{k}=d_k\Big[1+\big(\frac{Z_{ref}}{\beta_k+\epsilon}\big)\Big],~~k=0,...,l,
\end{equation}
where $Z_{ref}$ is the local high-order reference value. $\beta_{k}$ is the smooth indicator and defined as \cite{jiang}
\begin{equation}\label{smooth-indicator}
\beta_k=\sum_{r=1}^{r_k}\Delta x^{2r-1}\int_{x_{i-1/2}}^{x_i+1/2}\big(\frac{\text{d}^r}{\text{d}x^r}q_k(x)\big)^2dx,
\end{equation}
where $r_k$ is the order of $q_k(x)$.
$Z_{ref}$ is defined by $\beta_{k}$ as
\begin{align}\label{tau}
Z_{ref}^n=\left|3(\beta_{0}-\beta_{4}) + (\beta_{4}-\beta_{3})\right|.
\end{align}

In smooth region, the first two candidate polynomial $q_0(x)$ and $q_1(x)$ can be combined into a cubic polynomial
which is symmetric counterpart of $q_2(x)$. Then, current sub-stencils can become basically symmetric for interface $x_{i+1/2}$.
In order to maintain the symmetry for the nonlinear schemes,
the smooth indicator $\beta_1$ of $q_1(x)$ corresponding to $S_1$ is replaced by the indicator of the cubic polynomial
$\widetilde{q}_1(x)$ on $\widetilde{S}_1=\{Q_{i-1},Q_{i},Q_{i+1},Q^{'}_{i-1}\}$.
Even without showing in this paper, some tests demonstrate that the current choice $\beta_1$ can
present a slightly better resolution in the numerical results with excellent robustness.
The detailed formulae for all $\beta_k,~k=0,...,l_n$ are given in \cite{CGKSAIA}.

\begin{figure}[!htb]
\centering
\includegraphics[width=0.55\textwidth]{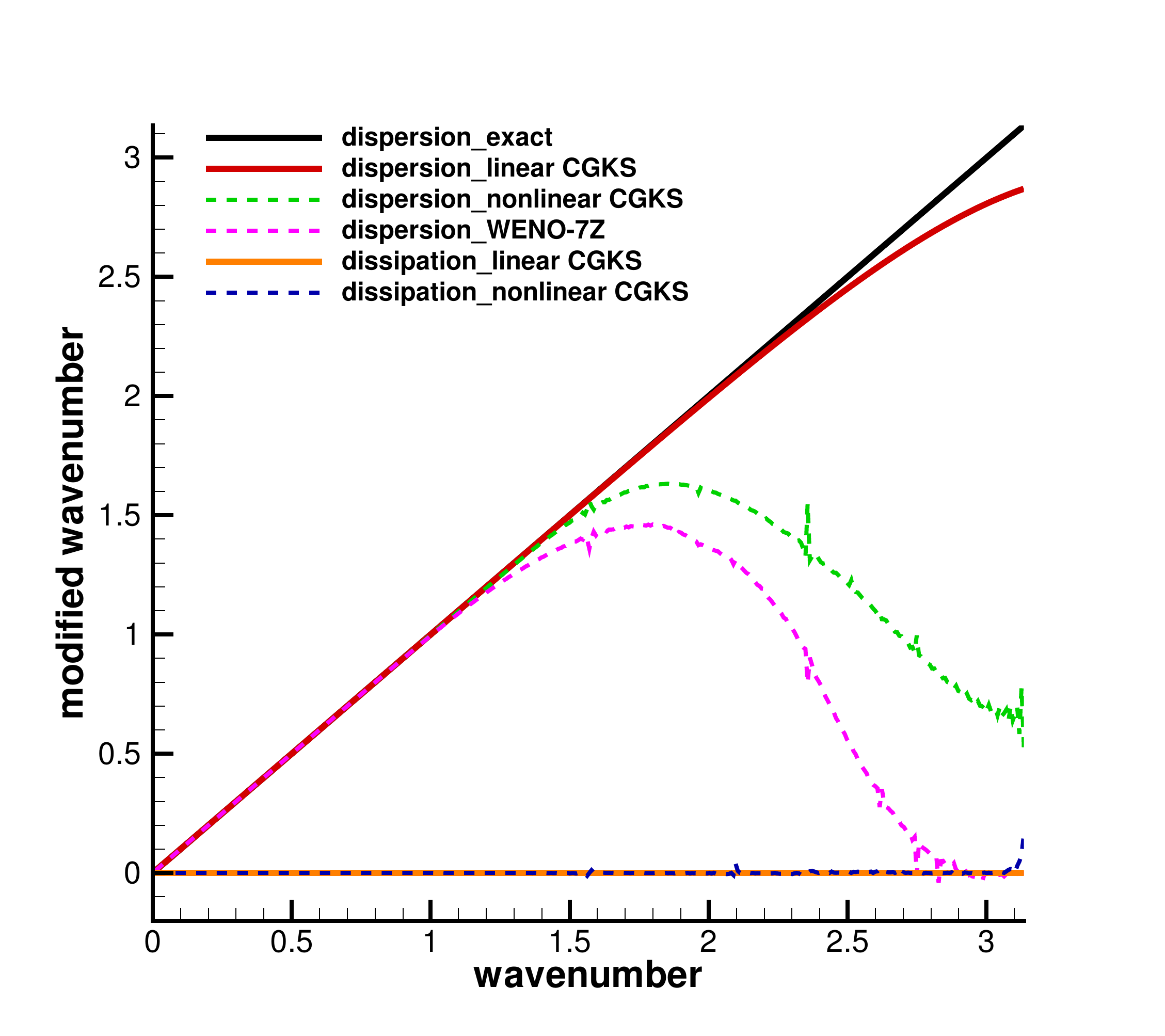}
\caption{\label{1d-resolution-nonlinear} Numerical dispersion and dissipation properties of compact 8th-order GKS with nonlinear reconstruction and WENO-7Z with GKS flux. }
\end{figure}

The numerical dispersion and dissipation properties of compact 8th-order GKS with nonlinear reconstruction are given in Fig.\ref{1d-resolution-nonlinear}. When the wavenumber is greater than $\pi/2$, the numerical dispersion curve of nonlinear scheme starts to deviate from the analytical curve obviously. In contrast, the non-compact nonlinear schemes, such as the $7$th-order WENO scheme with the GKS fluxes for the
updates of cell averaged flow variables only, are also tested. The dispersion of WENO-7Z begins to deviate from the analytical curve when the wave number is less than $\pi/2$ \cite{Numer-dispersion}. It shows that the compact scheme is better than the non-compact one.
In the case of large wavenumber (wave number greater than $\pi/2$), there is a large deviation between nonlinear and linear weights, and the sub-stencil plays a major role in reconstruction. The nonlinear scheme cannot maintain the high resolution as the linear one.
When the wavenumber is close to $\pi$, the modified wavenumber from compact GKS is not zero, because the smallest sub-stencil contains $3$ independent values. Even based on the non-linear reconstruction, the compact GKS will perform well once there are around $6$ to $8$ cells to resolve a wavelength.

\begin{figure}[!htb]
\centering
\includegraphics[width=0.85\textwidth]{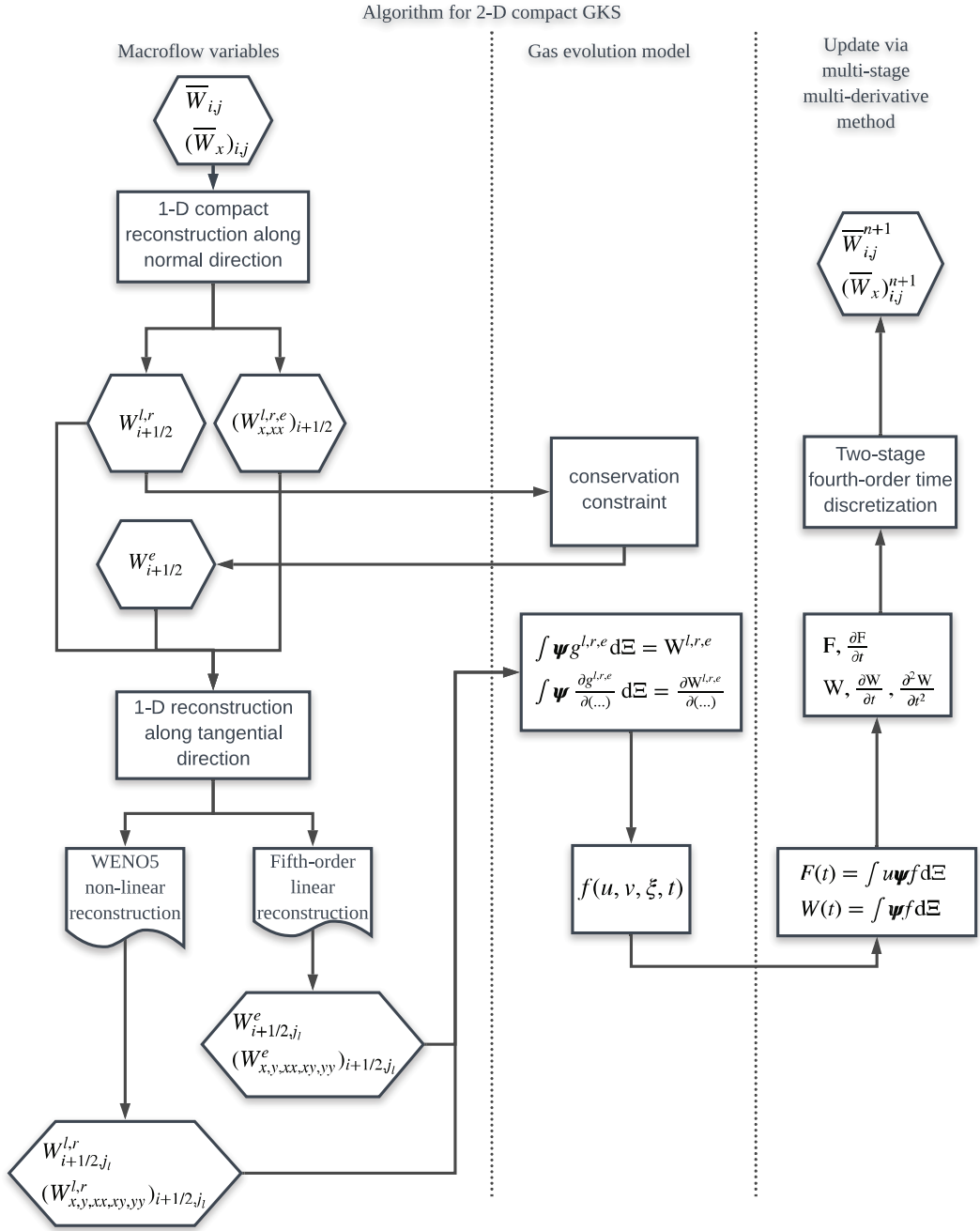}
\caption{\label{0-algorithm} Implementation algorithm of compact GKS in two-dimensional case.}
\end{figure}

The algorithm of the current compact 8th-order GKS in two-dimensional case is presented in Fig. \ref{0-algorithm}.
The overall framework of the algorithm can be used for other high-order GKS as well.

\section{Numerical examples}
In this section, the compact 8th-order GKS is used in flow simulations, including acoustic problems.
The tests include advection of small perturbation, one-dimensional acoustic waves propagation, one-dimensional solution with discontinuities,
two-dimensional pressure pulse evolution, propagation of sound, entropy and vorticity waves, shock acoustic wave interaction,
and two-dimensional high speed jet flow.
The mesh used in this paper is rectangular one and the time step is determined by the CFL condition with a CFL number ($\geq 0.3$) in all test cases if not specified.
In all tests, the same linear reconstruction is used for the equilibrium state and the nonlinear reconstruction for the non-equilibrium state.
There is no any additional "trouble cell" detection or any additional limiter designed for specific test.
The GKS  basically presents a dynamical process from non-equilibrium state with nonlinear reconstruction to equilibrium state with linear one, and
the rate is related to the relaxation process $\exp(-\Delta t/\tau)$.
The collision time $\tau$ for inviscid flow at a cell interface is defined by
\begin{align*}
\tau=\varepsilon \Delta t+C\displaystyle|\frac{p_l-p_r}{p_l+p_r}|\Delta
t,
\end{align*}
where $\varepsilon=0.01$, $C=1$, and $p_l$ and $p_r$ are the pressure at the left and right sides of a cell interface.
For the viscous flow, the collision time is related to the viscosity coefficient,
\begin{align*}
\tau=\frac{\mu}{p}+C \displaystyle|\frac{p_l-p_r}{p_l+p_r}|\Delta t,
\end{align*}
where $\mu$ is the dynamic viscosity coefficient and $p$ is the pressure at the cell interface. In
smooth flow region, it will reduce to $\tau=\mu/p$.
The reason for including pressure jump term in the particle collision time is to add artificial
dissipation in the discontinuous region to keep the non-equilibrium dynamics in the shock layer with $\tau \sim \Delta t$.

Most test cases presented below are coming from literatures and the references are given.
Based on the simulation solutions and comparison with the solutions provided in the references,
it confirms that the compact 8th-order GKS is one of the best schemes in terms of the accuracy, robustness, and efficiency.
It works in all cases and provides solution no worse than any other result in the reference papers.

\subsection{Advection of small perturbation}
To capture the propagation of small perturbation in a long period is necessary in computational acoustics.
The case of advection of small density perturbation is used to validate the order of accuracy. The initial
conditions are given as follows
\begin{align*}
\rho(x)=1+\epsilon \sin(\pi x),\ \  U(x)=1,\ \ \  p(x)=1, x\in[0,2].
\end{align*}
The periodic boundary condition is adopted, and the analytic solutions are
\begin{align*}
\rho(x,t)=1+\epsilon \sin(\pi(x-t)),\ \ \  U(x,t)=1,\ \ \  p(x,t)=1,
\end{align*}
where $\epsilon=1\times10^{-4}$ is used for a small magnitude. With the $r$th-order spatial reconstruction and 2-stage 4th-order
temporal discretization, the leading term of the truncation error is $O(\Delta x^r+\Delta t^4)$. To keep the
$r$th-order accuracy in the test, $\Delta t=C \Delta x^{r/4}$ needs to be used for the $r$th-order scheme.
In the computation, uniform meshes with $N$ mesh points are used. The $L^1$, $L^2$ and $L^\infty$ errors and
convergence orders at $t=200$ (100 periods) by the 8th-order compact GKS are presented in Table \ref{1d-accuracy-8}. As a
comparison, a 5th-order WENO-5Z scheme with GKS flux is also constructed. The main difference is about the initial reconstruction.
For the  5th-order WENO-5Z scheme, same as many other WENO-type schemes, only the cell averages are used in the initial reconstruction.
As a result, a large stencil has to be used and the scheme is not compact. But. in terms of flux function and the time-stepping evolution, the same
GKS flux and two-stage fourth-order method are adopted. So, the numerical results from the 5th-order WENO-5Z scheme is to identify the
differences from the compact and non-compact reconstruction. Similarly, a 7th-order WENO-7Z scheme is developed as well.
Table \ref{1d-accuracy-5} shows the numerical performance of the 5th-order WENO-5Z scheme.
The expected 8th-order of accuracy is obtained by the compact 8th-order scheme, while the WENO-5Z cannot obtain
5th-order of accuracy on coarse meshes because of the poor resolution.
The error of compact 8th-order scheme is very small even on the coarsest mesh,
which is about $1/100$ of the amplitude of the initial small perturbation.

\begin{table}[!h]
	\small
	\begin{center}
		\def\temptablewidth{0.90\textwidth}
		{\rule{\temptablewidth}{0.70pt}}
		\begin{tabular*}{\temptablewidth}{@{\extracolsep{\fill}}c|cc|cc|cc}
			
			mesh number & $L^1$ error & Order & $L^2$ error & Order& $L^{\infty}$ error & Order  \\
			\hline
            6  & 1.622861e-06 & ~    & 1.747588e-06 & ~    & 2.434333e-06 & ~    \\
            12 & 6.544449e-09 & 7.95 & 7.205669e-09 & 7.92 & 1.001831e-08 & 7.92 \\
            24 & 2.504852e-11 & 8.03 & 2.774863e-11 & 8.02 & 3.904232e-11 & 8.00 \\
            48 & 6.442763e-14 & 8.60 & 7.456438e-14 & 8.54 & 1.558753e-13 & 7.97 \\
		\end{tabular*}
		{\rule{\temptablewidth}{0.1pt}}
	\end{center}
	\vspace{-1mm} \caption{\label{1d-accuracy-8} 1-D accuracy test: errors and convergence orders of 8th-order
                  compact GKS with $\Delta t = 0.5 \Delta x ^{2}$ at $t=200$.}
	\small
	\begin{center}
		\def\temptablewidth{0.90\textwidth}
		{\rule{\temptablewidth}{0.70pt}}
		\begin{tabular*}{\temptablewidth}{@{\extracolsep{\fill}}c|cc|cc|cc}
			
			mesh number & $L^1$ error & Order & $L^2$ error & Order& $L^{\infty}$ error & Order  \\
			\hline
            6  & 6.368388e-05 & ~    & 6.754696e-05 & ~    & 9.552583e-05 & ~    \\
            12 & 1.584507e-05 & 2.01 & 1.785133e-05 & 1.92 & 2.522884e-05 & 1.92 \\
            24 & 5.945625e-07 & 4.74 & 6.633959e-07 & 4.75 & 9.381550e-07 & 4.75 \\
            48 & 1.902794e-08 & 4.97 & 2.115510e-08 & 4.97 & 2.991779e-08 & 4.97 \\
		\end{tabular*}
		{\rule{\temptablewidth}{0.1pt}}
	\end{center}
	\vspace{-1mm} \caption{\label{1d-accuracy-5} 1-D accuracy test: errors and convergence orders of WENO-5Z (GKS flux)
                  with $\Delta t=0.5\Delta x^{2}$ at $t=200$.}
\end{table}

\begin{table}[!h]
	\small
	\begin{center}
		\def\temptablewidth{0.90\textwidth}
		{\rule{\temptablewidth}{0.70pt}}
		\begin{tabular*}{\temptablewidth}{@{\extracolsep{\fill}}c|cc|cc|cc}
			
			mesh number& $L^1$ error & Order & $L^2$ error & Order& $L^{\infty}$ error & Order  \\
            \hline
            6  & 8.002597e-06 & ~    & 8.568390e-06 & ~    & 1.200379e-05 & ~    \\
            12 & 4.499126e-07 & 4.15 & 5.007476e-07 & 4.10 & 7.047701e-07 & 4.09 \\
            24 & 2.815459e-08 & 4.00 & 3.128317e-08 & 4.00 & 4.418423e-08 & 4.00 \\
            48 & 1.763045e-09 & 4.00 & 1.958421e-09 & 4.00 & 2.768812e-09 & 4.00 \\
            96 & 1.102605e-10 & 4.00 & 1.224735e-10 & 4.00 & 1.733280e-10 & 4.00 \\
            192& 6.878465e-12 & 4.00 & 7.643323e-12 & 4.00 & 1.109268e-11 & 3.97 \\
		\end{tabular*}
		{\rule{\temptablewidth}{0.1pt}}
	\end{center}
	\vspace{-1mm} \caption{\label{1d-accuracy-8-1} 1-D accuracy test: errors and convergence orders of
                  compact 8th-order GKS with $CFL=0.8$ at $t=200$.}
\end{table}

In order to validate the robustness of the compact 8th-order scheme, the accuracy test is also done with a large CFL number $CFL=0.8$.
For acoustics applications, a large CFL number is preferred.
The results in Table \ref{1d-accuracy-8-1} demonstrate that
the compact 8th-order scheme still has very small error on the coarse mesh even though the order of accuracy decreases to four, which is consistent with temporal convergence order.

\begin{figure}[!htb]
\centering
\includegraphics[width=0.45\textwidth]{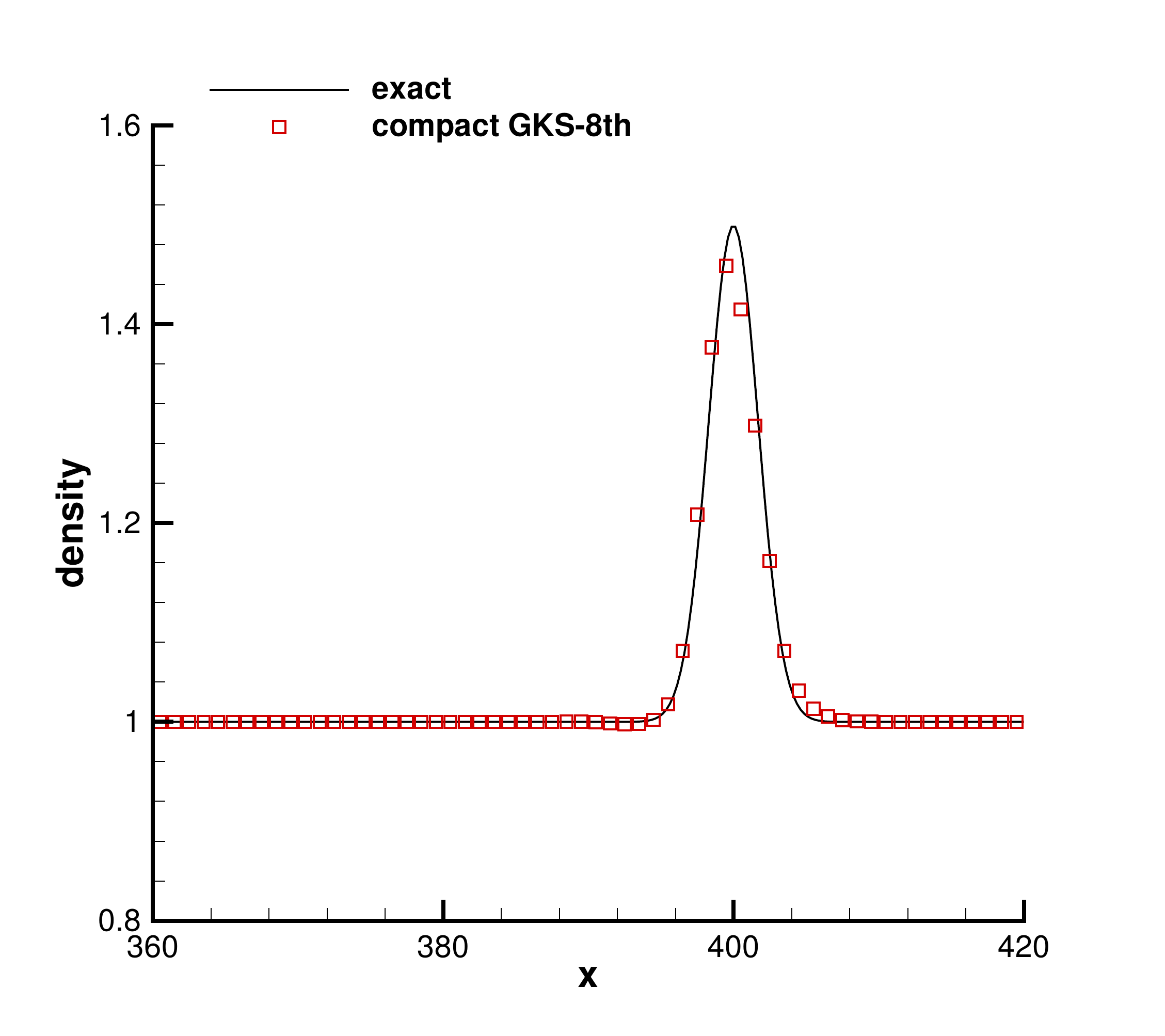}
\includegraphics[width=0.45\textwidth]{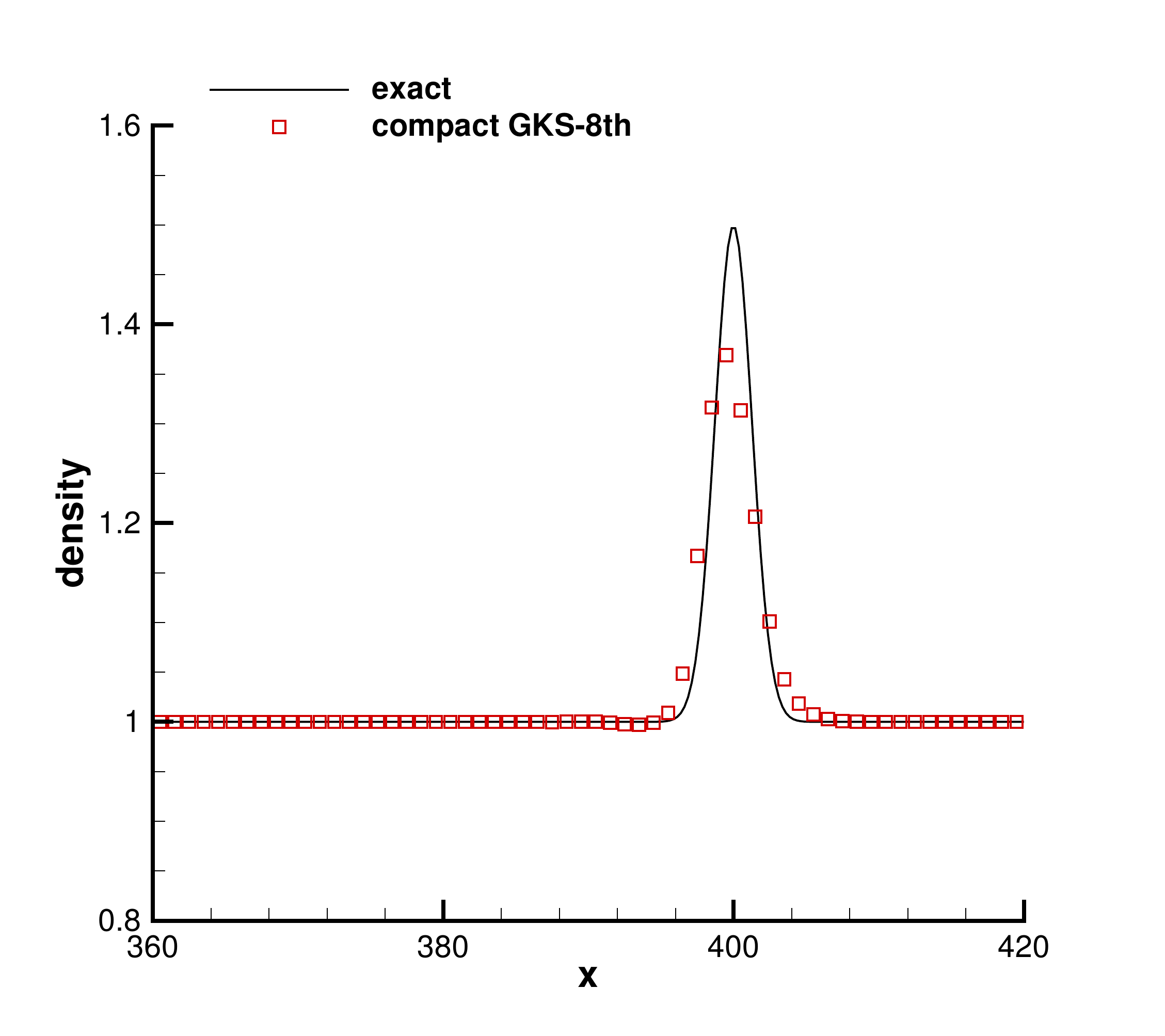}
\caption{\label{1d-advection} Linear advection problem: the results of compact 8th-order GKS at $t=400$ with $b=2.0$ and $b=1.5$. The mesh size is $1$.  }
\end{figure}

\subsection{Linear advection in a long distance}
The test is a problem of advection of 1-D entropy wave in a stationary mean flow. The initial condition is defined as
\begin{align*}
&\rho= 1 + 0.5 e^{-ln2\cdot x^2/b^2},\\
& p=1, \\
& U=1,
\end{align*}
where different $b=2.0$ and $b=1.5$ are tested respectively. The computational domain is $[-800,1000]$. Free flow boundary condition is adopted. The mesh size is $1$ and the CFL number is 0.3.
A similar test is also presented in \cite{DGacoustic2019}, where the linear advection equation is solved directly and a very small time step $(CFL<0.05)$ is adopted for their high-order finite difference schemes and DG schemes.
Fig. \ref{1d-advection} shows the results of compact GKS. In comparison with the results of high-order finite difference schemes, DRP scheme, and DG schemes \cite{DGacoustic2019}, the compact GKS obviously presents more accurate solutions.

\begin{figure}[!htb]
\centering
\includegraphics[width=0.425\textwidth]{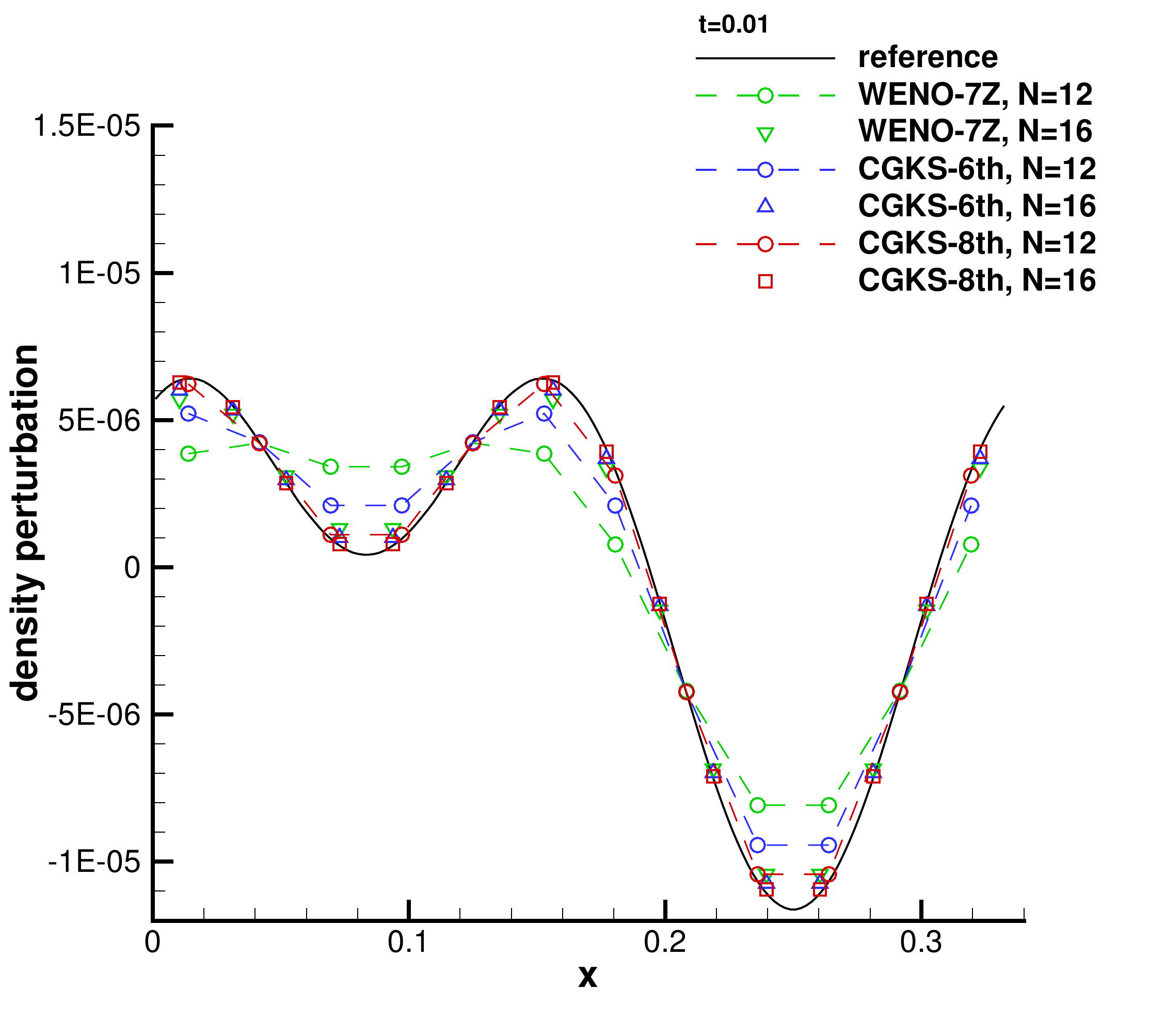}
\includegraphics[width=0.425\textwidth]{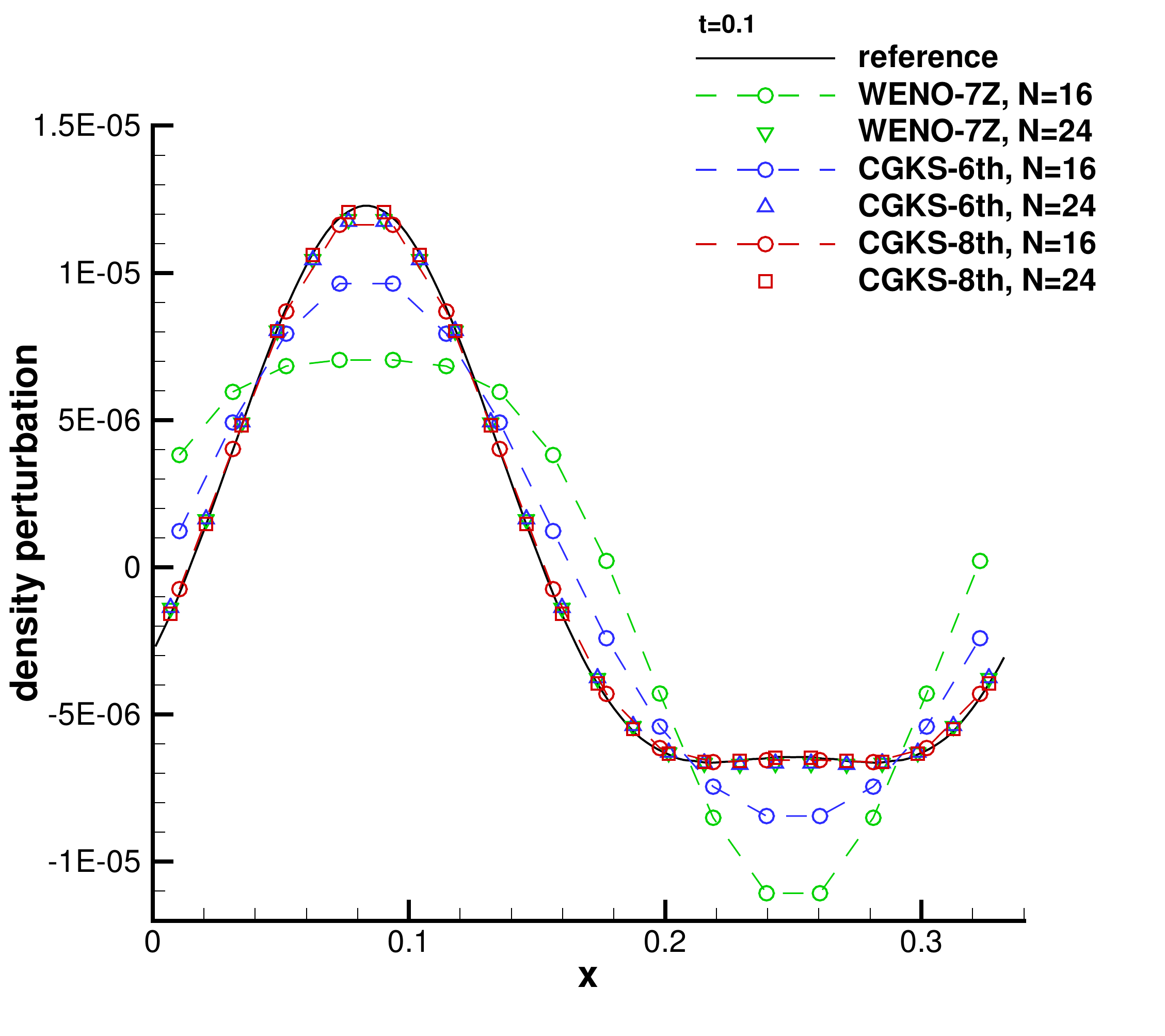}
\caption{\label{1d-acoustic-1} Propagation of 1-D acoustic wave: density perturbation of compact 8th-order GKS and non-compact WENO-7Z with GKS flux at t=0.01 (left) and t=0.1 (right).  The ratio
of non-dimensional initial density wavelength to sound speed is $\lambda_{\rho_0}/a_{\infty}=4.8\times 10^{-4}$.}
\end{figure}

\subsection{Propagation of 1-D acoustic wave}
The quantities concerned in acoustics simulation are the spectrum of the radiating sound waves in the far field.
The sound wave propagates in a very long distance. In order to get the accurate solution after long distance travel,
the numerical scheme should have minimal dispersion and dissipation error.
The 1-D acoustic wave traveling in a very long distance is computed to validate the low dispersion and dissipation error in the compact GKS.
The initial condition is given as follows \cite{bai}
\begin{align*}
&U=U_{\infty}+\delta U, \delta U=\epsilon a_{\infty} \cos(\omega x), U_{\infty}=0,\\
&\rho= \rho_{\infty}+\delta \rho,\delta \rho=\epsilon \rho_{\infty} \cos(2\omega x), \rho_{\infty}=1.1771,\\
&\frac{p}{p_{\infty}}=(\frac{\rho}{\rho_{\infty}})^r, p_{\infty}=101325.0,\\
&a_{\infty}=\sqrt{\gamma\frac{p_{\infty}}{\rho_{\infty}}},
\end{align*}
where $\epsilon=10^{-5}$ is the magnitude of initial perturbation, and $\omega=6\pi$ is the wavenumber of initial perturbations in velocity. The computational domain is $[0,1/3]$. Periodic boundary conditions on both sides are adopted.
For comparison, a non-compact WENO-7Z is constructed with the reconstruction method in \cite{WENO-Z}, and the same GKS flux and temporal discretization are used, where $8$ cells must be used in the reconstruction to get the point-wise value at a cell interface in the
non-compact WENO-7Z method.

The coarse meshes are set to use 6 to 8 mesh points per wavelength. In particular, the CFL number $CFL=0.2$ is used in the test in order to
avoid the accumulation of temporal discrete error in the long time propagation. The current CFL number $CFL=0.2$ is much larger than those commonly used by the high-resolution finite difference methods in \cite{bai, tam1993}.
Fig. \ref{1d-acoustic-1} shows  the results of the compact GKS and non-compact WENO-7Z \cite{WENO-Z} at $t=0.01$ (left) and $t=0.1$ (right) with  long propagating distance of $10L$ and $100L$ respectively ($L=1/3$). The compact 8th-order GKS can give accurate results without obvious dispersion and  dissipation error on coarse meshes ($N=12$ and $N=16$). While the compact 6th-order GKS and WENO-7Z give results with large numerical dissipation at $t=0.01$ and large error at $t=0.1$ on the coarse mesh case. When refining the mesh to a total of $24$ mesh points, the solutions from 6th-order GKS and WENO-7Z are much improved. Comparing the results of compact 6th-order GKS and non-compact WENO-7Z, the compact 6th-order GKS has better resolution than the non-compact WENO-7Z. It shows that the compact stencil with local independent cell averages and slopes is reliable to
get accurate physical solution rather than the use of a large stencil with the collection of information which are physically irrelevant to the local cell, especially in the coarse mesh case.

\subsection{Flow with discontinuities}
In order to verify the shock capturing property of the compact GKS, the tests with discontinuous solutions are computed.
The first test is about linear advection of discontinuities \cite{tam1995AIAA,cheng2019assessment}. The initial condition is
\begin{align*}
&\rho=1+0.5(H(x+50)-H(x-50)), \\
& p=1, \\
& U=1.
\end{align*}
The computation domain is $[-420,800]$. Free flow boundary condition is adopted.
The second test is the one-dimensional shock-tube problem.
The initial condition is
\begin{equation*}
(\rho,U,p) = \begin{cases}
(0.445, 0.698, 3.528),   0\leq x<0.5,\\
(0.5, 0, 0.571), 0.5\leq x\leq1.
\end{cases}
\end{equation*}
The shock wave, contact discontinuity, and rarefaction wave emerge from initial condition.

\begin{figure}[!htb]
\centering
\includegraphics[width=0.45\textwidth]{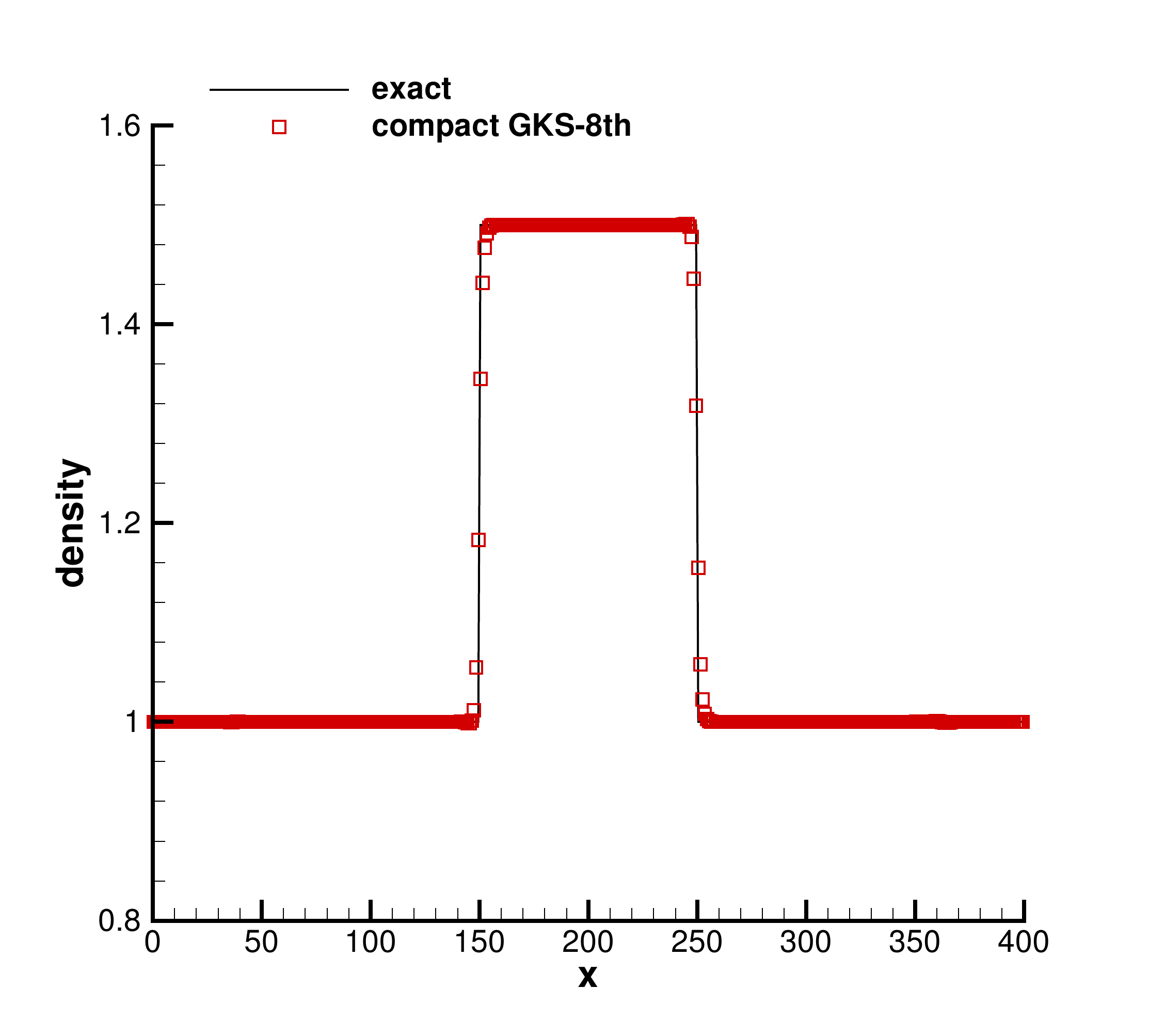}
\includegraphics[width=0.45\textwidth]{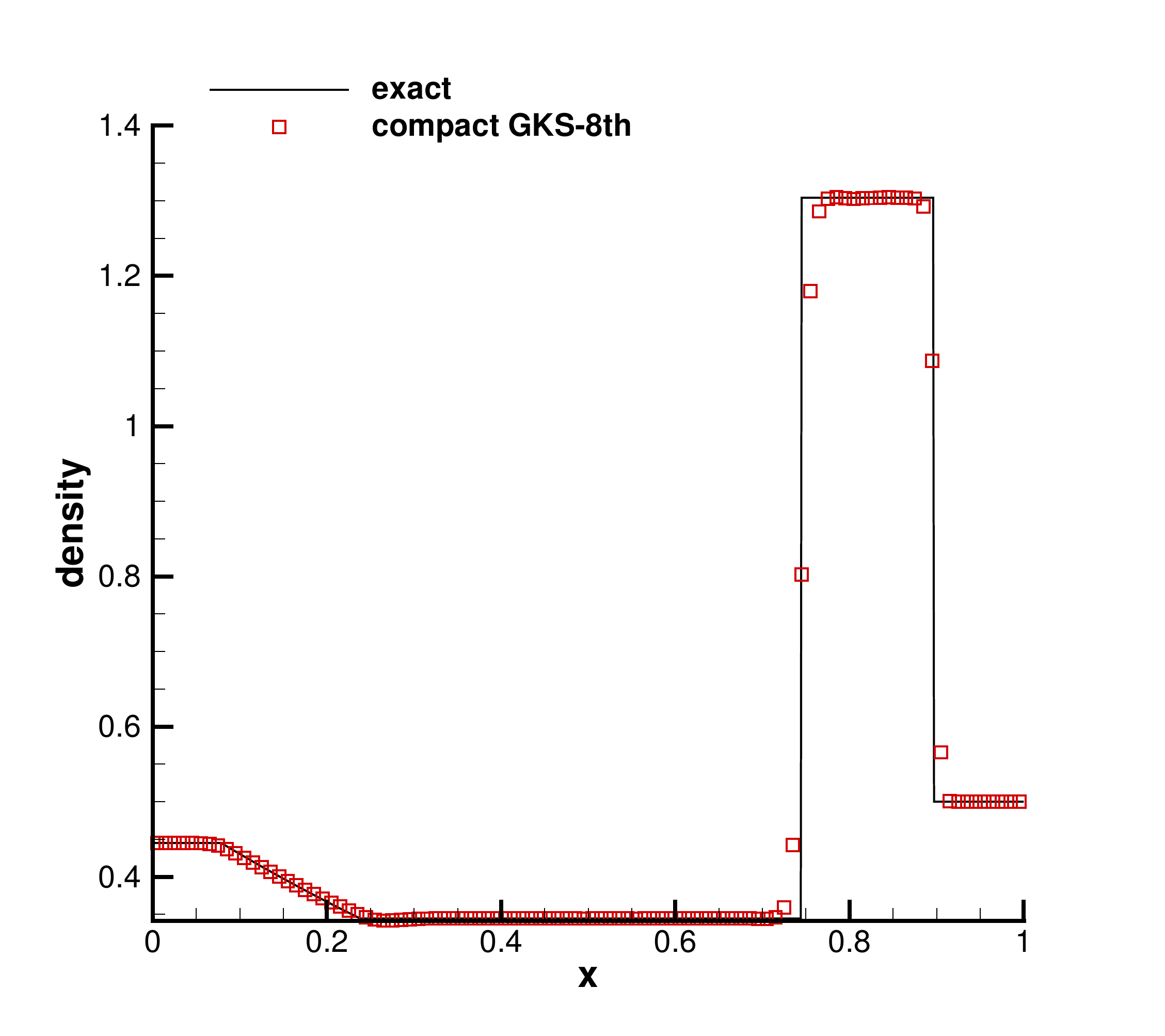}
\caption{\label{1d-discon-capture} 1-D test with discontinuous solutions: the left is the result of linear advection of discontinuities at $t=200$ with mesh size $\Delta x=1$. The right is the result of one-dimensional shock tube test at $t=0.16$ with mesh size $\Delta x=0.01$. }
\end{figure}

The results by compact 8th-order GKS are shown in Fig. \ref{1d-discon-capture}. The left figure in Fig. \ref{1d-discon-capture} shows the local enlargement of discontinuity at $t=200$ with $\Delta x=1$. The right figure in Fig. \ref{1d-discon-capture} is the numerical solution at $t=0.16$ with $\Delta x=0.01$.
There is no obvious oscillation near discontinuities in both tests. In comparison with the results in the reference papers, the compact GKS
can give very accurate discontinuous transition.

\begin{figure}[!htb]
\centering
\includegraphics[width=0.495\textwidth]{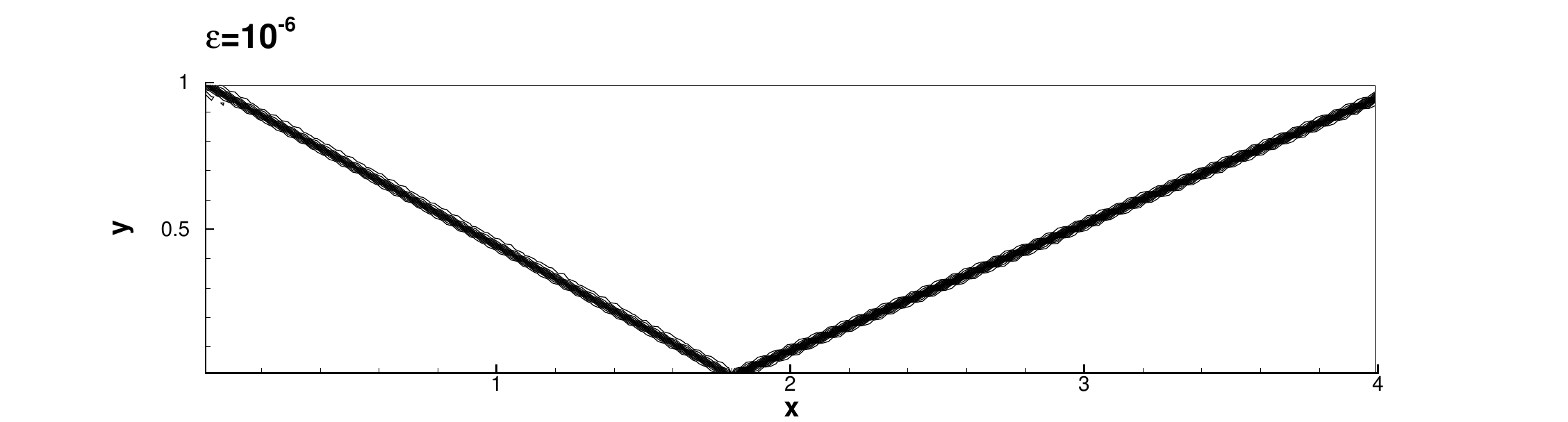}
\includegraphics[width=0.495\textwidth]{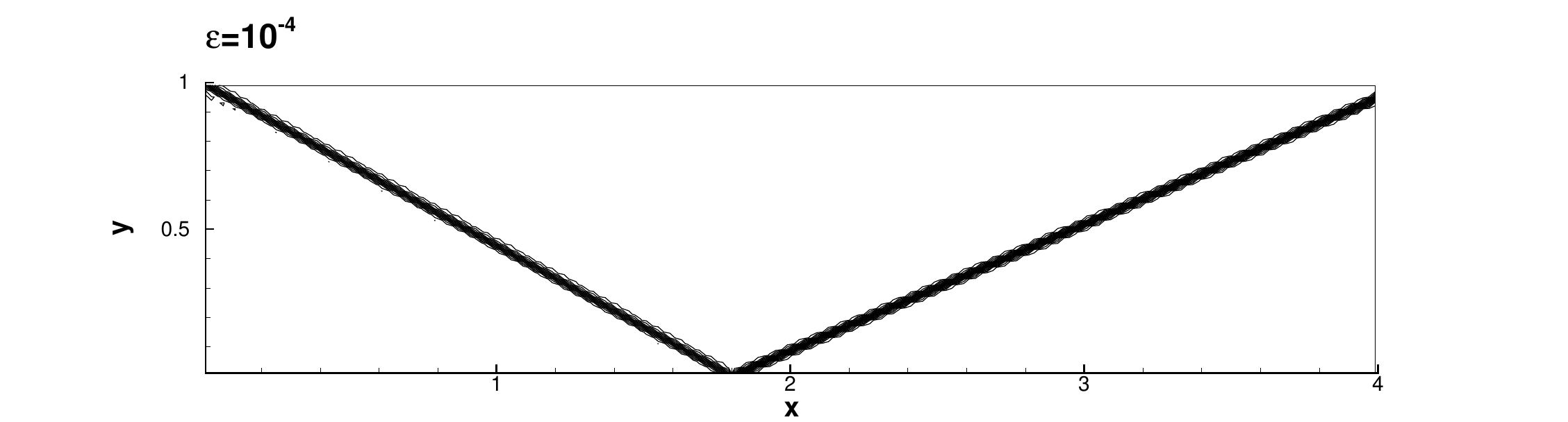}
\caption{\label{shock-reflection-1} Regular shock reflection with $Ma_{\infty}=2.9$: density of compact 8th-order GKS at $t=50$ with $200\times 50$ mesh points. Here 30 equal-spaced contours from $0.999$ to $2.7$ are plotted. }
\end{figure}

\begin{figure}[!htb]
\centering
\includegraphics[width=0.425\textwidth]{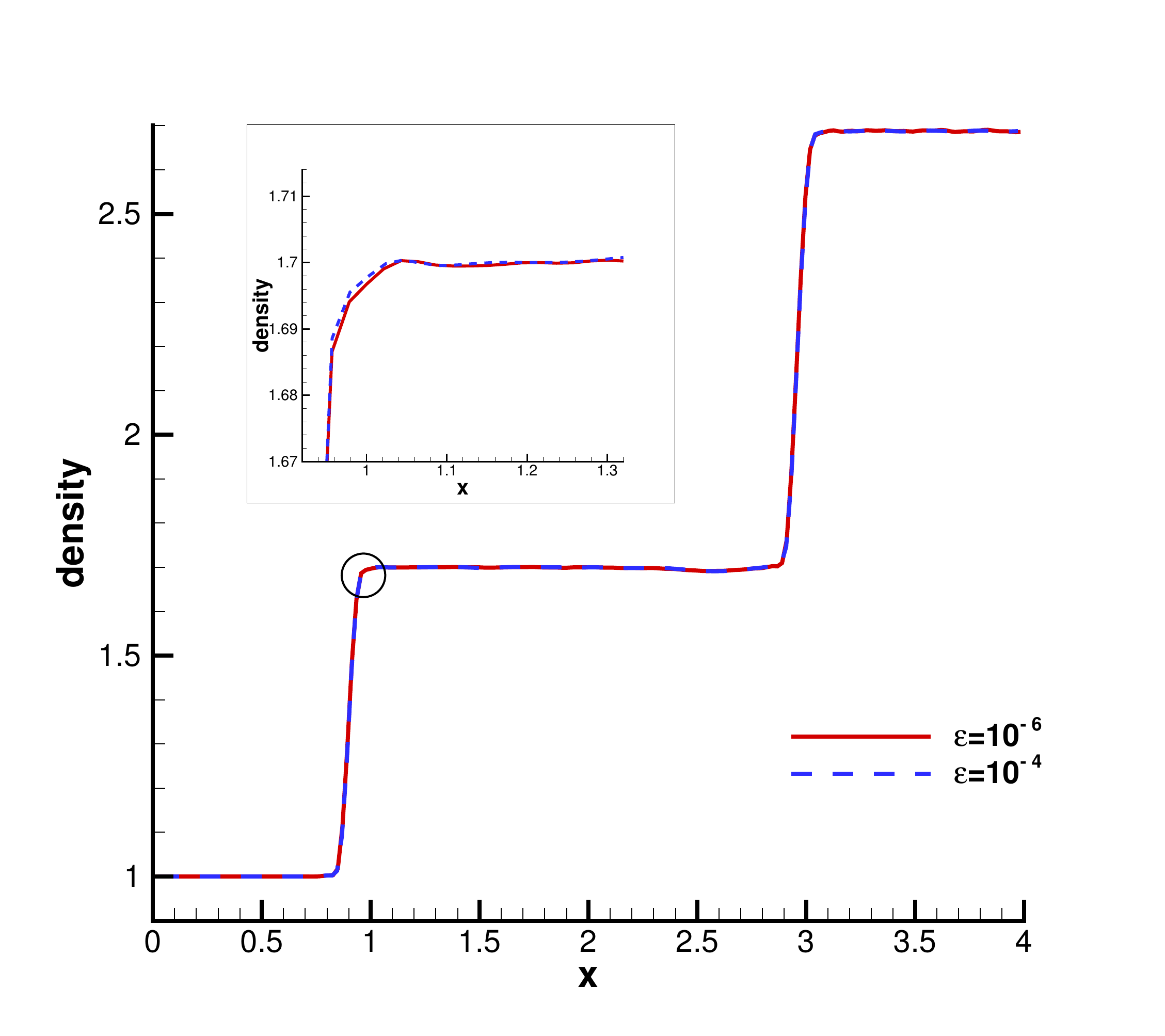}
\includegraphics[width=0.425\textwidth]{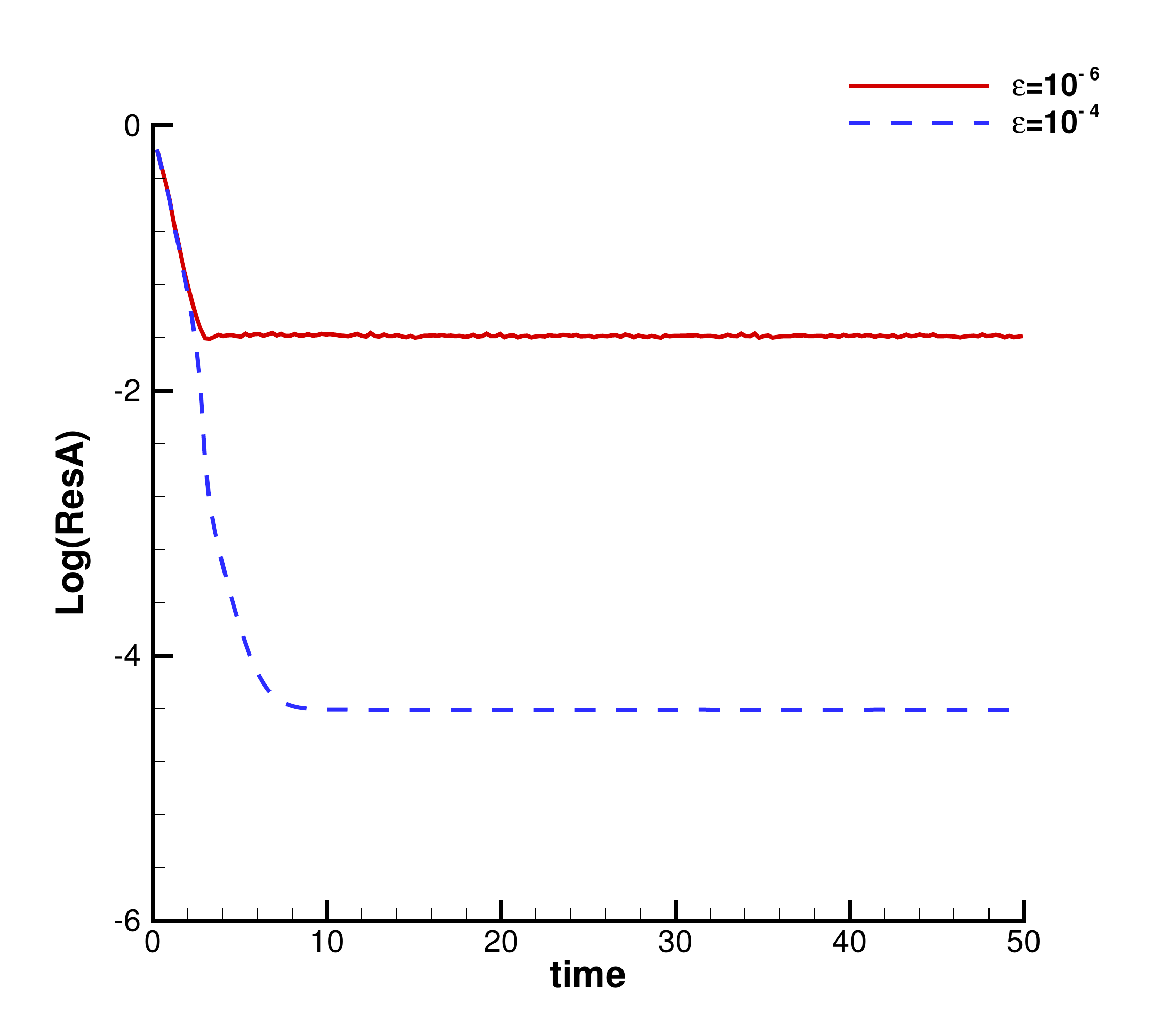}
\caption{\label{shock-reflection-2} Regular shock reflection with $Ma_{\infty}=2.9$: the left is density distribution along $y=0.5$ and local enlargement of compact 8th-order GKS along y=0.5 at $t=50$ with $200\times 50$ mesh points; the right is the evolution of the average residue. }
\end{figure}

The two-dimensional regular shock reflection problem \cite{shockreflection2011} is also tested to validate the compact GKS for shock waves.
The computational domain is $[0,4]\times[0,1]$. The impinging shock wave and the reflected shock wave separate the domain into three parts.
The impinging angle is $\theta=29^\circ$ on the bottom wall. The Mach number of the flow entering from the left boundary is $Ma_{\infty}=2.9$,
and the flow from the top boundary can be determined by the Rankine-Hugoniot relationship.
In our computation, the mesh with $300\times 30$ points is used. The inflow boundary condition is adopted for the left and top boundaries.
The reflecting boundary condition is imposed on the bottom wall.

Fig. \ref{shock-reflection-1} shows the density contours of compact GKS, where two different $\epsilon$ in nonlinear weights are tested.
There is no significant spurious oscillation on both sides of the impinging shock wave and reflected shock wave from the two $\epsilon$.
The impinging location at the bottom wall is the same as it in \cite{shockreflection2011}.
In order to confirm in more detail that there is no significant numerical oscillation on either side of the impinging shock wave,
density distribution along the line $y=0.5$ and local enlargement near the impinging shock wave are also presented.
The average residue from  $\epsilon=10^{-6}$ is similar to the results using WNEO-JS and WENO-Z weighting functions in \cite{shockreflection2011}.
The average residue from $\epsilon=10^{-4}$ becomes smaller than that from $\epsilon=10^{-6}$.
In order to further reduce the residue for steady state calculation, the high-order WENO-type reconstruction can be modified \cite{zhu}.

\begin{figure}[!htb]
\centering
\includegraphics[width=0.425\textwidth]{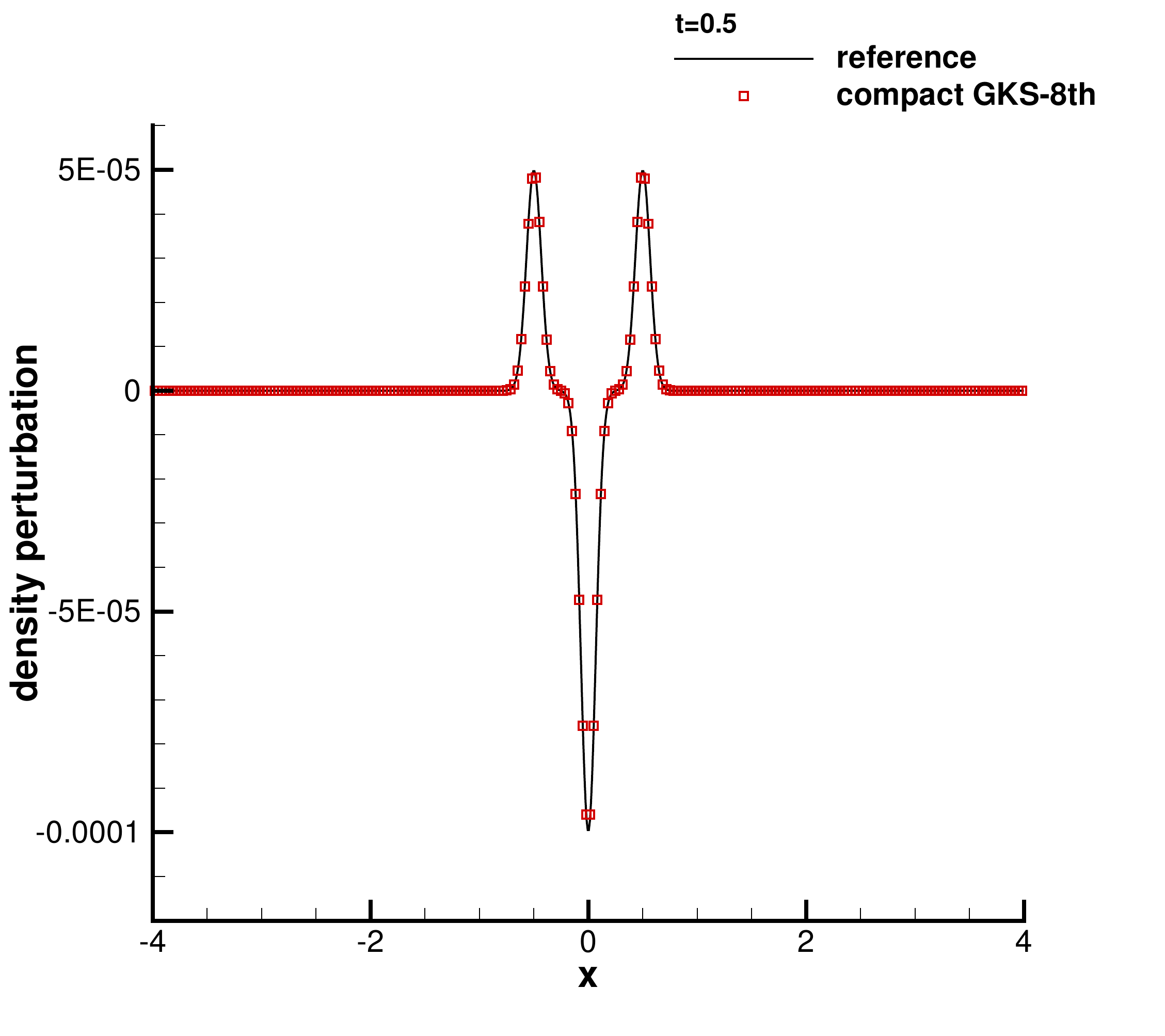}
\includegraphics[width=0.425\textwidth]{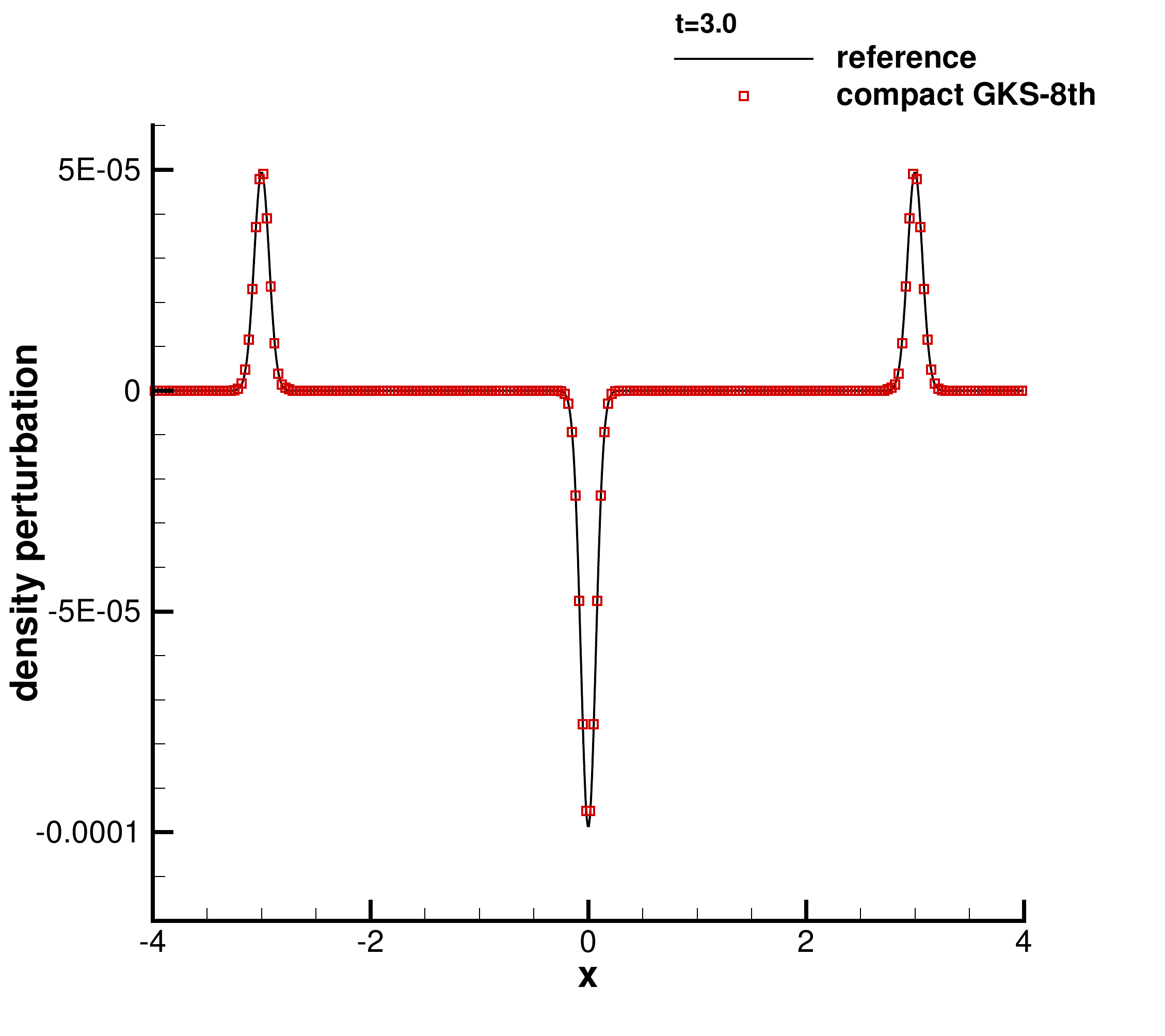}\\
\includegraphics[width=0.425\textwidth]{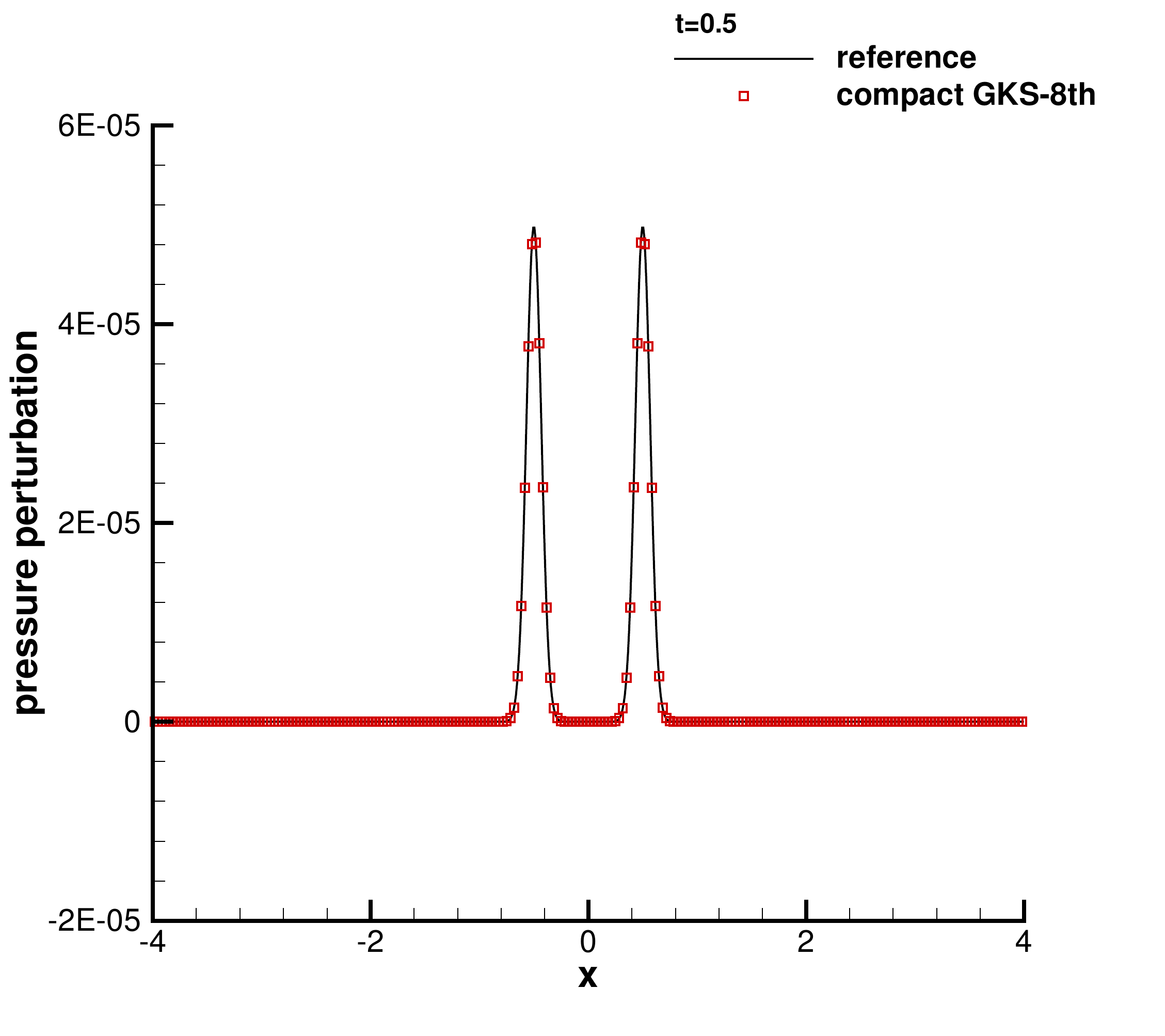}
\includegraphics[width=0.425\textwidth]{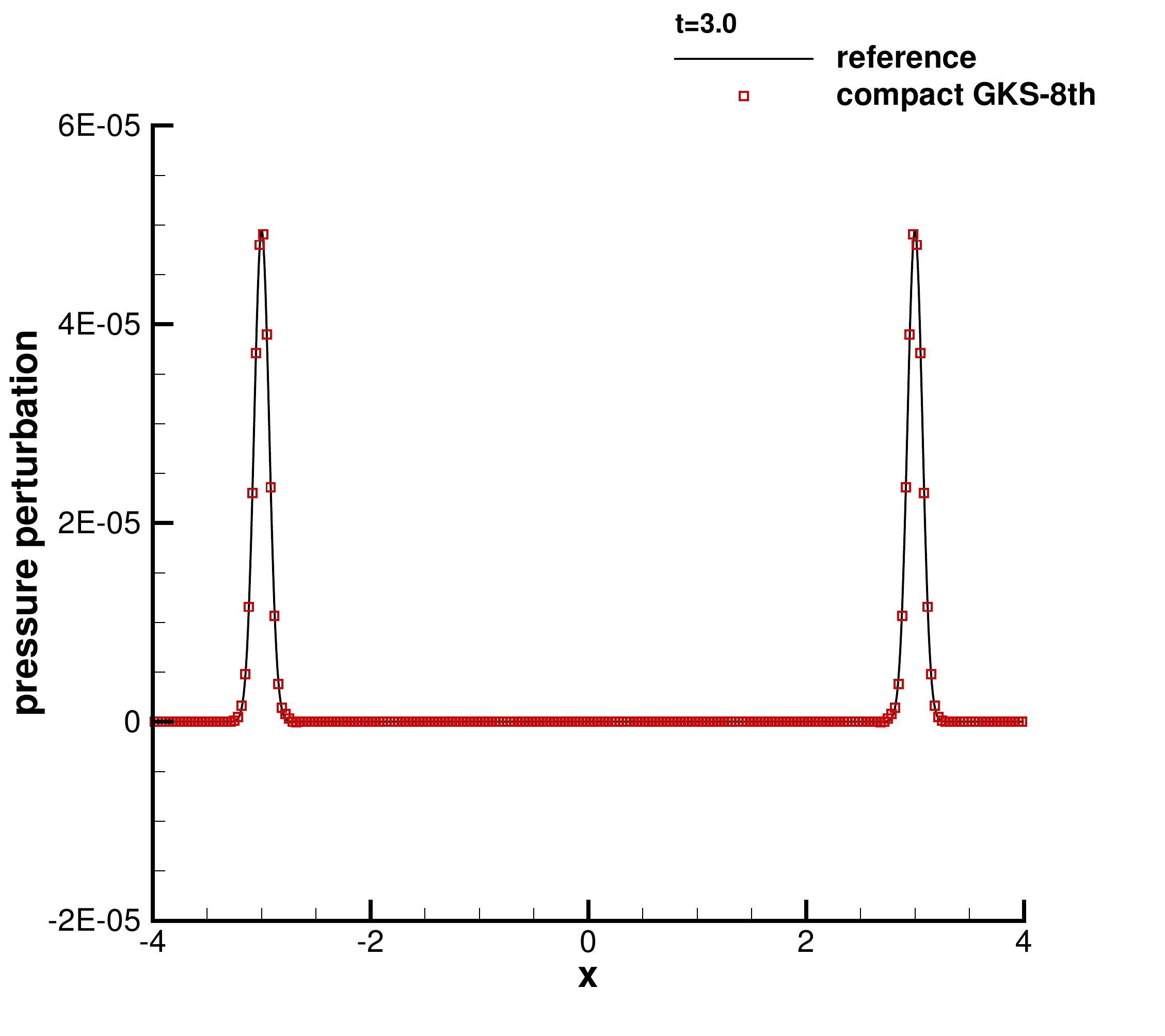}
\caption{\label{plane-prepulse} Propagation of plane pressure pulse: density and pressure perturbation of compact 8th-order GKS along y=0.5 at $t=0.5$ and $t=3.0$ with $300\times 30$ mesh points. }
\end{figure}

\subsection{Propagation of a plane pressure pulse}
The test is that a plane pressure pulse induces the sound waves and entropy waves  in a stationary mean flow \cite{li2006}. The initial condition is defined as
\begin{align*}
&\rho= \rho_{\infty},\\
& p=p_{\infty} +\epsilon e^{-ln2\cdot (x/0.08)^2} ,  \\
& U=V=0, \\
\end{align*}
where $\epsilon=1\times 10^{-4}$ is the magnitude of initial perturbation. The density and pressure of the mean flow are $\rho_{\infty}=1.0$ and $p_{\infty}=1.0/\gamma$. Reynolds number is $Re=5000$, and $Re$ is defined by $Re=\rho_{\infty}a_{\infty}L_{\infty}/\mu, L_{\infty}=0.08$.
The computation domain is $[-5,5]\times[0,1]$. The non-reflection boundary condition is adopted for all four boundaries, and $300\times 30$ mesh points are used.
Fig. \ref{plane-prepulse} shows the density and pressure perturbations from the compact 8th-order GKS along y=0.5 at $t=0.5$ and $t=3.0$.
There are about 10 mesh points in the core region of the sound and entropy waves.
The results demonstrate the high-resolution property of compact GKS. Almost exact solutions can be obtained.

\begin{figure}[!htb]
\centering
\includegraphics[width=0.325\textwidth]{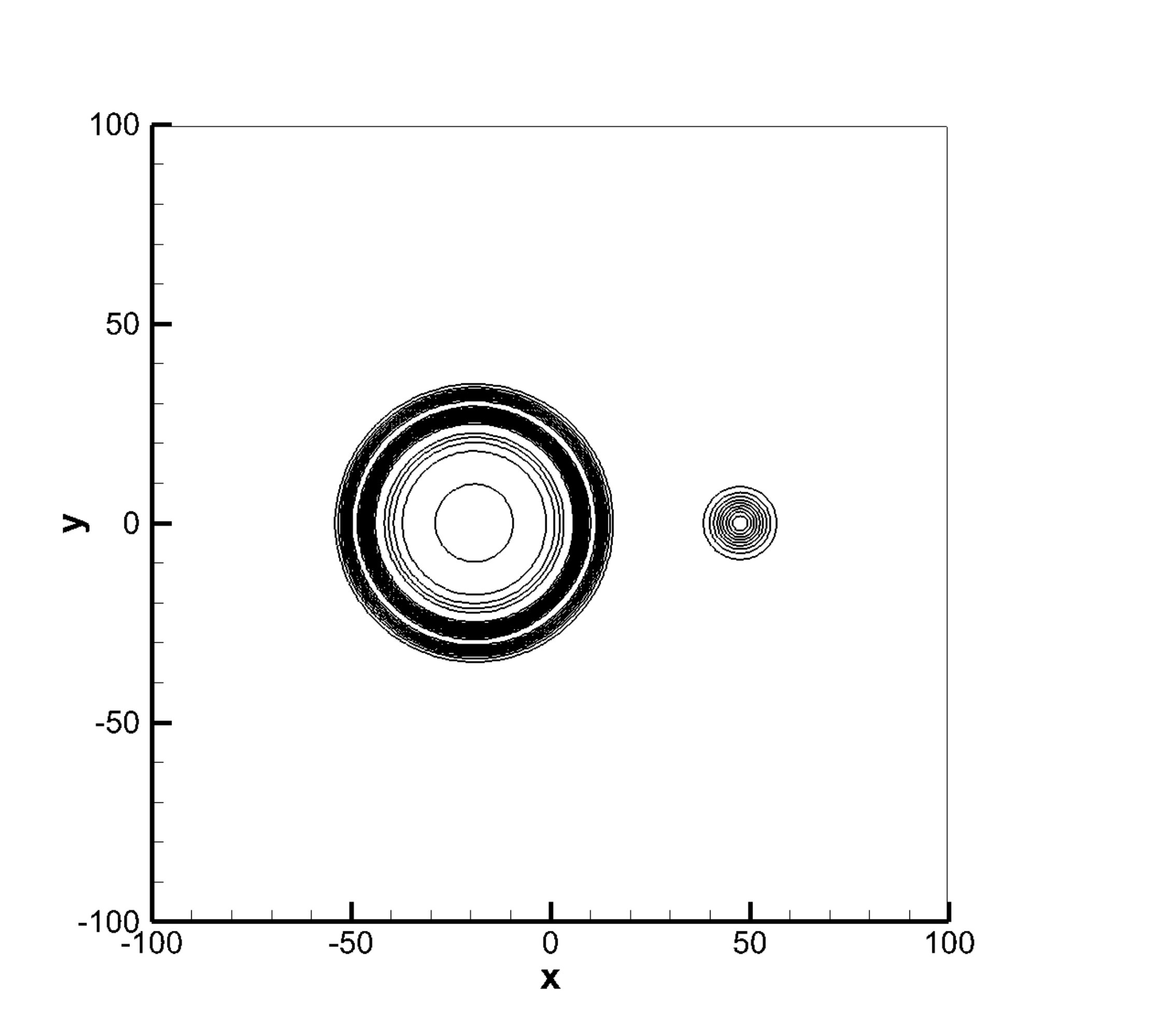}
\includegraphics[width=0.325\textwidth]{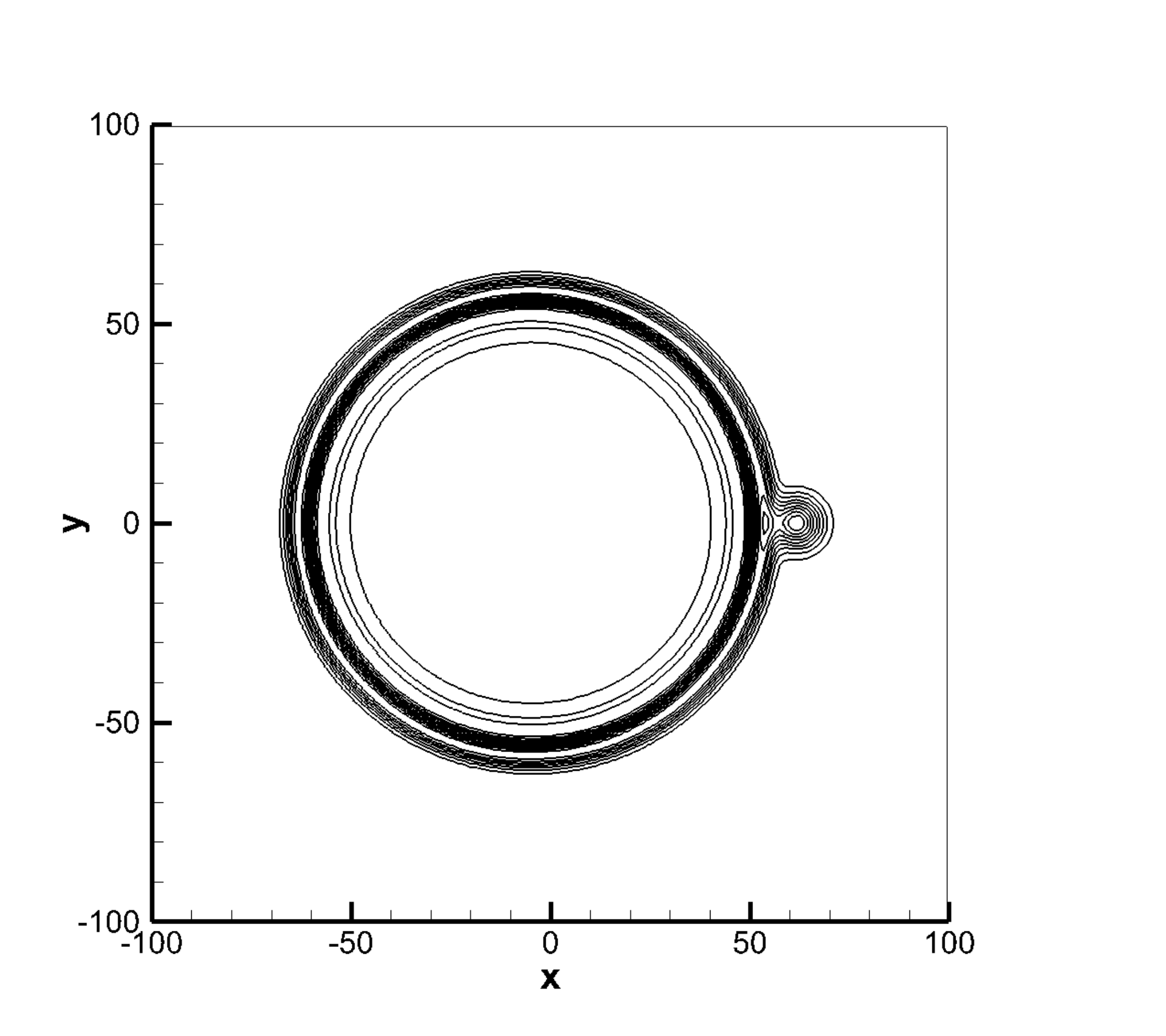}
\includegraphics[width=0.325\textwidth]{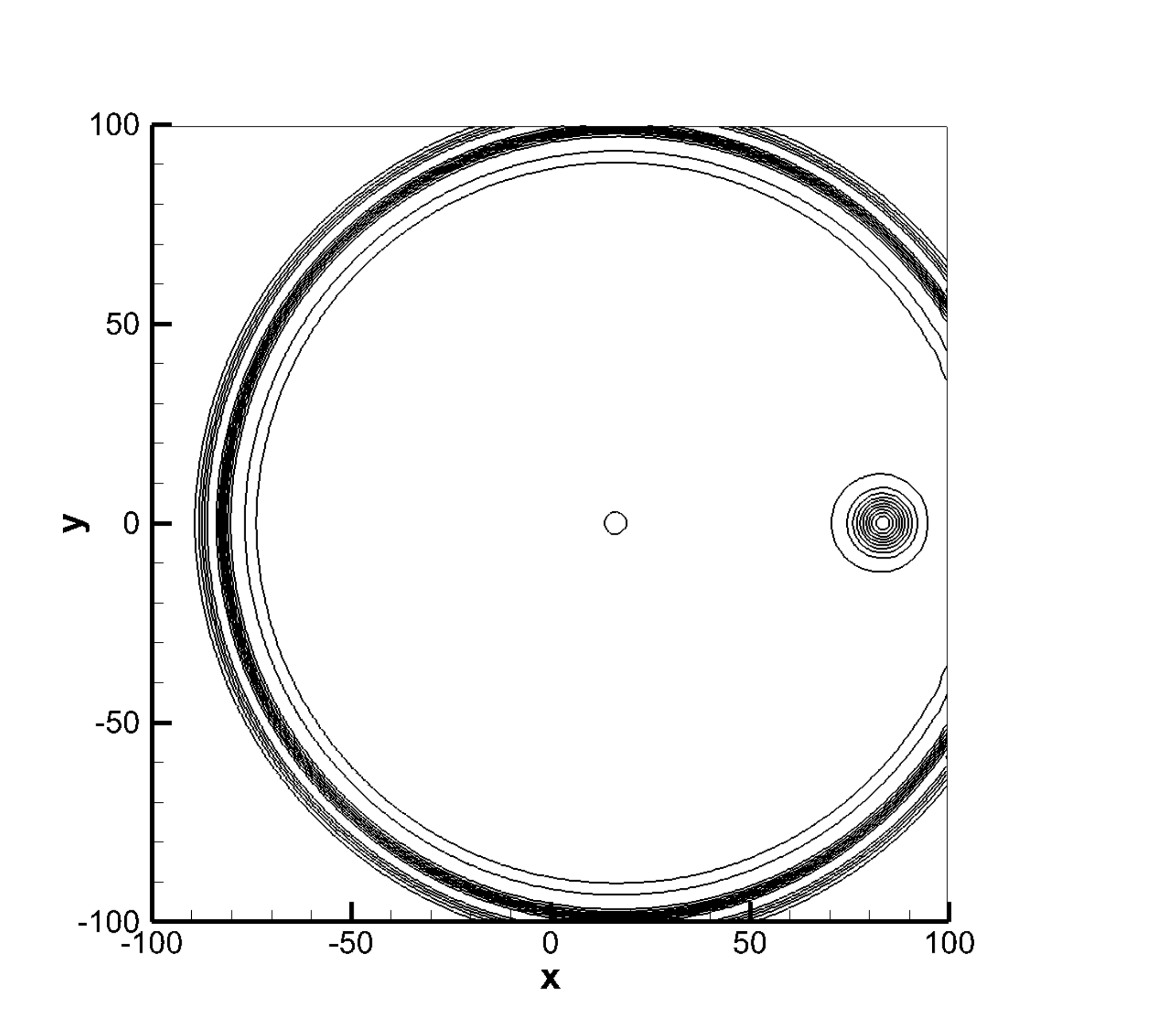}
\caption{\label{multiwave-1} Sound, entropy and vorticity waves test: density perturbation contours in $[-0.0006,0.001]$ with 0.0001 increment obtained by compact 8th-order GKS at $t=28.45, t=56.9$ and $t=100$ with $200\times 200$ mesh points. $t=28.45$ and $t=56.9$ corresponds to 500 and 1000 time steps in \cite{tam1993} respectively, while the compact 8th-order GKS just needs 143 and 286 time steps to get solutions at
the same output times with a $CFL=0.3$.}
\end{figure}

\begin{figure}[!htb]
\centering
\includegraphics[width=0.325\textwidth]{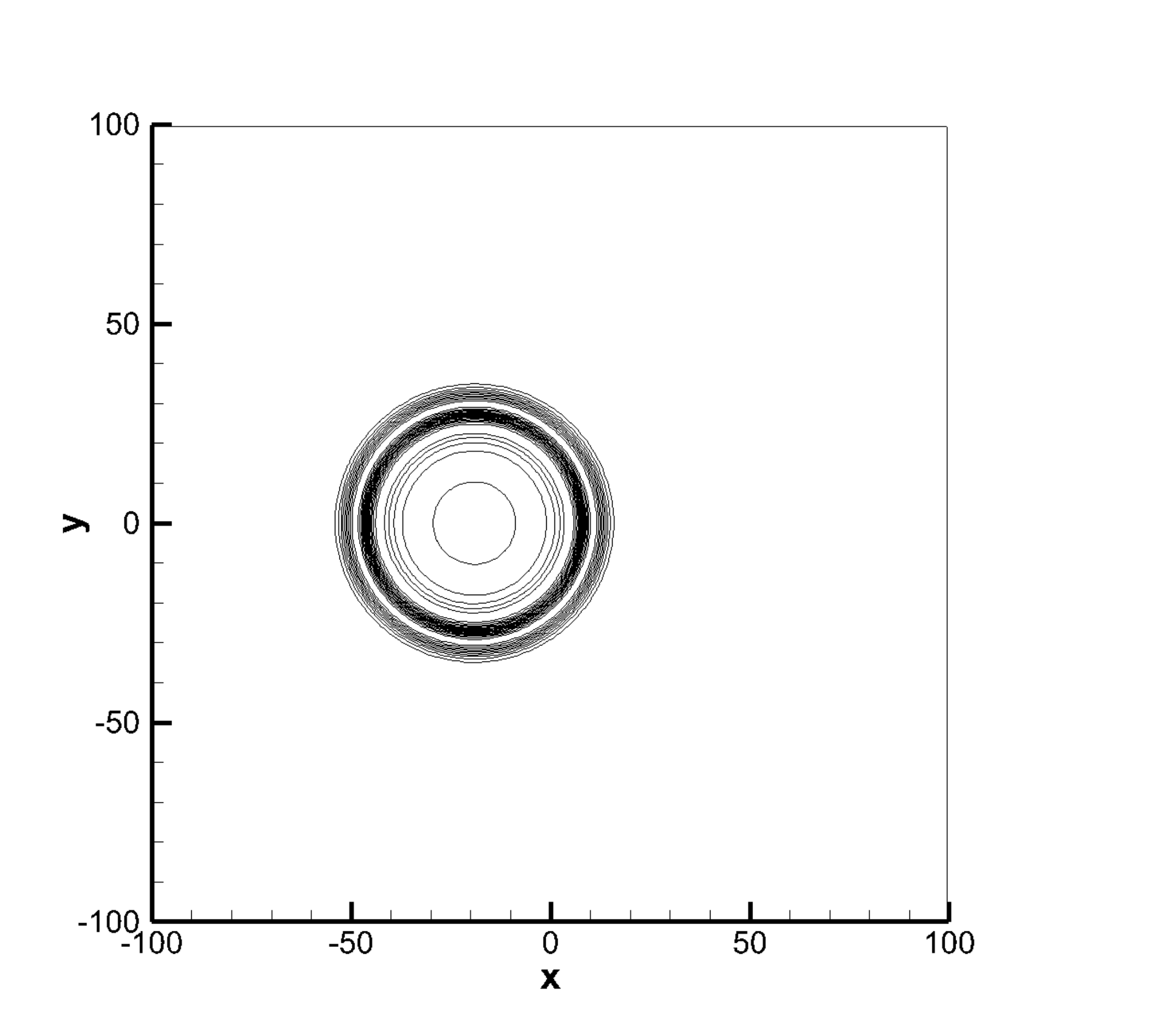}
\includegraphics[width=0.325\textwidth]{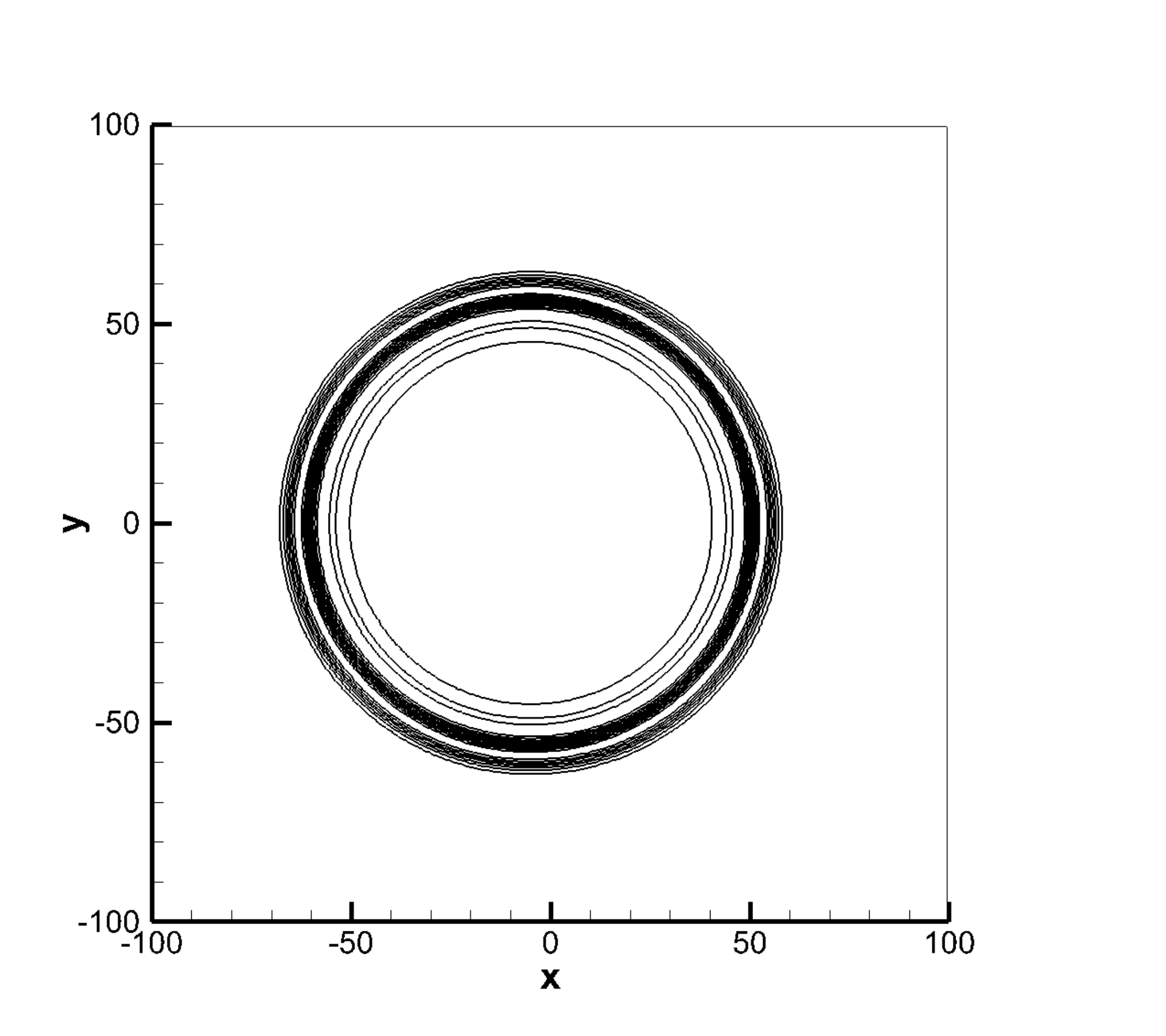}
\includegraphics[width=0.325\textwidth]{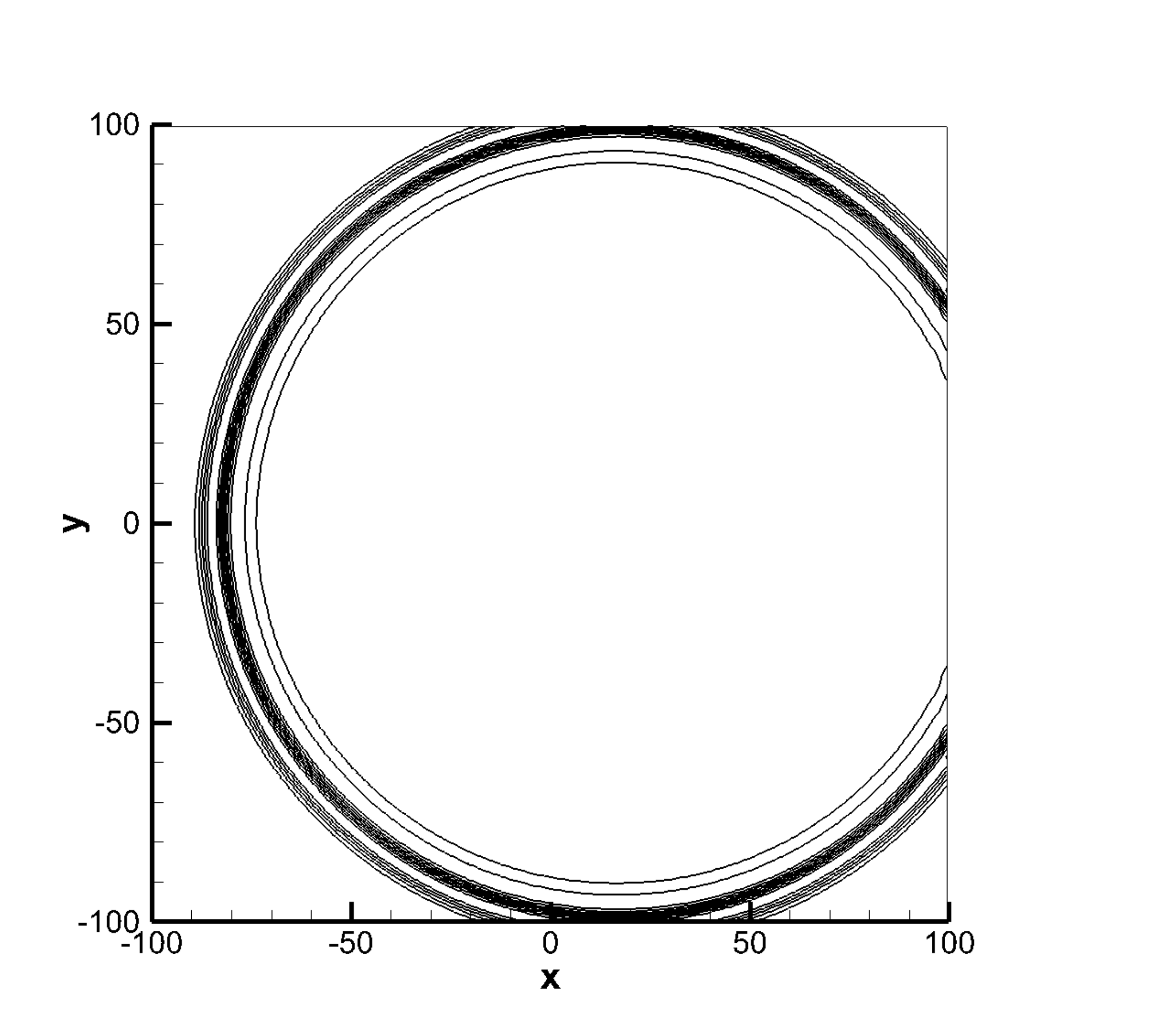}
\caption{\label{multiwave-2} Sound, entropy and vorticity waves test: pressure perturbation contours in $[-0.0006,0.001]$ with 0.0001 increment obtained by compact 8th-order GKS at $t=28.45, t=56.9$ and $t=100$ with $200\times 200$ mesh points.}
\end{figure}

\begin{figure}[!htb]
\centering
\includegraphics[width=0.425\textwidth]{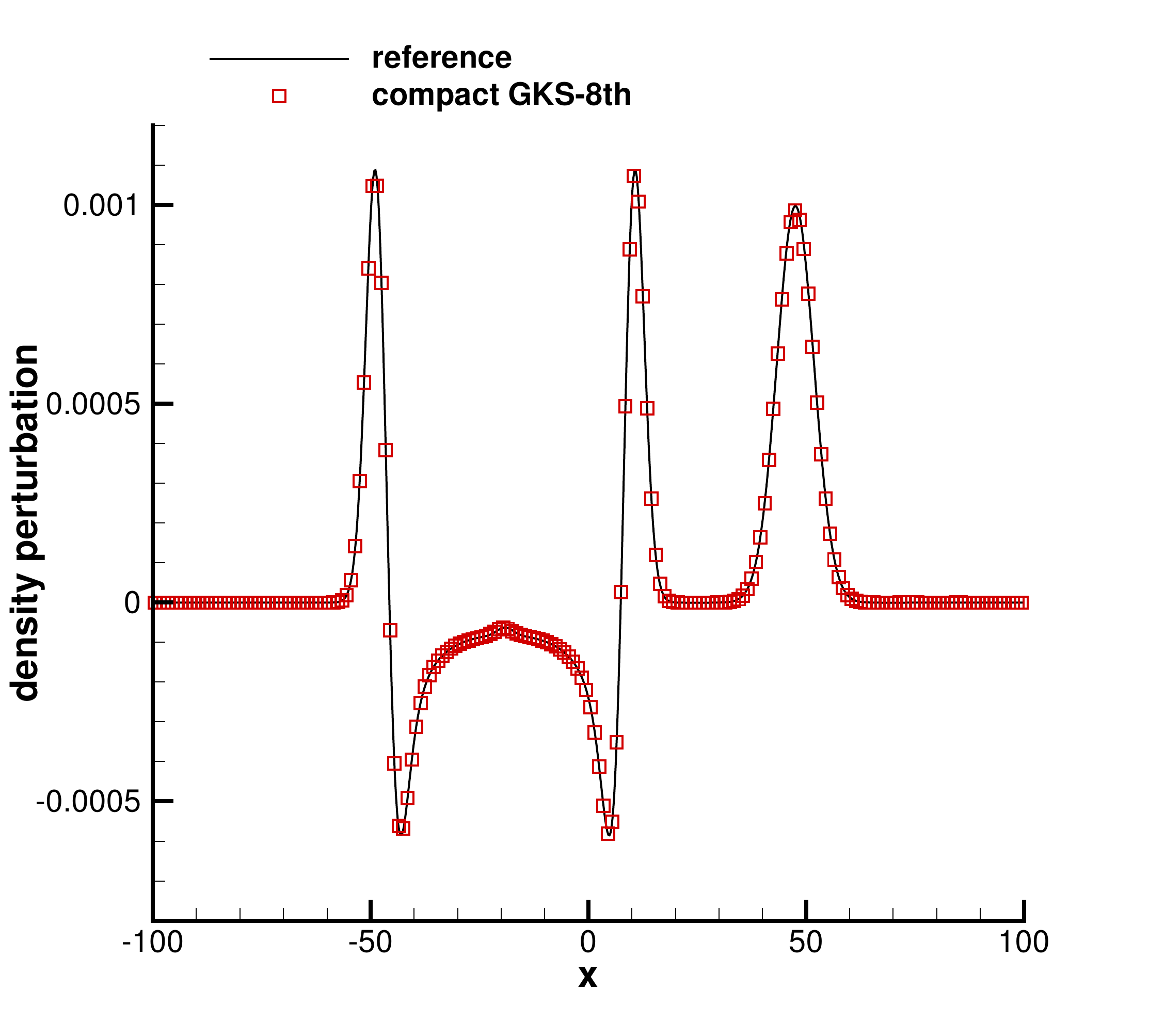}
\includegraphics[width=0.425\textwidth]{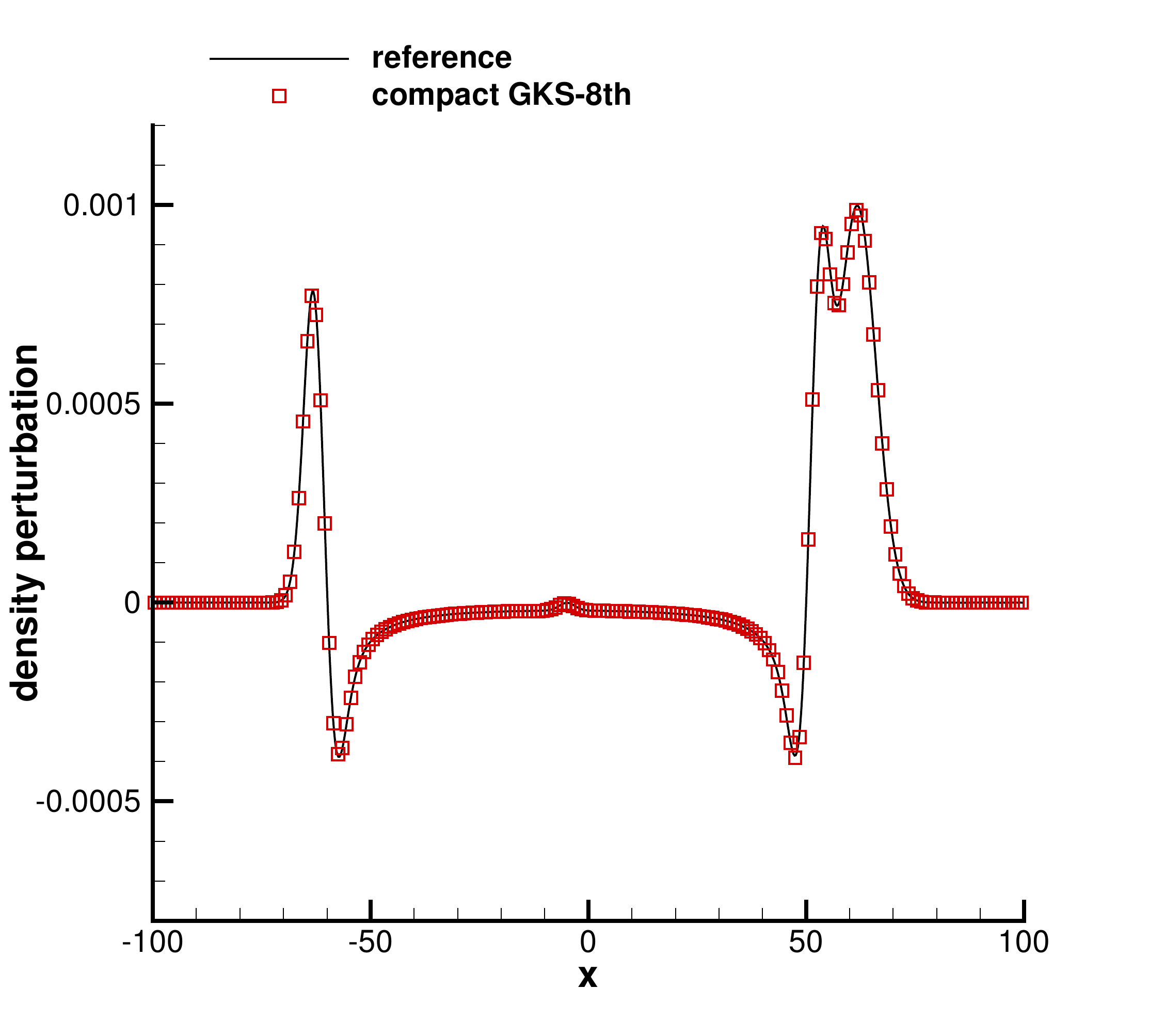}
\includegraphics[width=0.425\textwidth]{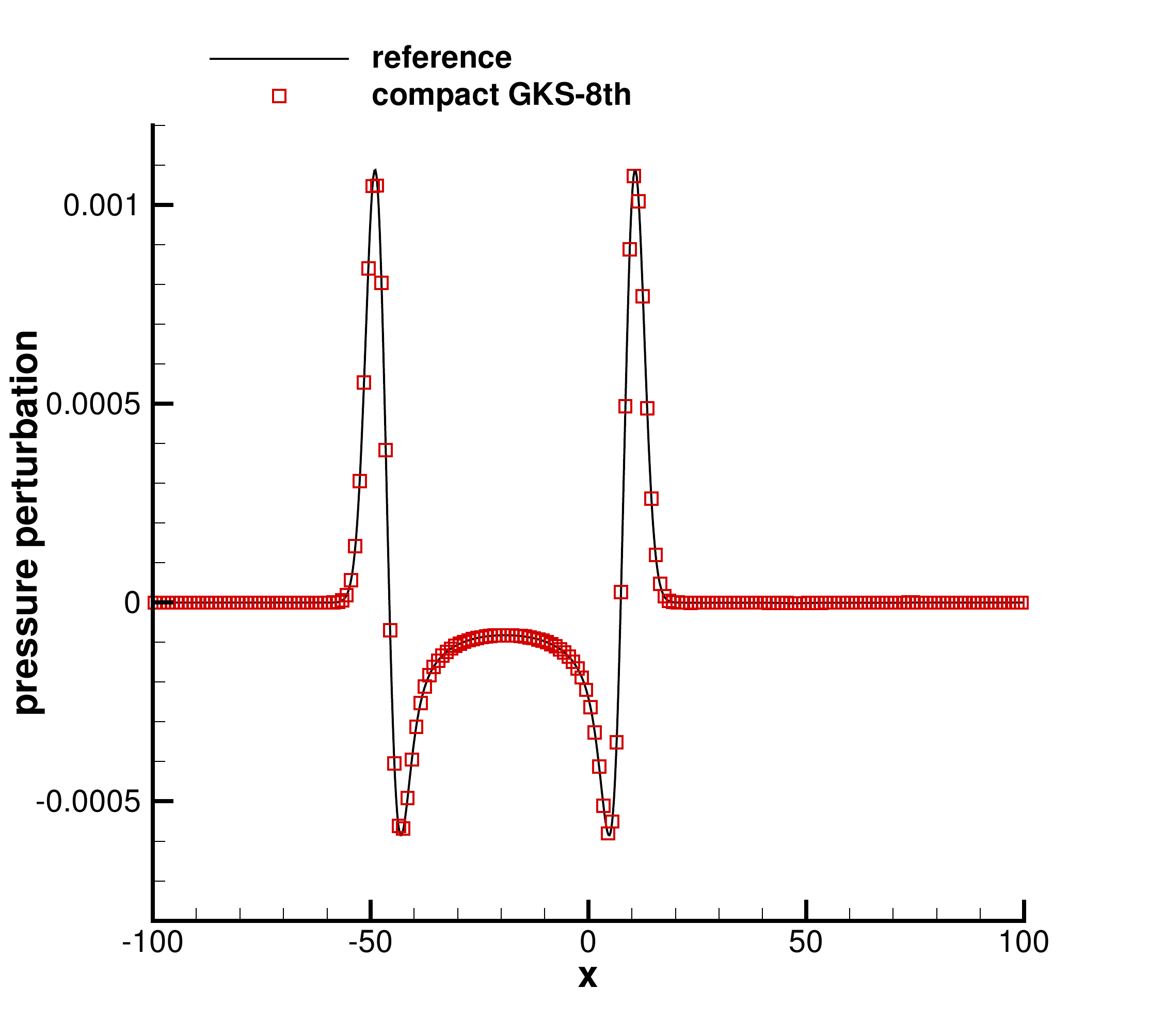}
\includegraphics[width=0.425\textwidth]{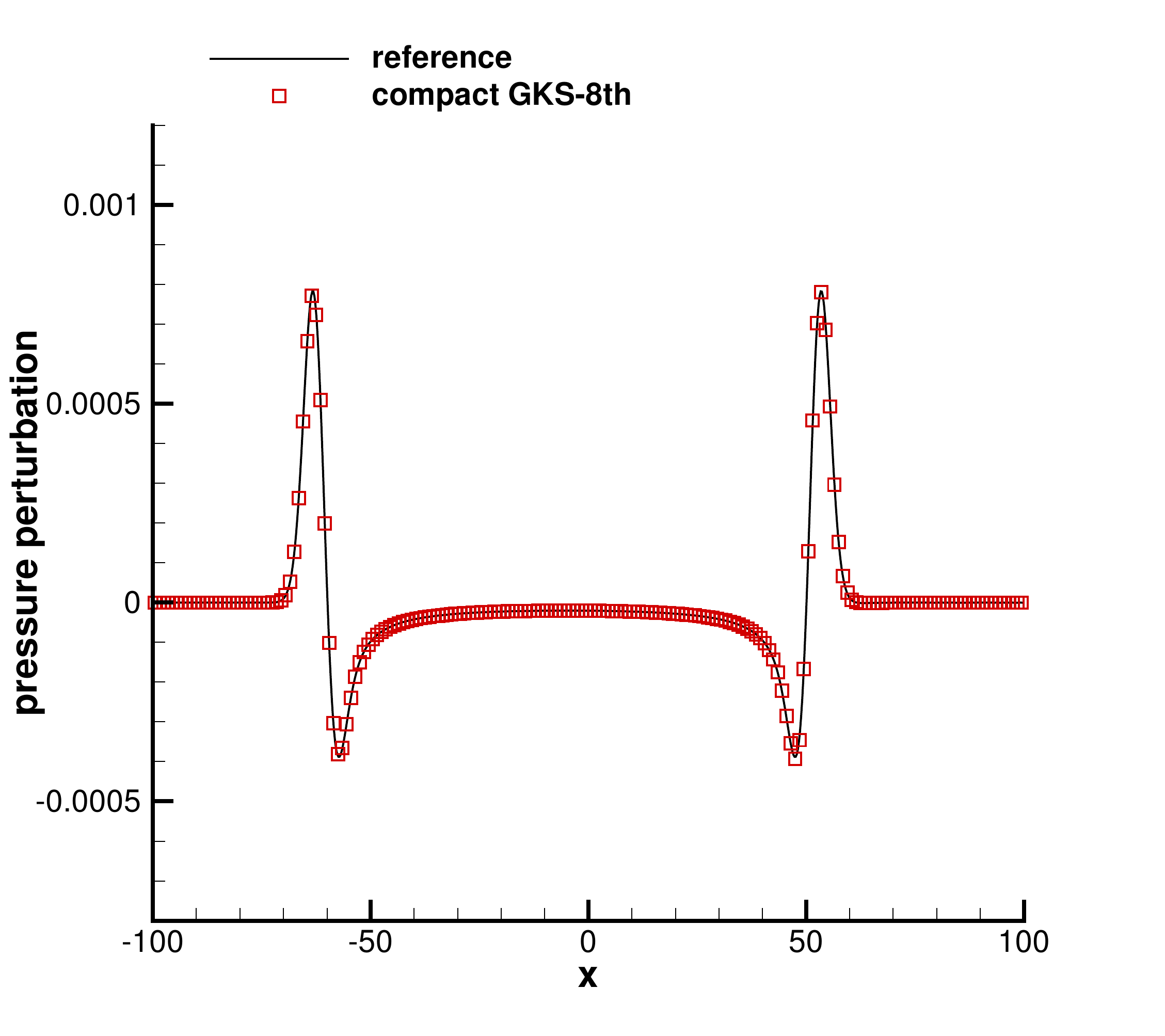}
\caption{\label{multiwave-3} Sound, entropy and vorticity waves test: density and pressure perturbation of compact 8th-order GKS along x axis at $t=28.45$ and $t=56.9$ with $200\times 200$ mesh points.}
\end{figure}

\subsection{Propagation of sound, entropy and vorticity waves}
The case was studied by Tam and Webb \cite{tam1993} to demonstrate the high resolution of traditional compact finite difference schemes.
Initially, an acoustic, entropy, and vorticity pulses are set on a uniform mean flow. The wave front of acoustic pulse expands radially, and the wave pattern is convected downstream with the mean flow. The entropy and vorticity pulses are convected downstream with the mean flow without any distortion.
The initial condition of the uniform mean flow with three pulses is the following,
\begin{align*}
&\rho= \rho_{\infty} +\epsilon_1 e^{-\alpha_1\cdot r_1^2} + \epsilon_2 e^{-\alpha_2\cdot r_2^2}, \\
& p=p_{\infty} +\epsilon_1 e^{-\alpha_1\cdot r_1^2},  \\
& U=U_{\infty} +\epsilon_3 e^{-\alpha_3\cdot r_3^2}(y-y_3), \\
& V=V_{\infty} -\epsilon_3 e^{-\alpha_3\cdot r_3^2}(x-x_3),
\end{align*}
where $\rho_{\infty}=1.0$, $U_{\infty}=0.5$ and $V_{\infty}=0$. The Mach number is $Ma=0.5$. $\alpha_l$ $ (l=1,2,3)$ is $\alpha_l=ln2/b_l^2$ and $b_l$ is the half-width of the Gaussian perturbation. The parameters of these initial pulses are $\epsilon_1=1\times 10^{-2}$, $\epsilon_2=1\times 10^{-3}$, $\epsilon_3=4\times 10^{-4}$, $b_1=3$, $b_2=b_3=5$. $r_l $ $ (l=1,2,3)$ is $r_l=\sqrt{(x-x_l)^2+(y-y_l)^2}$, where $(x_1,y_1)=(-100/3,0)$ and $(x_2,y_2)=(x_3,y_3)=(100/3,0)$. The computation domain is $[-100,100]\times[-100,100]$.

The results in Fig. \ref{multiwave-1} and Fig. \ref{multiwave-2} are the density and pressure perturbation contours in $[-0.0006,0.001]$ with 0.0001 increment obtained by compact 8th-order GKS with $200\times 200$ uniform mesh points. The results don't show obvious numerical oscillations near domain boundary at $t=100$.
The computational time step used in the compact 8th-order GKS is larger than that used in the traditional compact finite difference scheme.
 It requires 500 and 1000 time steps in \cite{tam1993} to get times $t=28.45$ and $t=56.9$ respectively, while the compact 8th-order GKS just needs 143 and 286 time steps to attain the same output times with $CFL=0.3$.
 The merging of acoustic wave and entropy wave at $t=28.45$ is similar to that in \cite{tam1993}, and the waves separate afterwards.
 In order to evaluate the performance of numerical scheme, the density and pressure perturbations along $y=0$ at $t=28.45$ and $t=56.9$ are plotted in Fig. \ref{multiwave-3}. The reference solution is the result from a refined mesh with $800\times 800$ mesh points, which is also identical to the analytical solution in \cite{tam1993}.

\begin{figure}[!htb]
\centering
\includegraphics[width=0.425\textwidth]{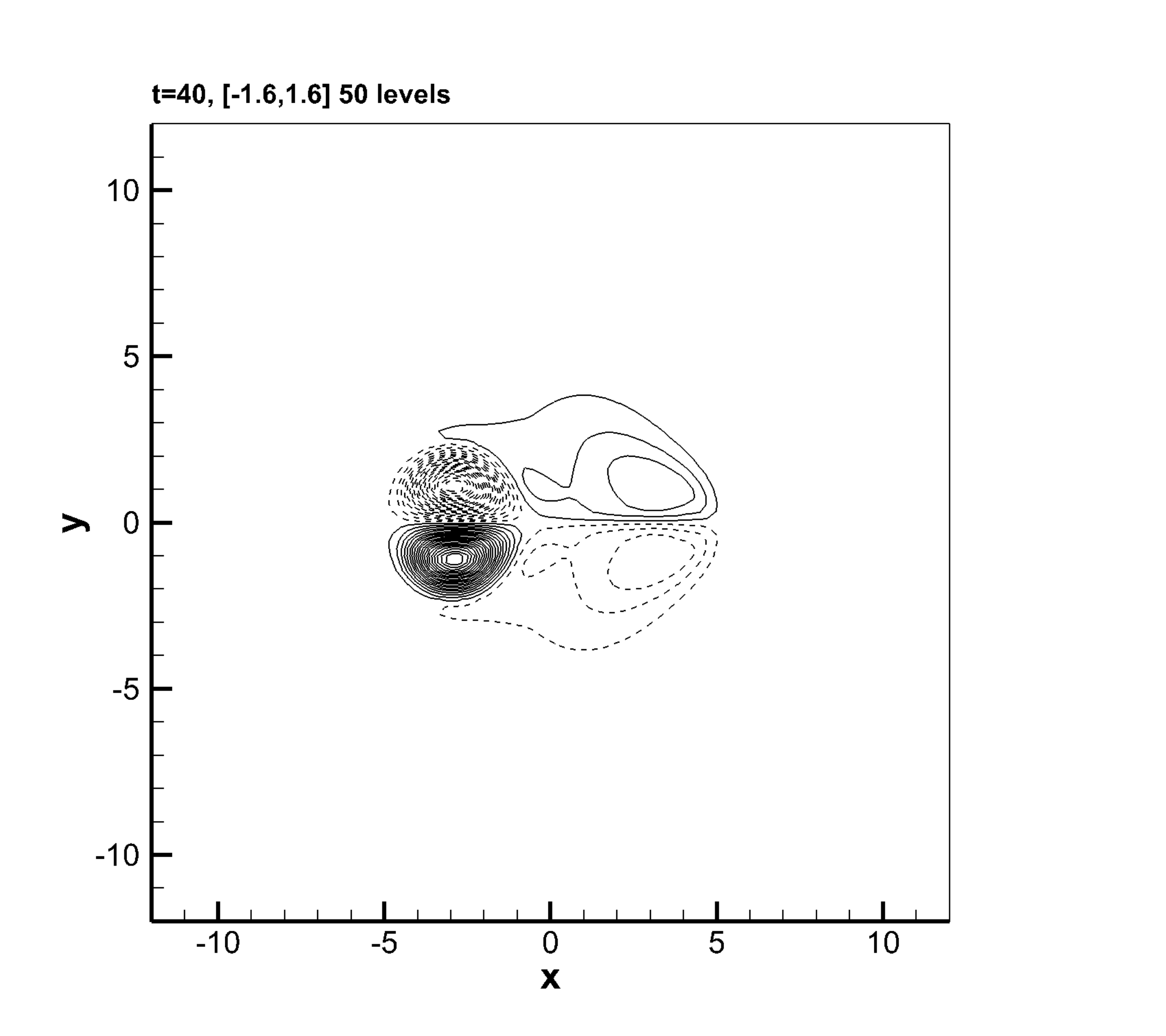}
\includegraphics[width=0.425\textwidth]{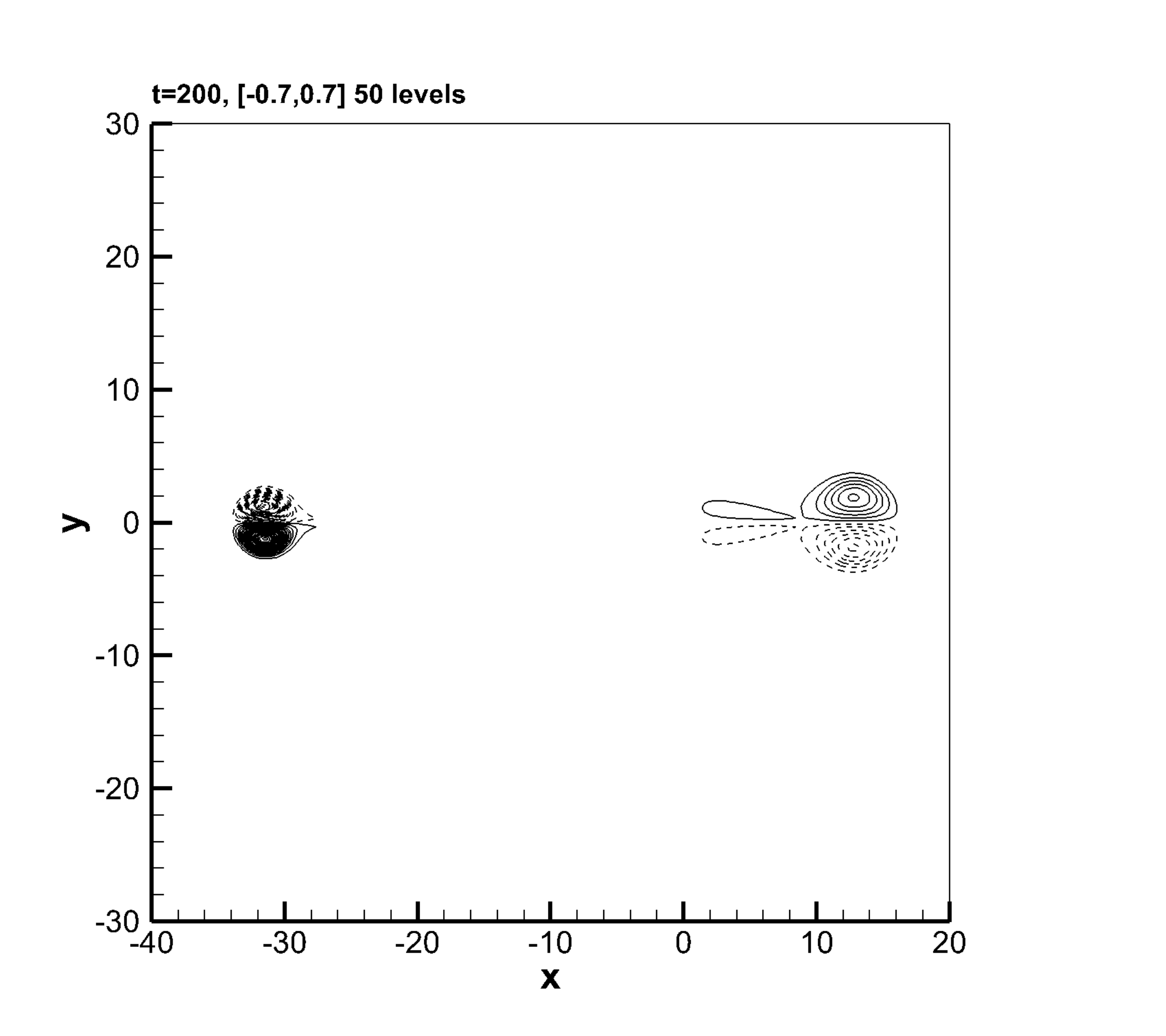}\\
\caption{\label{vortices-inter-vor} Sound generation by the interaction of viscous Taylor vortex pair: vorticity of counter-rotating vortices interaction obtained by compact 8th-order GKS with $2000 \times 2000$ mesh points at $t=40$ and $t=200$.}
\end{figure}

\begin{figure}[!htb]
\centering
\includegraphics[width=0.425\textwidth]{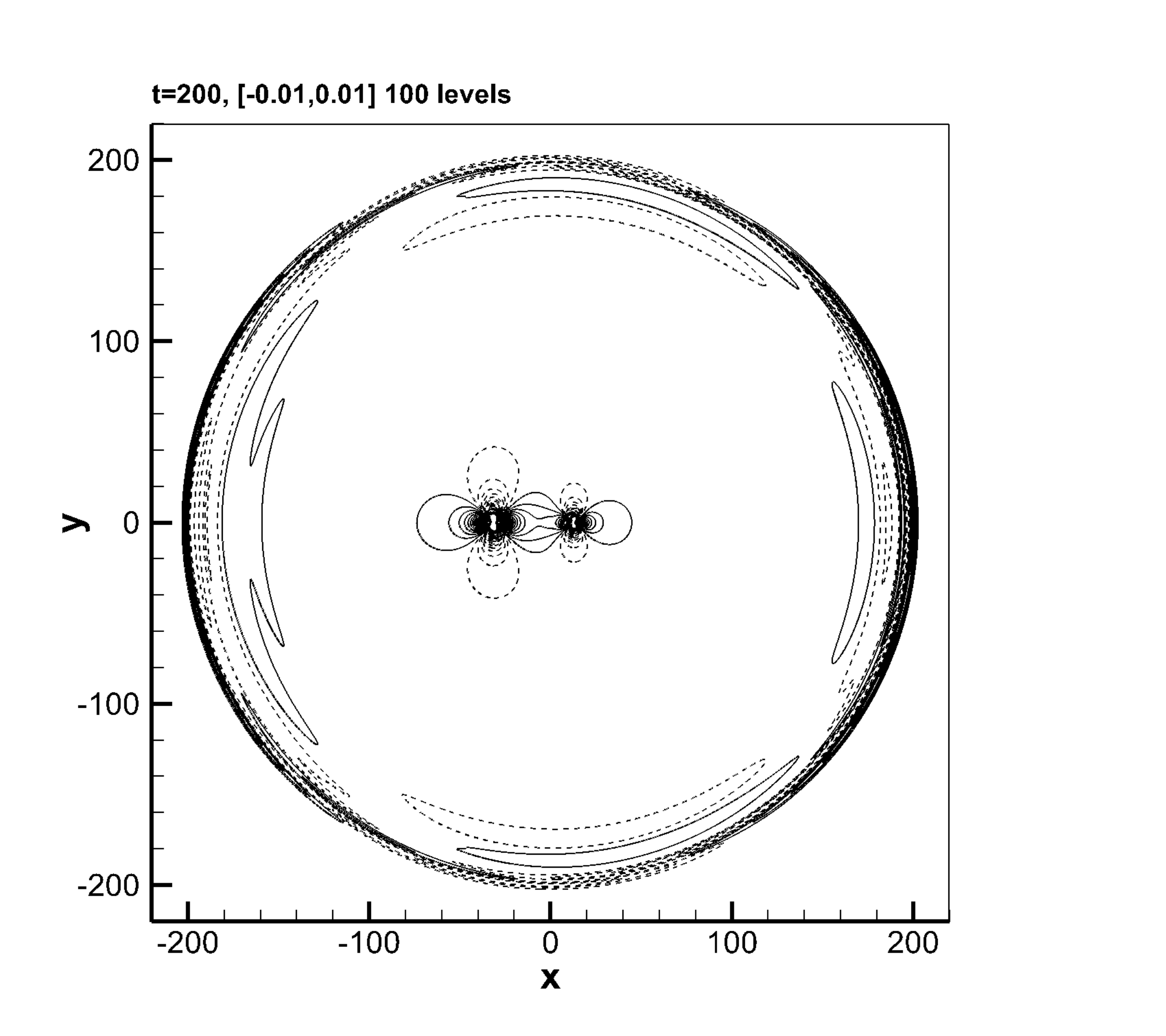}
\includegraphics[width=0.425\textwidth]{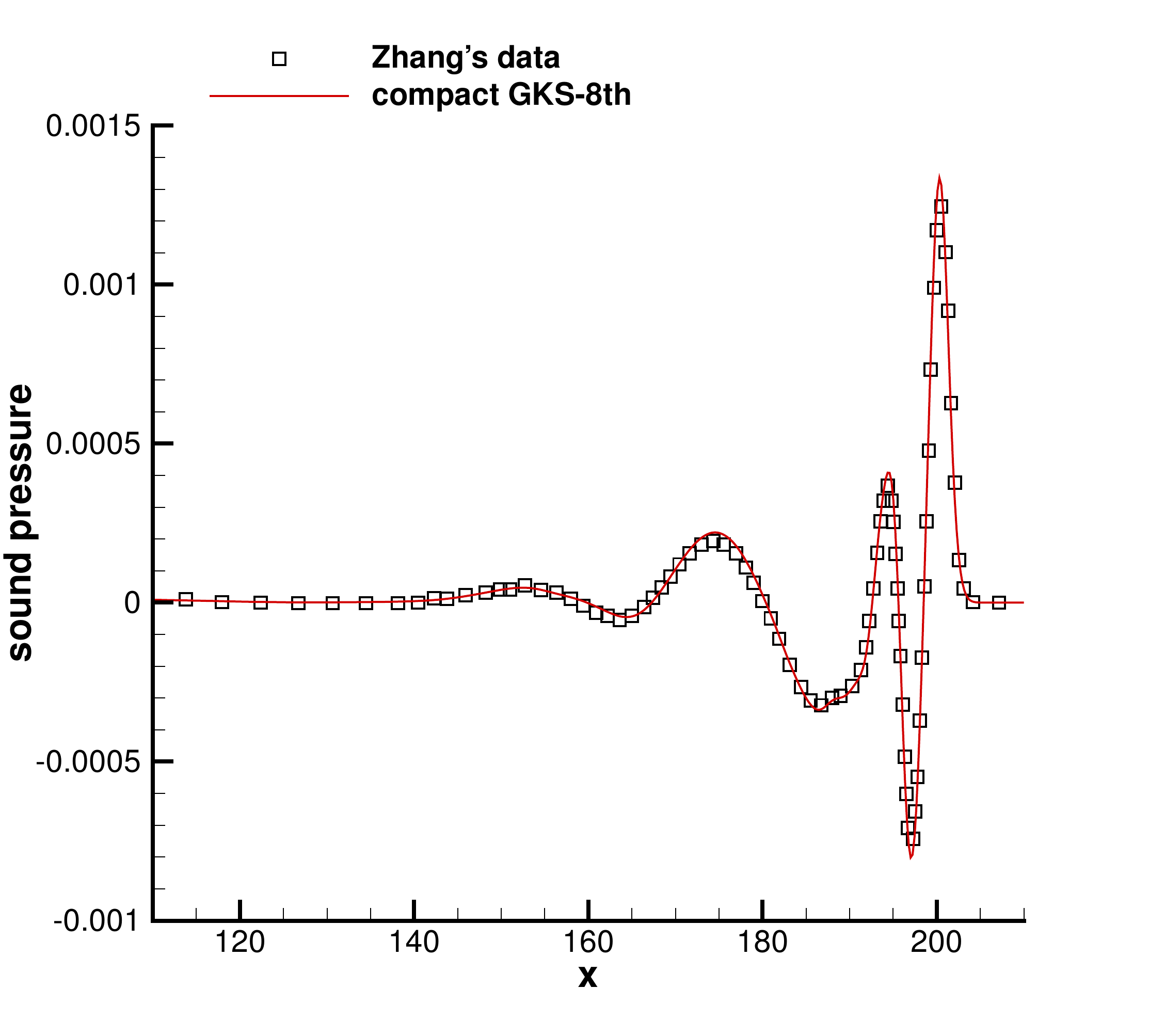}
\caption{\label{vortices-inter-ps} Sound generation by the interaction of viscous Taylor vortex pair: pressure perturbation and the distribution of sound pressure along $y=0$ of compact 8th-order GKS with $2000 \times 2000$ mesh points at $t=200$.}
\end{figure}

\subsection{Sound generation by interaction of viscous Taylor vortex pair}
Vortices interaction is one of the sources of sound generation in turbulence flows. The interaction of vortex pair is usually used to numerically investigate the mechanism of vortices induced sound. For numerical computation, the main difficulties are acoustic wave/mean flow disparity and large computational domain. Here, the interaction of two Taylor vortices presented in \cite{zhang2013vor} is computed to validate the compact GKS. The single Taylor vortex is set as,
\begin{align*}
U_{\theta}(r)&=M_v r e^{(1-r^2)/2},~~U_r=0, U_{r}=0, \\
p(r)&=\frac{1}{\gamma}[1-\frac{\gamma-1}{2}M_v^2e^{(1-r^2)}]^{\gamma/(\gamma-1)}, \\
\rho(r)&=[1-\frac{\gamma-1}{2}M_v^2e^{(1-r^2)}]^{1/(\gamma-1)},
\end{align*}
where $U_{\theta}$ and $U_r$ are the tangential and radial velocity, respectively. $r^2=((x-x_v)^2+(y-y_v)^2)/R_c$. $(x_v,y_v)$ is the center of the initial vortex. $R_c$ is the critical radius of a single vortex for which the vortex has the maximum strength. $M_v$ is the parameter determining the strength of the single vortex. Reynolds number is $Re=800$. The initial flow of vortex pairs interaction is given by the superposition of two vortices. The current test is about two counter-rotating vortices with similar strengths.
The vortex centers are set as $(x_{v1},y_{v1})=(0,2)$ and $(x_{v2},y_{v2})=(0,-2)$, and the same critical radius are $R_{c1}=R_{c2}=1.0$. The strength of two vortices are $M_{v1}=-0.5$ and $M_{v2}=0.5$.
The computational domain is $[-220,220]\times[-220,220]$, the non-reflection boundary condition is adopted for all boundaries.

 A uniform mesh of $2000\times2000$ points is used. Fig. \ref{vortices-inter-vor} shows the vorticity from the compact 8th-order GKS at $t=40$ and $t=200$, where the dash line represents negative vorticity and the solid line represents positive vorticity.
 The current results are similar to that calculated by $6000\times6000$ mesh points in \cite{zhang2013vor}. The pressure perturbation is plotted in Fig. \ref{vortices-inter-ps}, and the distribution of sound pressure along $y=0$ is also presented in order to evaluate the numerical solution quantitatively. Compared with the distribution of sound pressure in \cite{zhang2013vor}, the compact 8th-order GKS gives similar result using about $1/10$ of  total cells of \cite{zhang2013vor}.

\begin{figure}[!htb]
\centering
\includegraphics[width=0.325\textwidth]{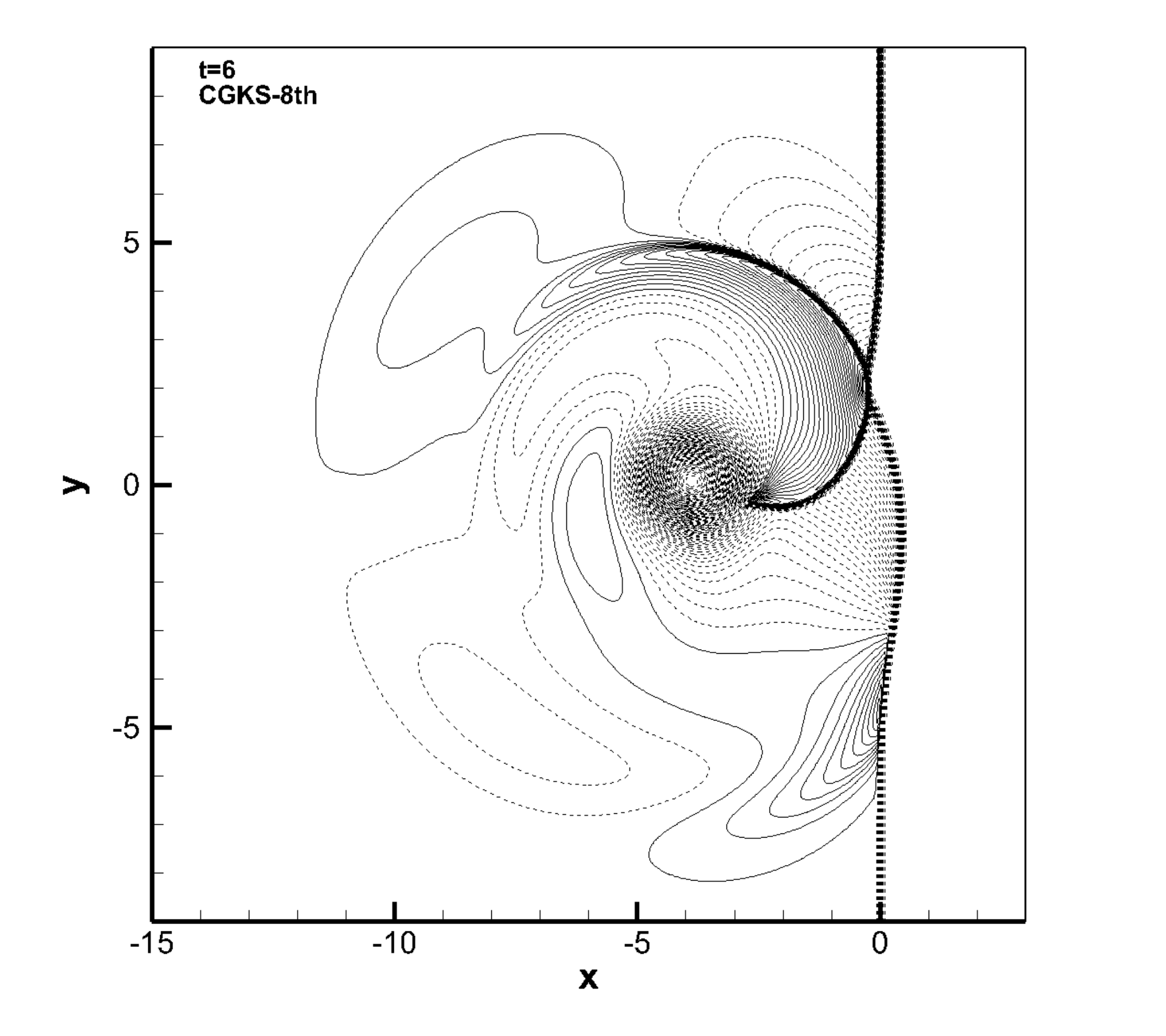}
\includegraphics[width=0.325\textwidth]{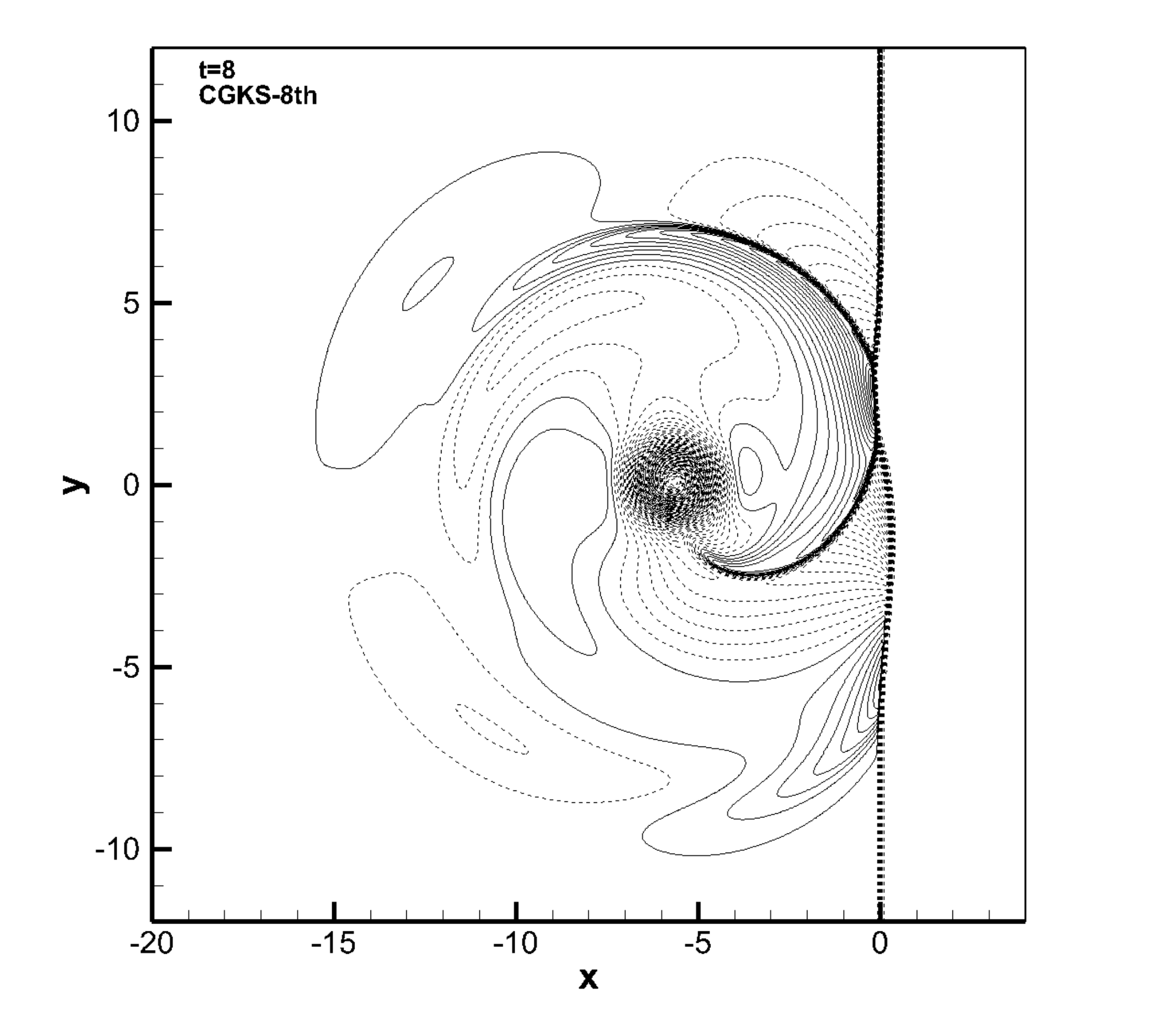}
\includegraphics[width=0.325\textwidth]{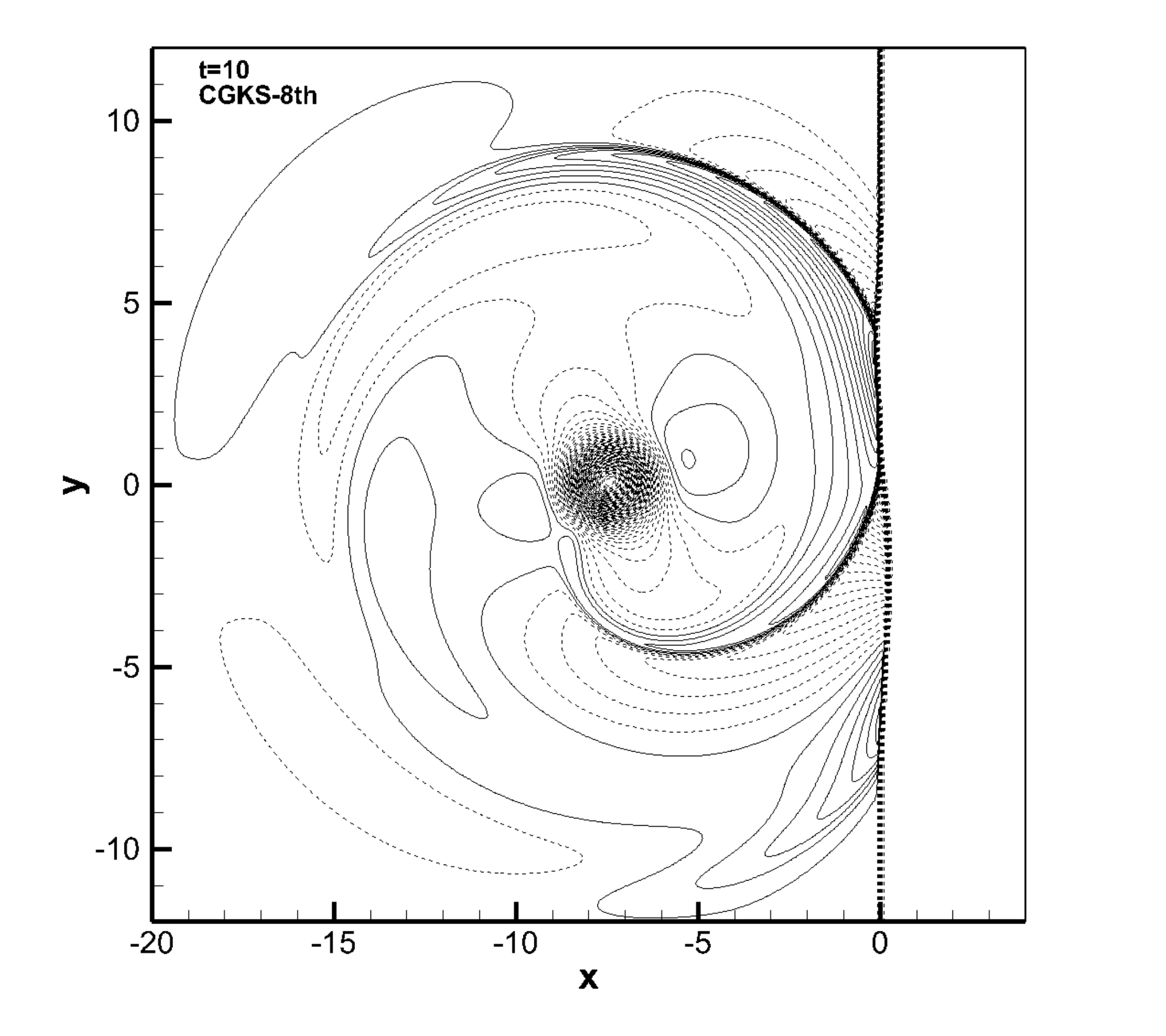}
\caption{\label{vis-shockvor-1} Sound generation by viscous shock-vortex interaction: $M_v=0.5, M_s=1.2$, 100 equal-spaced sound pressure contours from $\Delta p_{min}=-0.400$ to $\Delta p_{max}=0.152$ with $700 \times 600$ mesh points. The dash line represents rarefaction region, and the solid line represents the compression region.}
\end{figure}

\begin{figure}[!htb]
\centering
\includegraphics[width=0.325\textwidth]{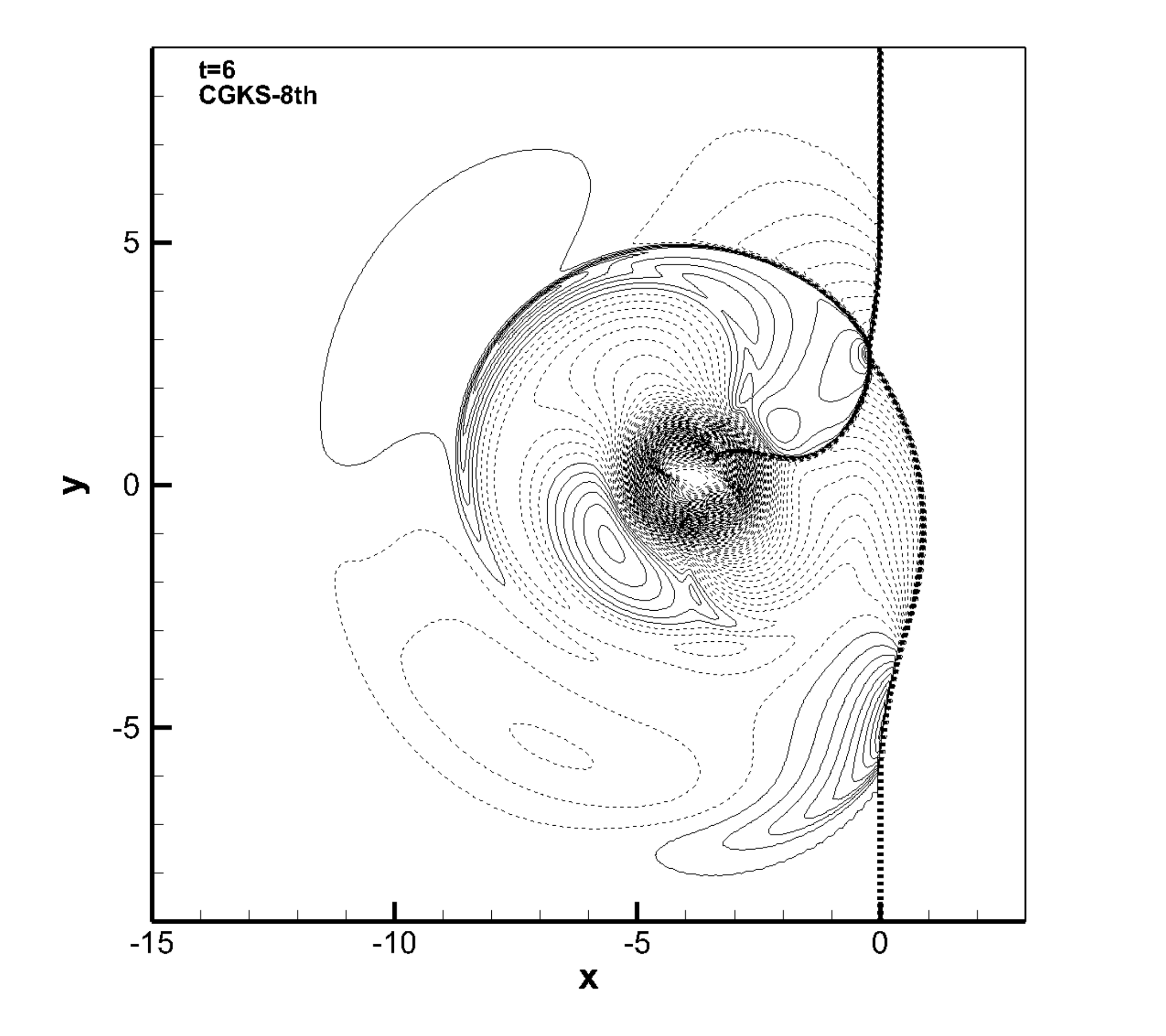}
\includegraphics[width=0.325\textwidth]{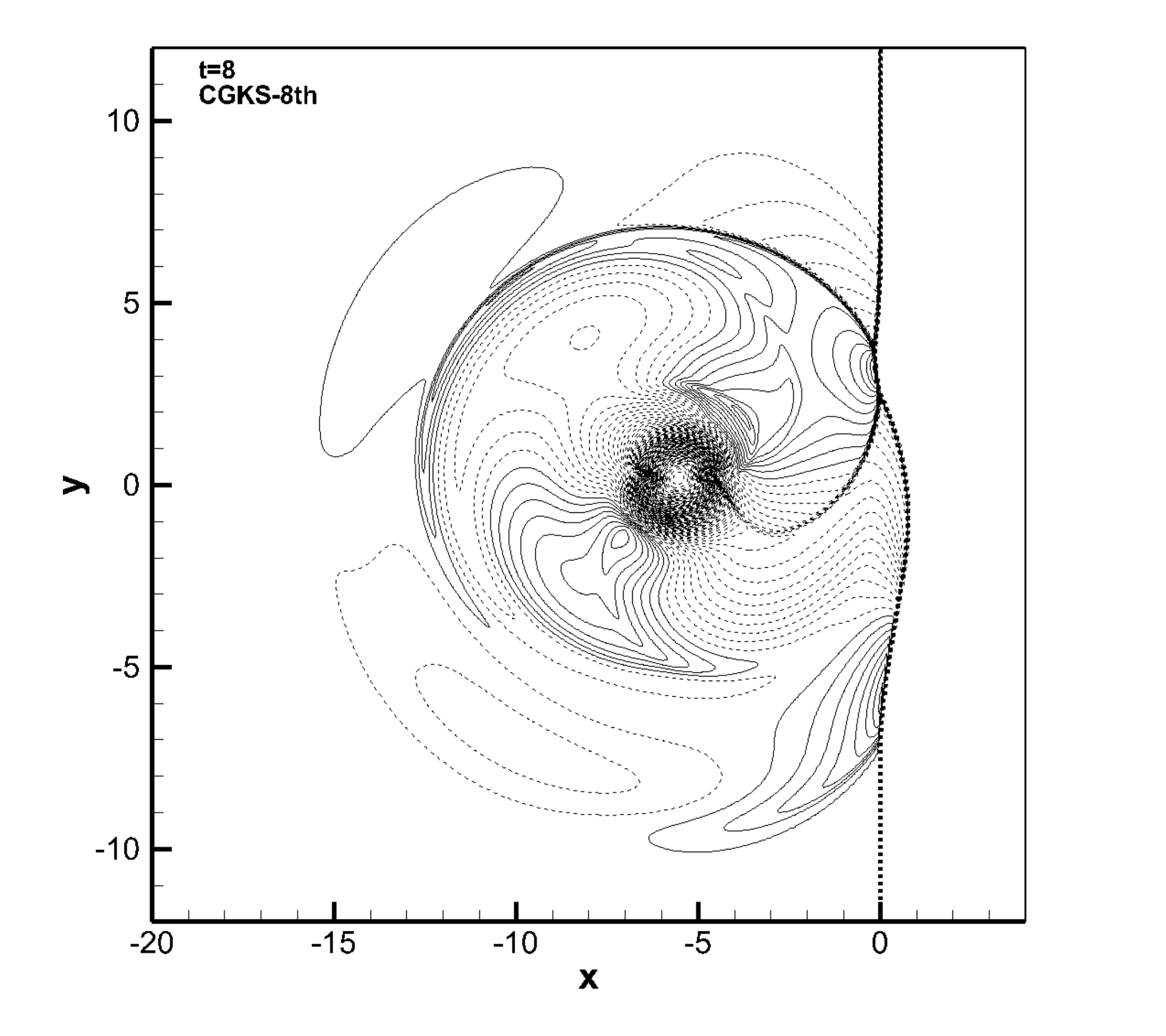}
\includegraphics[width=0.325\textwidth]{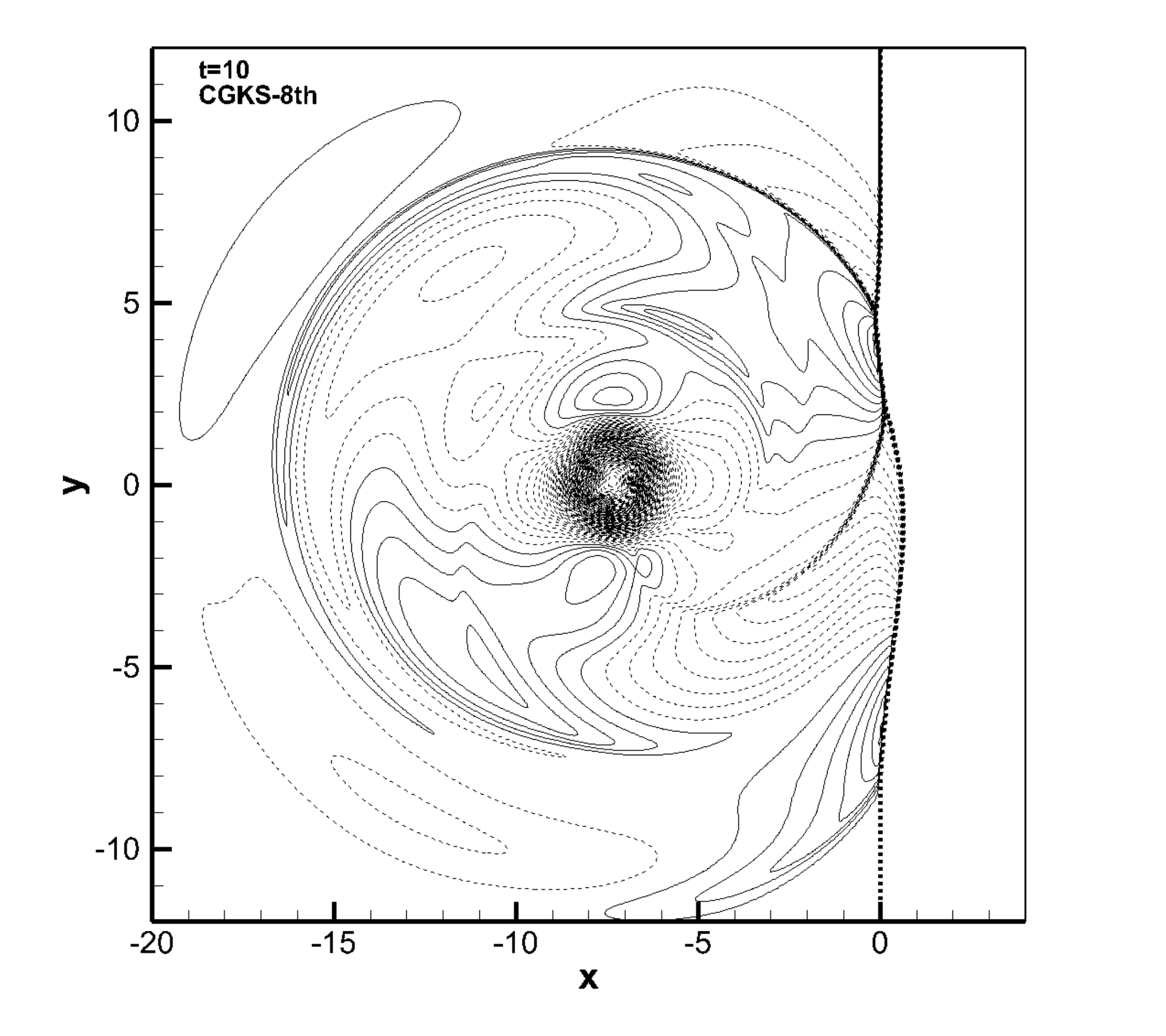}
\caption{\label{vis-shockvor-2} Sound generation by viscous shock-vortex interaction: $M_v=1.0, M_s=1.2$, 100 equal-spaced sound pressure contours from $\Delta p_{min}=-0.876$ to $\Delta p_{max}=0.114$ with $700 \times 600$ mesh points. }
\end{figure}

\begin{figure}[!htb]
\centering
\includegraphics[width=0.485\textwidth]{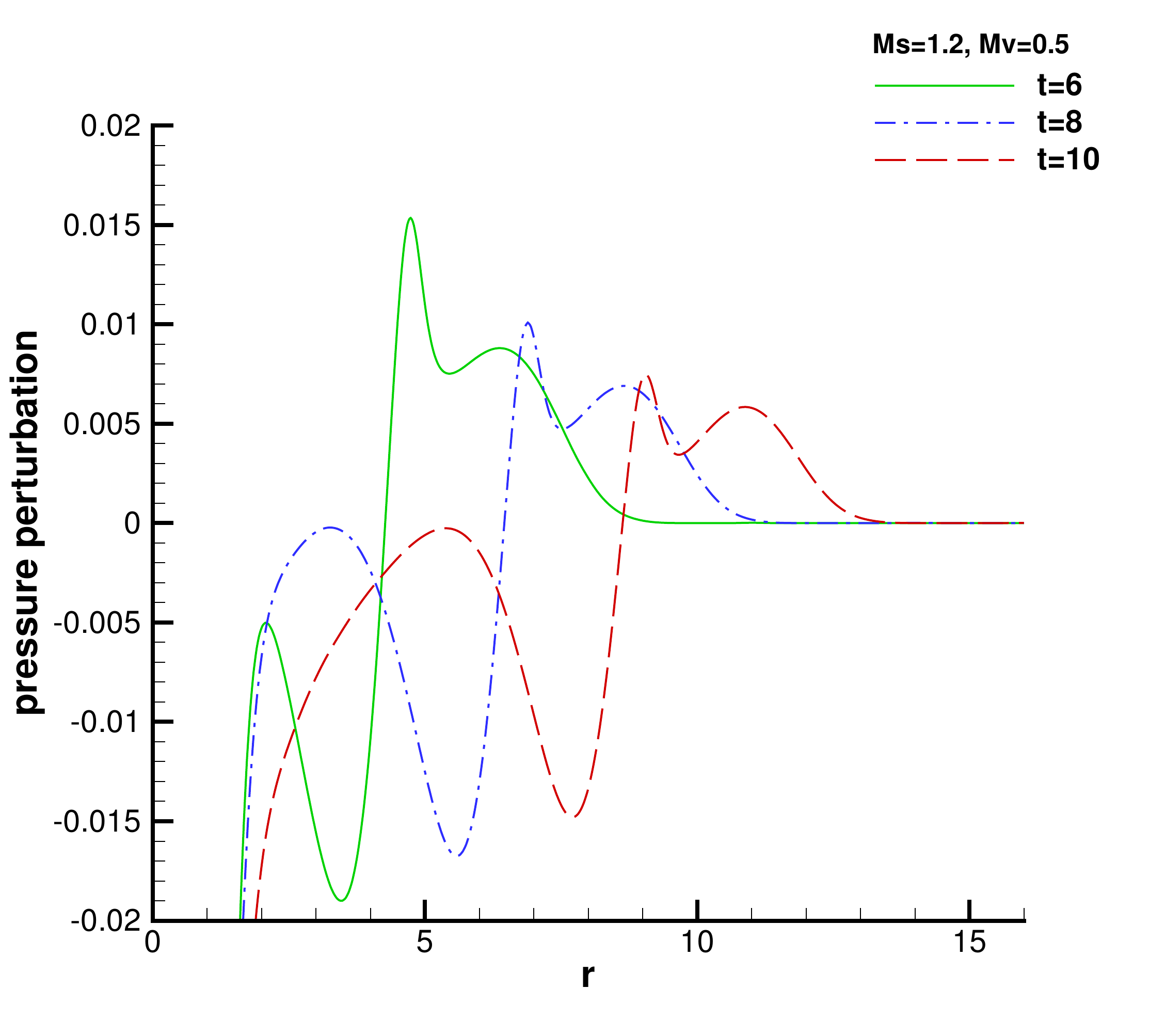}
\includegraphics[width=0.485\textwidth]{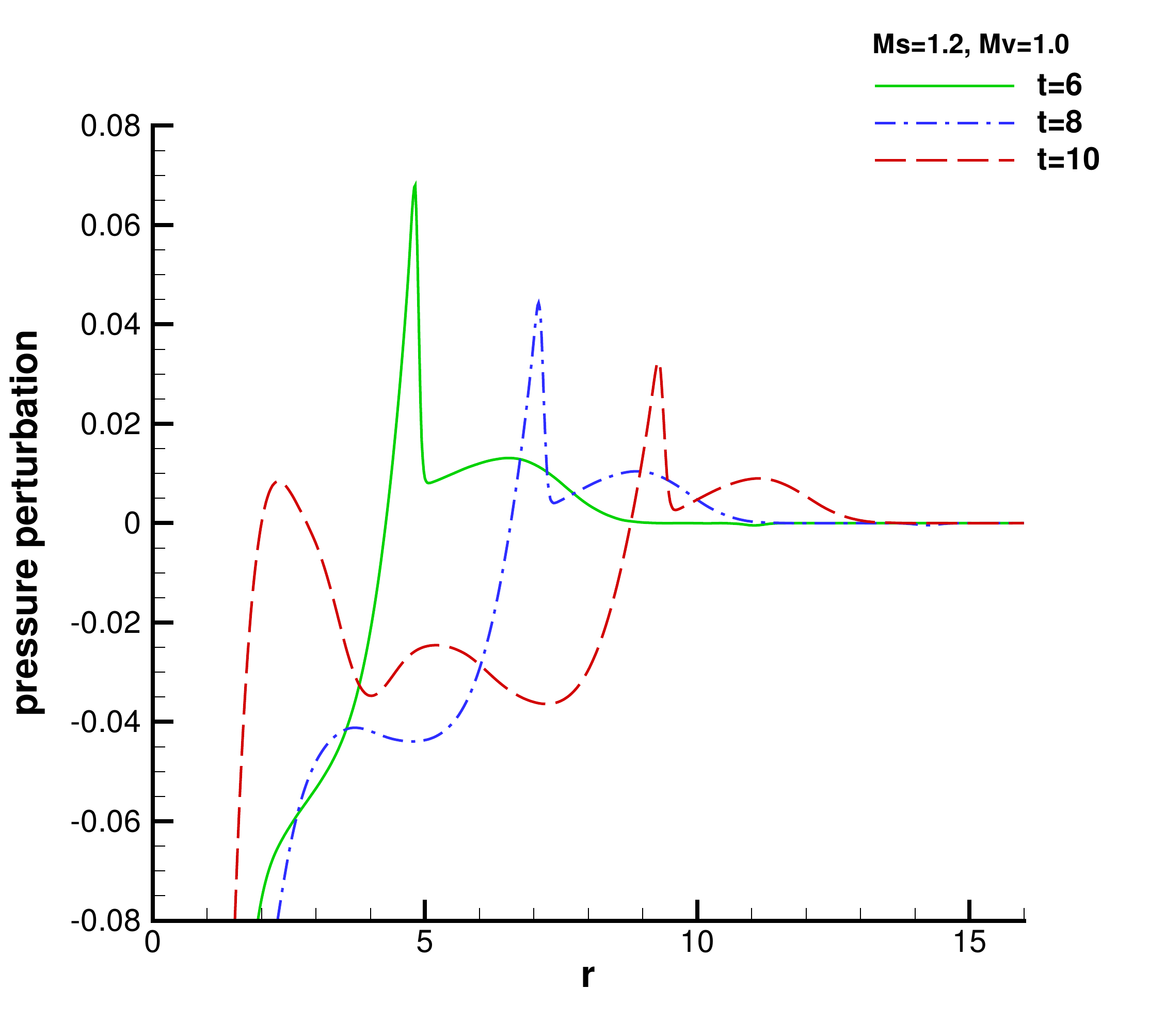}
\caption{\label{vis-shockvor-3} Sound generation by viscous shock-vortex interaction: radial distribution of the pressure perturbation $\triangle p$ obtained by 8th-order compact GKS with $700\times600$ mesh points.
The vortex centers at $t=6, t=8$ and $t=10$ are approximately located at $(-3.87,0.08)$, $(-5.67,0.07)$ and $(-7.44,0.07)$ for $M_v=0.5$, and $(-3.88,0.18)$, $(-5.53,0.18)$ and $(-7.35,0.10)$ for  $M_v=1.0$.}
\end{figure}

\begin{figure}[!htb]
\centering
\includegraphics[width=0.485\textwidth]{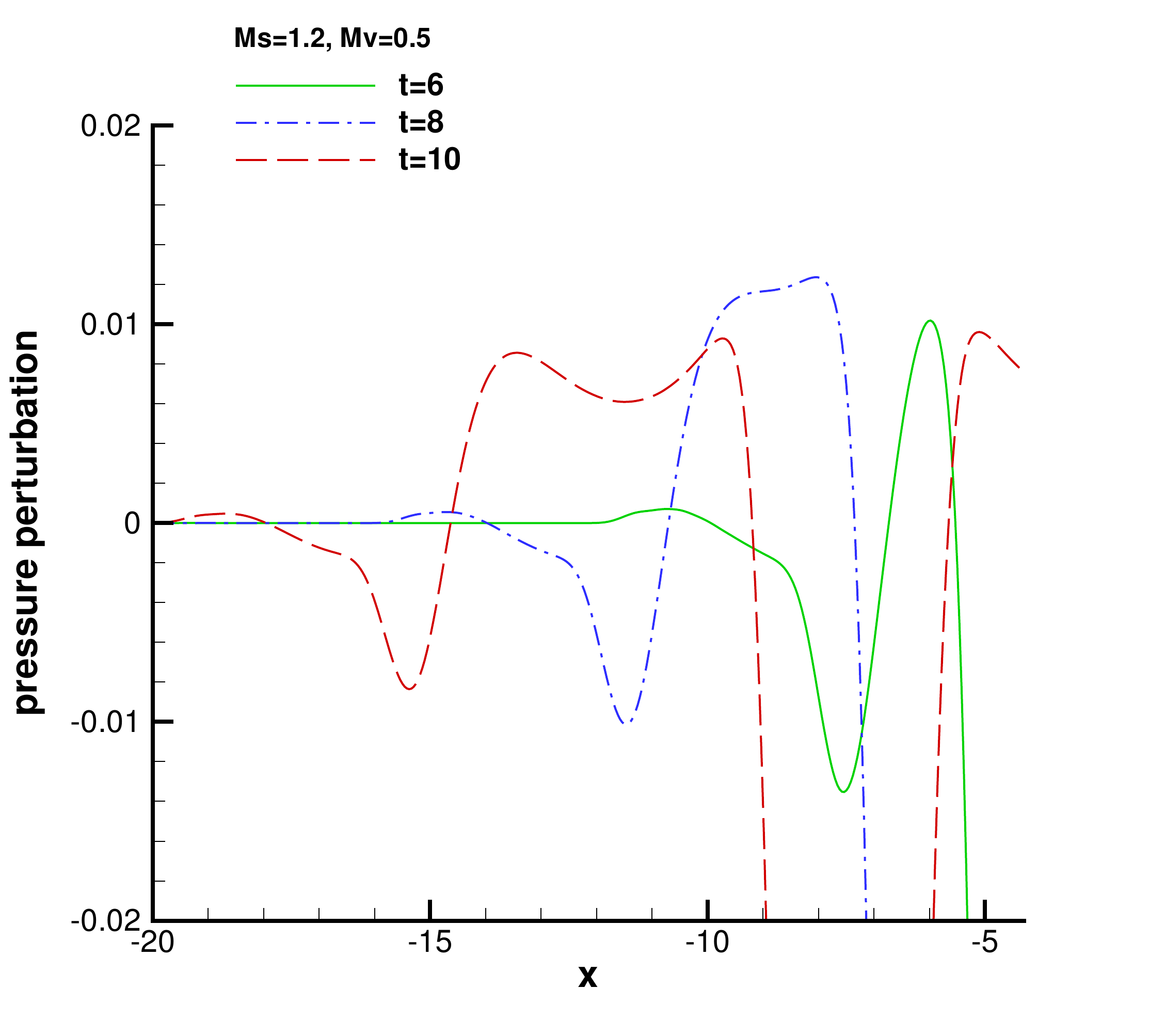}
\includegraphics[width=0.485\textwidth]{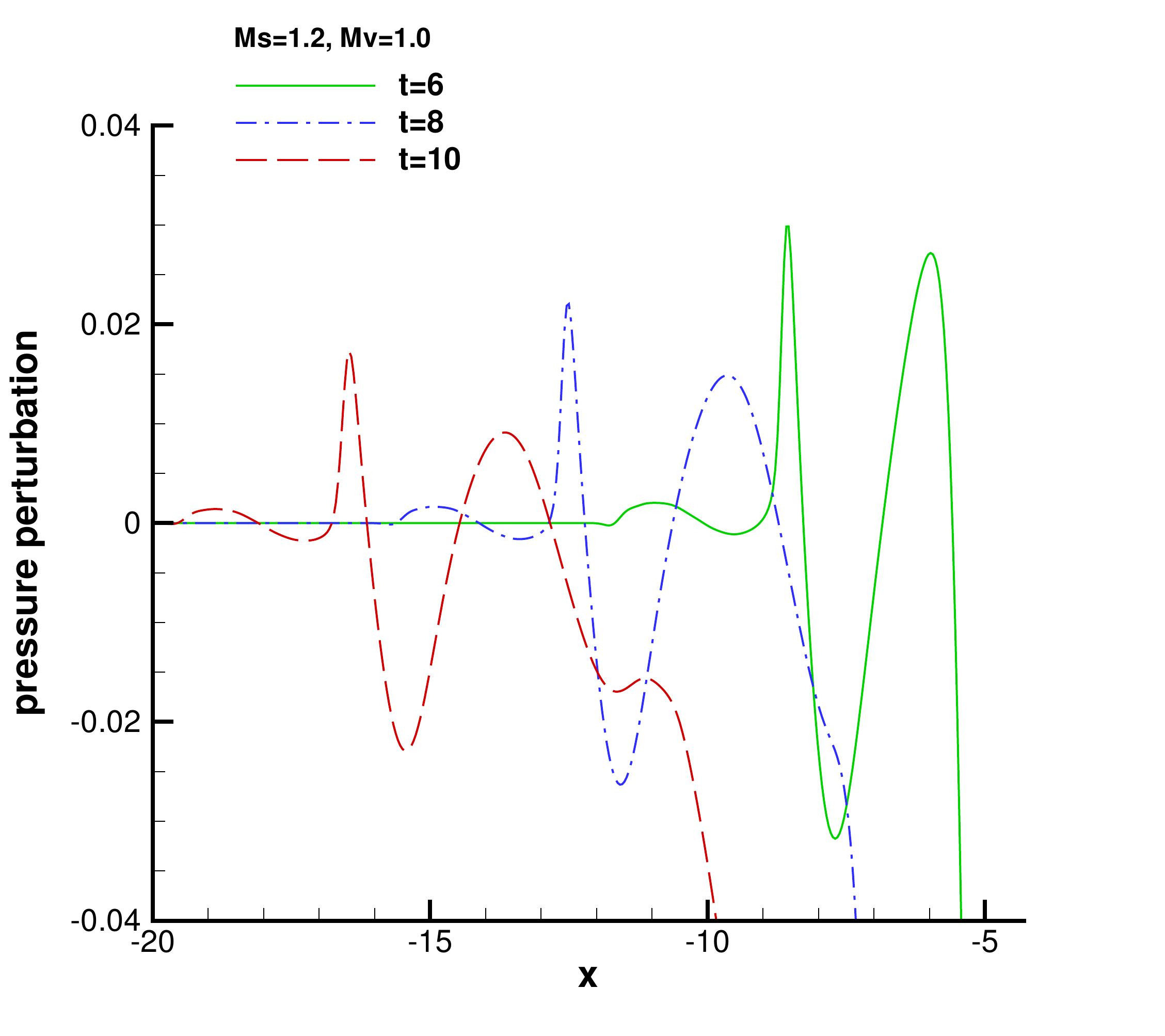}
\caption{\label{vis-shockvor-4} Sound generation by viscous shock-vortex interaction: the distribution of the pressure perturbation $\triangle p$ along $y=0$ obtained by 8th-order compact GKS with $700\times600$ mesh points.}
\end{figure}

\subsection{Sound generation by viscous shock-vortex interaction}
The case is the interaction of a shock wave with a single vortex in a viscous flow \cite{shock-vor}.
The computational domain is $[-20,8]\times[-12,12]$.
The velocity of the initial counterclockwise vortex is
\begin{align*}
U_{\theta}(r)&=M_v r e^{(1-r^2)/2},~~U_r=0,
\end{align*}
where $U_{\theta}$ and $U_r$ are the tangential and radial velocity respectively.
The pressure and density distribution superposed by the isentropic vortex downstream of shock wave are
\begin{align*}
p(r)&=\frac{1}{\gamma}[1-\frac{\gamma-1}{2}M_v^2e^{(1-r^2)}]^{\gamma/(\gamma-1)},\\
\rho(r)&=[1-\frac{\gamma-1}{2}M_v^2e^{(1-r^2)}]^{1/(\gamma-1)}.
\end{align*}
Two cases of the vortex Mach number $M_v=0.5$ and $M_v=1.0$ are computed. The Mach number of shock wave is $M_s=1.2$.
The Reynolds number is $Re=800$ defined by $Re=\rho_{\infty} a_{\infty}/\mu_{\infty}$, where the
subscript $\infty$ denotes the quantity upstream of the shock wave. The initial location of vortex
is $(x_v,y_v)=(2,0)$, and the stationary shock is at $x=0$.
In the computation, the supersonic inflow boundary conditions at $x=8$ as well as the periodic boundary
conditions at $y=\pm 12$ are imposed. The non-reflective boundary conditions are adopted at $x=-20$.
A mesh with $700\times600$ points is used in the current computation.

The sound pressure contours of the vortex Mach number $M_v=0.5$ and $M_v=1.0$ are given in Fig. \ref{vis-shockvor-1} and Fig. \ref{vis-shockvor-2} respectively. The sound pressure is defined as $\triangle p=(p-p_{\infty})/p_{\infty}$
The multiple sound waves with quadrupolar structure are generated. The sound pressure are smoothly distributed without apparent spurious oscillations.
As $M_v$ increases, the strength of the reflected shock wave increases and extends to the core region of the vortex with interaction.
As a result, more complicated flow patterns are formed around the vortex.
Fig. \ref{vis-shockvor-3} is the radial distribution of the sound pressure at different times. The Mach waves are generated in both cases,
and the Mach waves of the case with $M_v=1.0$ is stronger than those of the case with $M_v=0.5$.
Fig. \ref{vis-shockvor-4} is the distribution of the sound pressure $\triangle p$ along $y=0$ at different times.
The reflected shock wave produces a pressure jump between the precursor and the second sound wave.
The jumps can be seen in the case with vortex Mach number $M_v=1.0$ due to the stronger reflected shock waves.
The solutions obtained by the current scheme uses a much coarse mesh in comparison with the results in the reference paper \cite{shock-vor}.

\begin{figure}[!htb]
\centering
\includegraphics[width=0.485\textwidth]{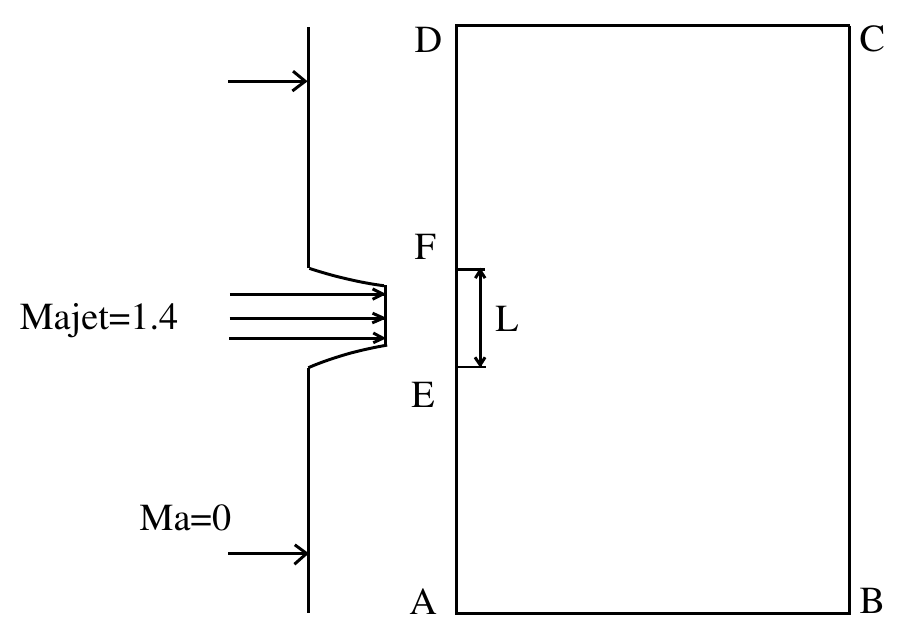}
\caption{\label{2d-jet-0} Schematic of 2-D planar jet test.}
\end{figure}

\begin{figure}[!htb]
\centering
\includegraphics[width=0.325\textwidth]{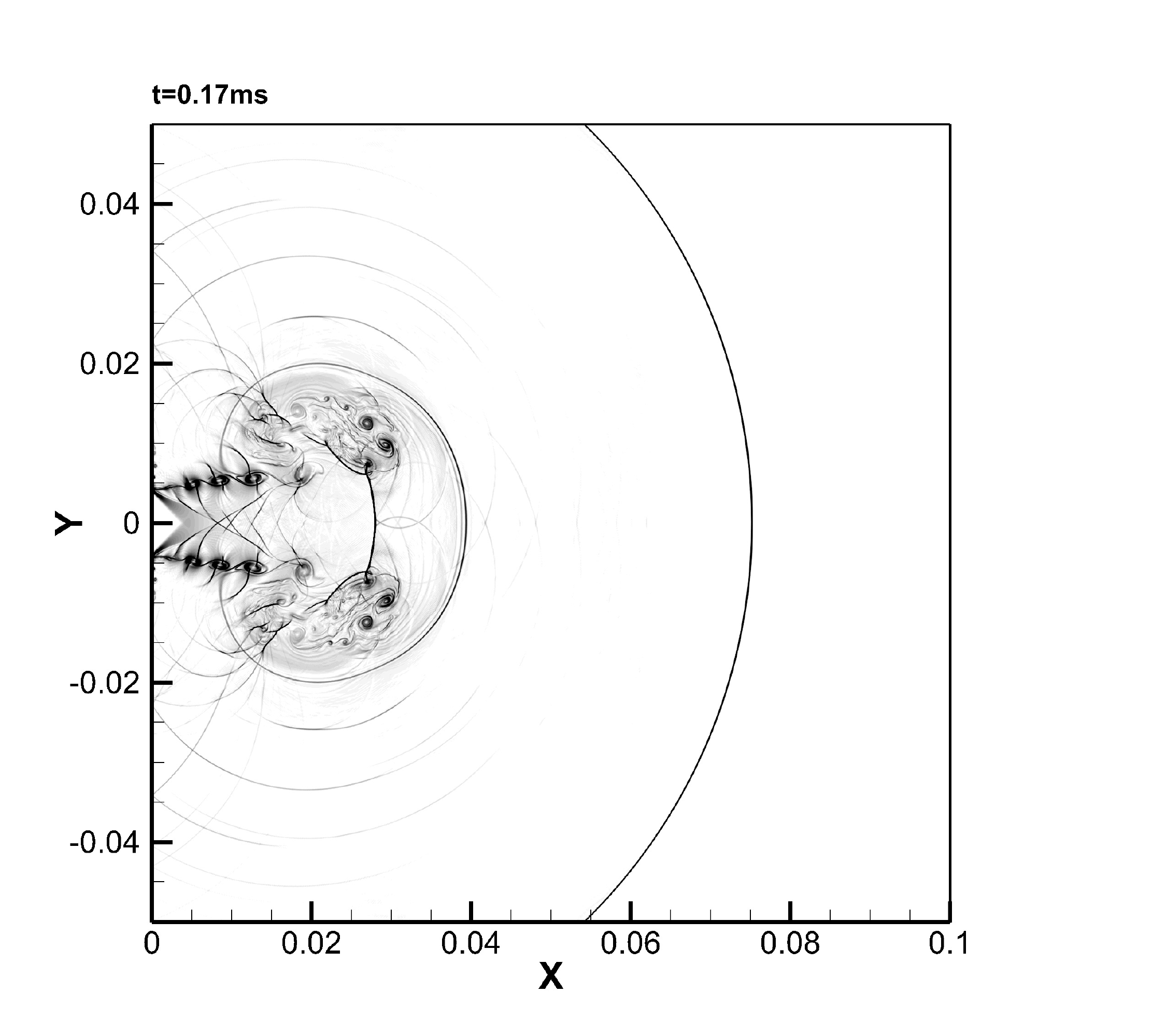}
\includegraphics[width=0.325\textwidth]{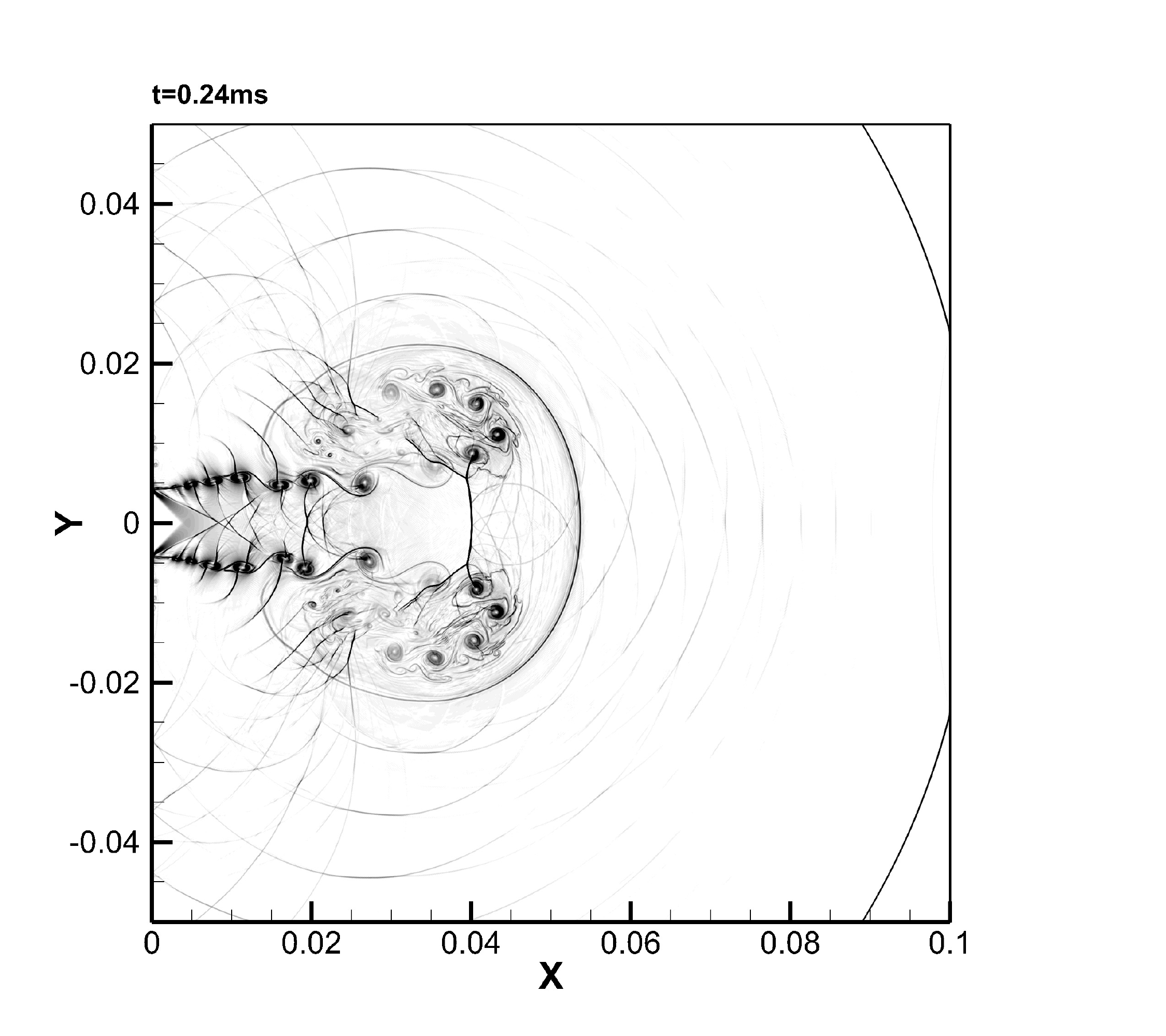}
\includegraphics[width=0.325\textwidth]{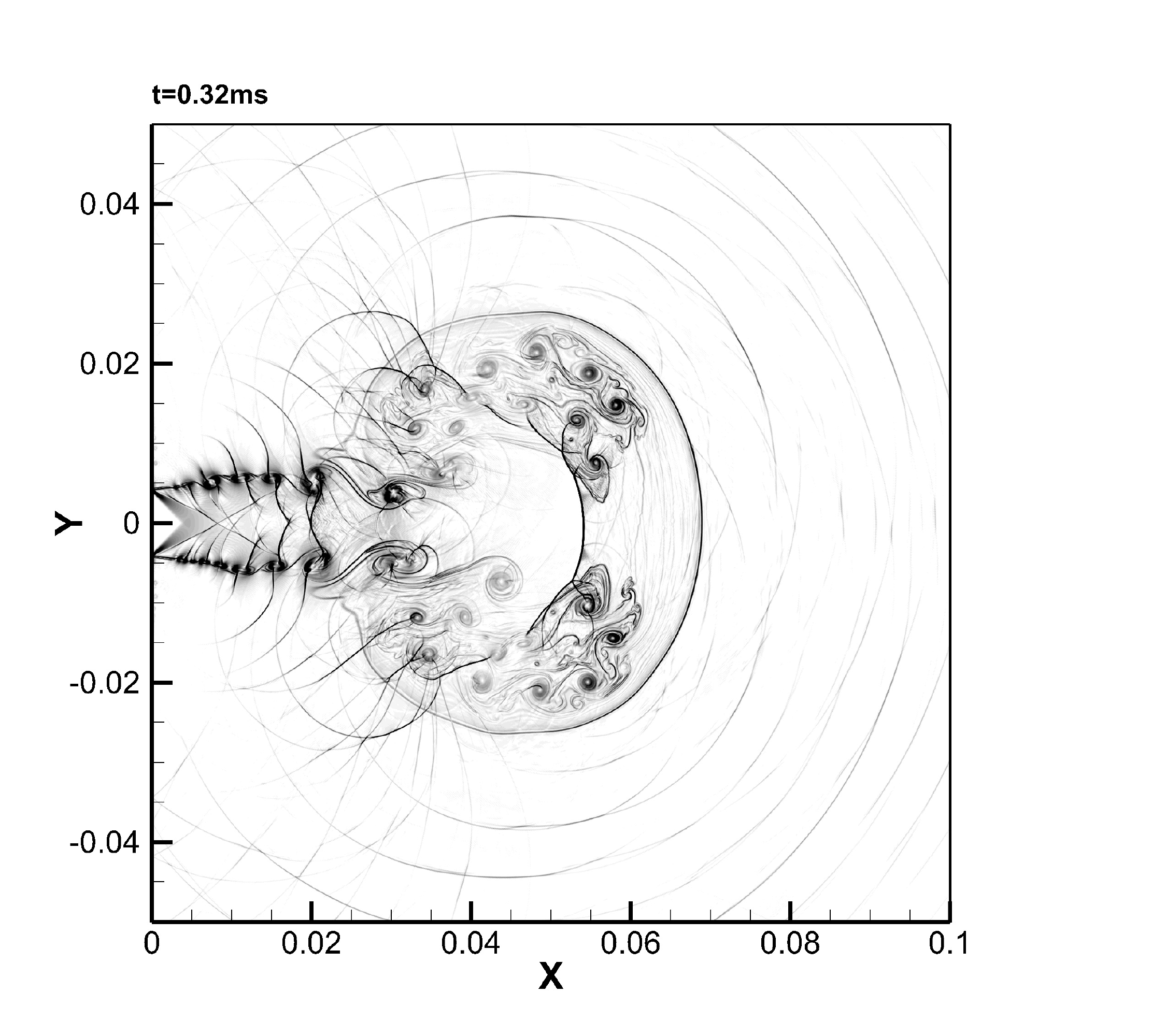}
\caption{\label{2d-jet-1} Planar jet: schlieren image of the supersonic jet obtained by compact 8th-order GKS with $1500\times2250$ mesh points at $t=0.17ms, t=0.24ms$ and $t=0.32ms$.}
\end{figure}

\begin{figure}[!htb]
\centering
\includegraphics[width=0.325\textwidth]{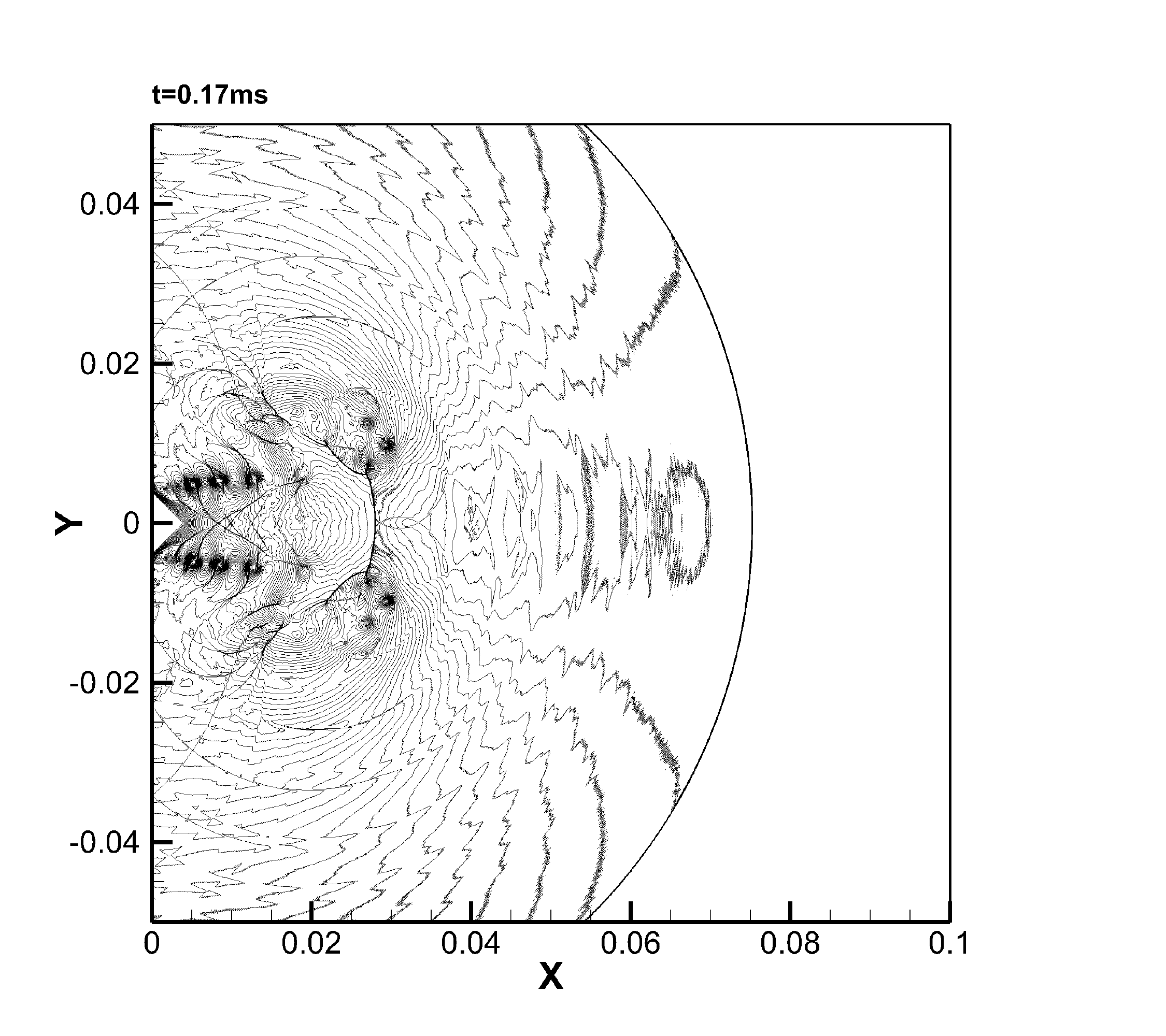}
\includegraphics[width=0.325\textwidth]{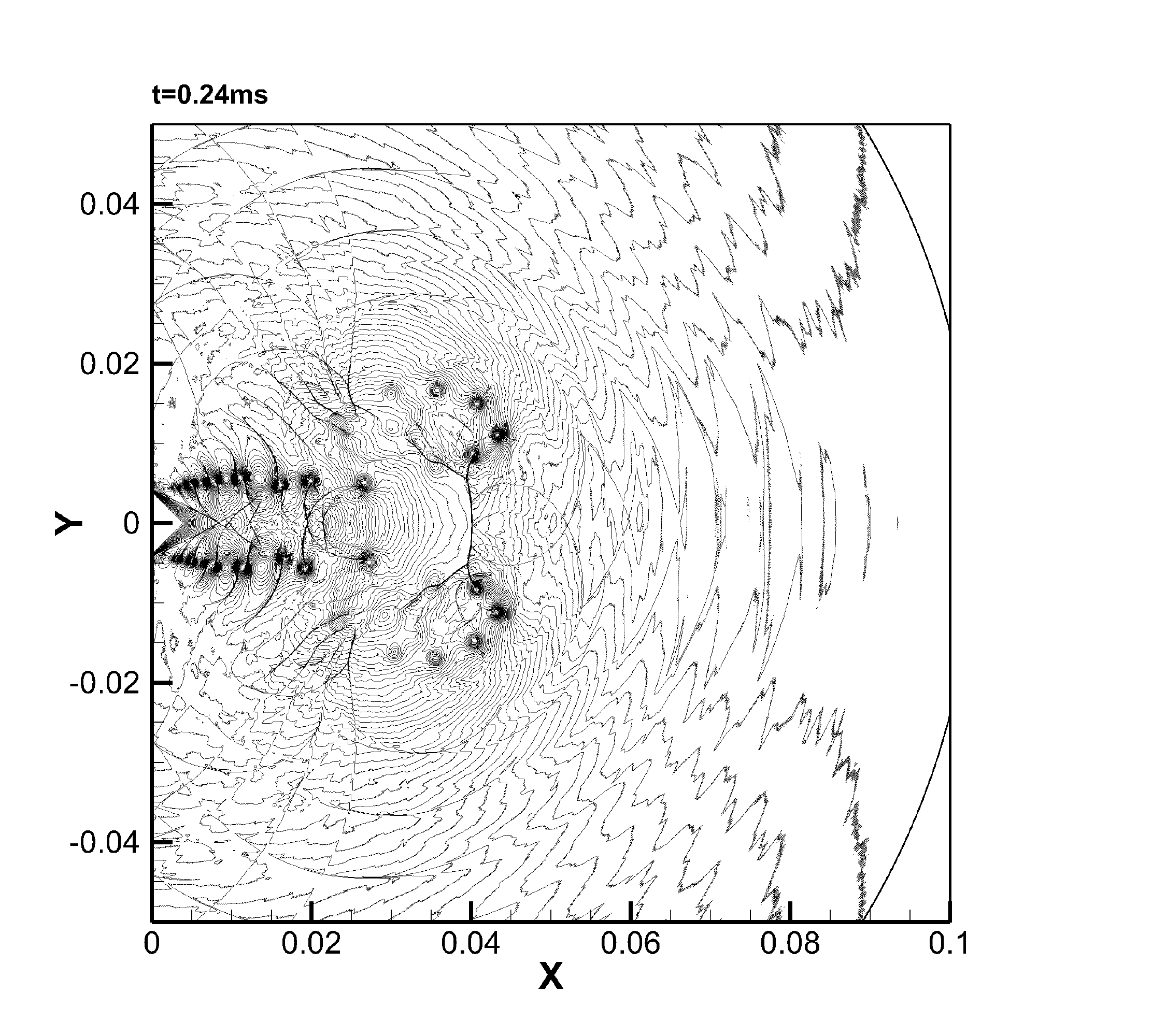}
\includegraphics[width=0.325\textwidth]{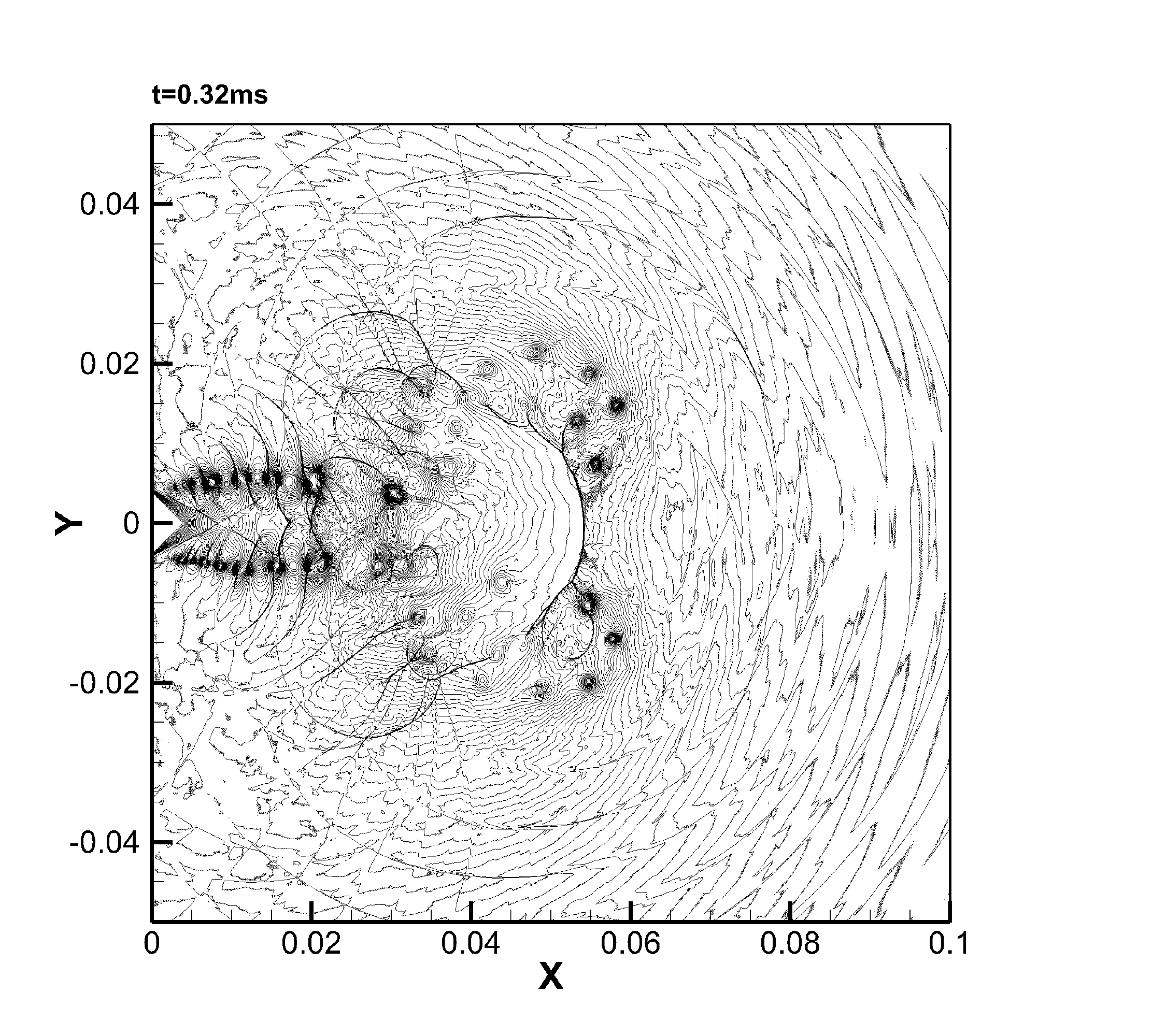}
\caption{\label{2d-jet-2} Planar jet: pressure perturbation contours of the supersonic jet obtained by compact 8th-order GKS with $1500\times2250$ mesh points at $t=0.17ms, t=0.24ms$ and $t=0.32ms$.}
\end{figure}

\begin{figure}[!htb]
\centering
\includegraphics[width=0.425\textwidth]{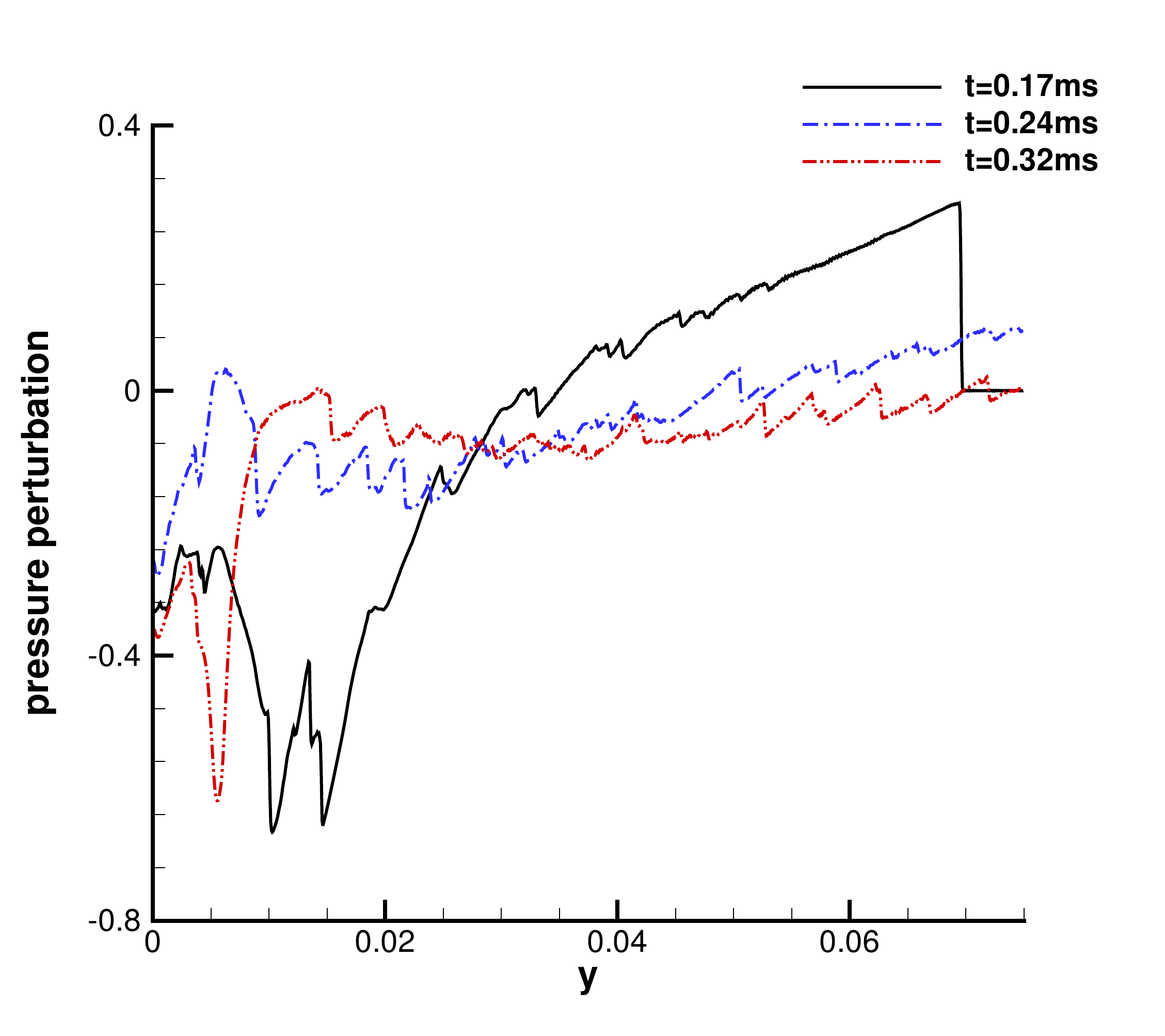}
\includegraphics[width=0.425\textwidth]{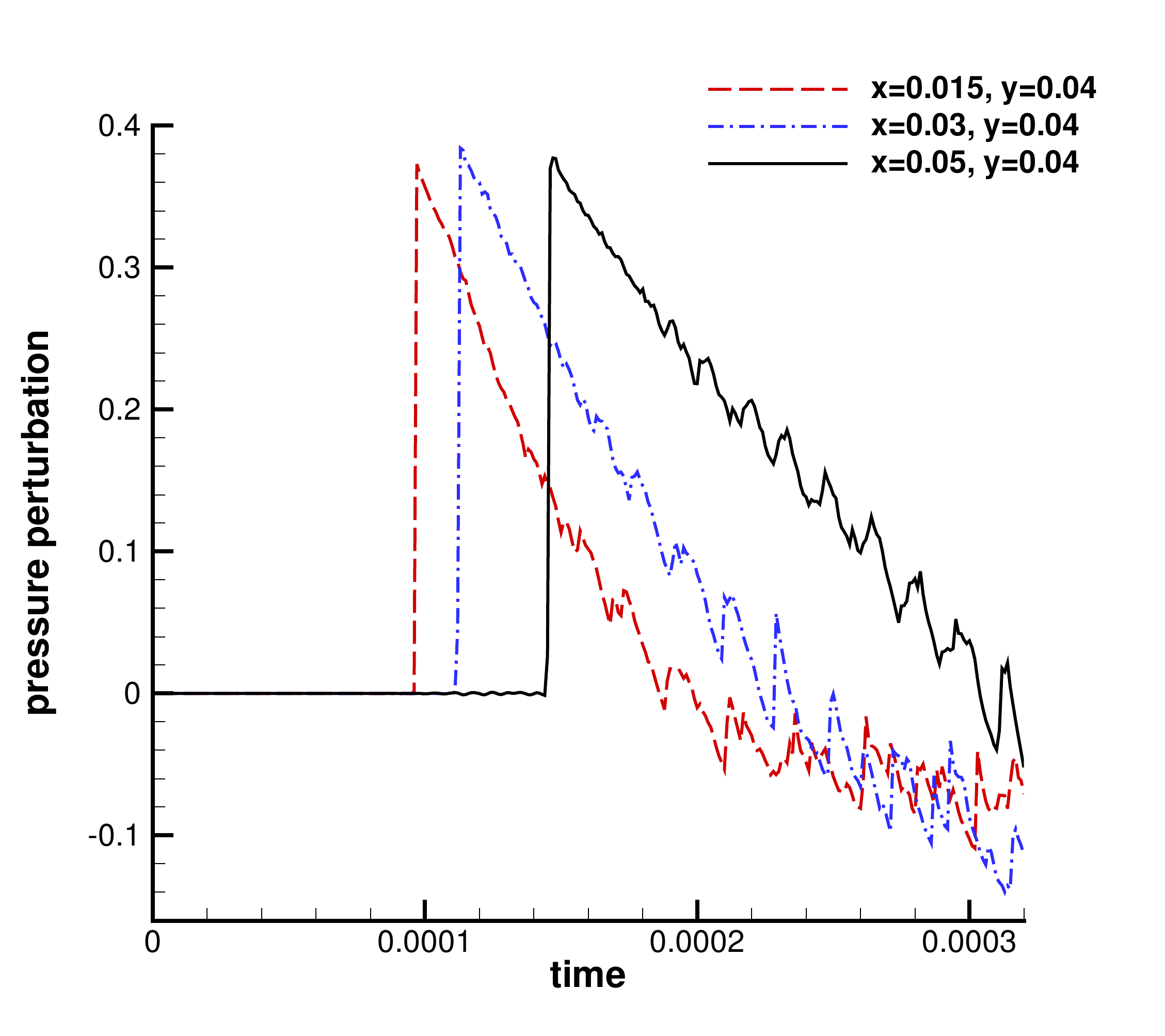}
\caption{\label{2d-jet-3} Planar jet: pressure perturbation of the supersonic jet obtained by compact 8th-order GKS with $1200\times1200$ mesh points. The left result is the pressure perturbation distribution along $x=0.015$ at $t=0.17ms, t=0.24ms$ and $t=0.32ms$. The right result is the pressure perturbation changes over time at $(0.015,0.04), (0.03,0.04)$ and $(0.05,0.04)$. }
\end{figure}

\subsection{Two-dimensional planar jet }
Supersonic jet flow is widely studied. A simplified 2-D planar jet, which was computed by Zhang et al. \cite{zhang2015jet}, is used here to test the compact high-order GKS. A Mach $1.4$ ($Ma_{jet}=1.4$) jet is injected through a width $L=0.01m$ entrance into a rectangular computational domain with the size $10L\times 15L$. The pressure of inlet flow is $p_{jet}=1.4 atm$. Initially the gas within the computational domain is static ($U=V=0$). The specific ratio, pressure and temperature are $\gamma=1.4$, $p_{0}=1.0 atm$ and $T_{0}=300K$, respectively. The Reynolds number is set as $Re_{jet}=U_{jet} L/\nu=2.8\times10^5$, and the dynamic viscosity coefficient $\nu=1.73\times 10^{-5}$.
A laminar boundary layer is set for the inlet flow, and the profile of velocity, density and pressure of inlet flow \cite{zhao2000jet} is
\begin{align*}
U(y)&=0.5U_{jet}[1-\tanh(\frac{b}{L}(y-L/2+2\delta))],\\
\rho(y)&=0.5\rho_{jet}[1-\tanh(\frac{b}{L}(y-L/2+2\delta))],\\
p(y)&=0.5p_{jet}[1-\tanh(\frac{b}{L}(y-L/2+2\delta))], 0\leq y \leq L/2,
\end{align*}
where $b$ is a nondimensional parameter, $\delta$ is the vorticity thickness of the inlet profile. There is a symmetric inlet flow for $-L/2\leq y \leq0$. The other regions ($-L/2 > y$ and $y > L/2$) on the left boundary of the computational domain are rigid walls.
Define the vorticity thickness of boundary layer by the profile function as
\begin{align*}
\delta=\frac{U_{jet}}{|(\partial U(y)/\partial y)_{max}|}.
\end{align*}
Thus, there is
\begin{align*}
\frac{\delta}{L}=\frac{2}{b}.
\end{align*}
In this test, $\delta$ is $4\times10^{-4}$. The free-boundary condition is applied to the boundaries AB, BC and CD, and the adiabatic viscous wall boundary condition is used for the boundaries AE and FD. A mesh with $1500\times2250$ points is used in the current computation.

The schlieren image and pressure perturbation contours at $t=0.17ms,t=0.24ms$ and $t=0.32ms$ obtained by compact 8th-order GKS are given in Fig. \ref{2d-jet-1} and Fig. \ref{2d-jet-2} respectively. The pressure perturbation is defined as $\triangle p=(p-p_{0})/p_{0}$.
The shear layer, vortices, and shock waves are developed. There are three kinds of shock waves which are the precursor shock wave, the secondary shock wave, and the vortex-induced shock wave. The precursor shock wave diffracts at the corner of the entrance, and the shock diffraction causes the misalignment of the pressure and density gradients. As a result, the vortex rings like a mushroom rolls up. The vortices in the jet flow are generated by the shear layer instability and the interaction between secondary shock wave and main vortex rings.
A lot of jumps are formed from the supersonic flow region and radiate outward. In Fig. \ref{2d-jet-2}, the waves with high wavenumbers after the precursor shock wave are generated by the interaction between the shock wave and vortices.
In the current computation, The large-scale structure of the jet flow is symmetric with respect to $y=0$.
Pressure perturbation distributions along $x=0.015$ at $t=0.17ms, t=0.24ms$ and $t=0.32ms$ are given in Fig. \ref{2d-jet-3}. With the development of shear layer instability and formation of vortex-induced shock waves, more pressure perturbations with significant amplitude appear along $x=0.015$.
The pressure perturbations over time at locations $(0.015,0.04), (0.03,0.04)$ and $(0.05,0.04)$ are given in Fig. \ref{2d-jet-3}. The results demonstrate the characteristics of waves in the supersonic jet flow. Low-frequency waves dominated at $(0.03,0.04)$, while higher-frequency waves exist at $(0.015,0.04)$ and $(0.05,0.04)$.

\section{Conclusion}
In this paper, the compact 8th-order GKS is presented and used in the acoustic wave simulation.
Based on the time-accurate gas evolution model at a cell interface, both cell averages and cell averaged slopes can be updated in GKS.
As a result, a compact stencil of a second-order scheme can be used to get a 8th-order linear and nonlinear spatial reconstructions.
The dispersion analysis for spatial discretizaition demonstrates that the compact scheme has a spectral-like resolution.
The GKS unifies the nonlinear and linear reconstructions through a relaxation process from the initial non-equilibrium state to the
final equilibrium one in the flow evolution around a cell interface.
For the linear acoustic wave, the linear reconstruction of the equilibrium state will contribute mainly for its evolution.
For the nonlinear shock, the nonlinear reconstruction for the initial non-equilibriums state will provide the numerical dissipation needed for the
shock capturing. As a result, the compact 8th-order GKS can capture discontinuous shock without generating numerical oscillation
and maintain high-order accuracy for smooth acoustic wave.
The compact GKS is particularly suitable for flow simulations with shock and acoustics wave interactions.
Due to the high-order evolution model beyond the first-order Riemann solution, the compact GKS has great advantages in capturing waves with a large wavenumber under a large CFL number.
Numerical results demonstrate that the 8th-order compact GKS provides the state-of-art solutions in comparison with the
existing high-order schemes targeting on both the shock and acoustic wave simulations.
The construction of high-order compact GKS on unstructured mesh and apply it to flow simulation with complex geometry is
under investigation.

\section*{Acknowledgements}
The current research is supported by National Numerical Windtunnel project, Hong Kong research grant council 16206617, and  National Science Foundation of China 11772281, 91852114, and 11701038.

\section*{References}

\bibliographystyle{ieeetr}
\bibliography{AIAbib}

\end{document}